\documentclass[aps,prx,reprint,twocolumn,amsmath,amssymb,amsfonts,footinbib,superscriptaddress,longbibliography,notitlepage]{revtex4-2}
\usepackage[table]{xcolor}
\usepackage{graphicx}

\usepackage{booktabs}
\usepackage{newtxtext,newtxmath}
\usepackage{braket,chemformula}
\usepackage{tcolorbox}
\usepackage{array}    

\usepackage{mathrsfs,wasysym}
\usepackage{tikz}
\usepackage{makecell}

\usepackage{tabularx,ragged2e}
\usepackage{multirow}
\newcolumntype{L}[1]{>{\hsize=#1\hsize\raggedright\arraybackslash}X}%
\newcolumntype{R}[1]{>{\hsize=#1\hsize\raggedleft\arraybackslash}X}%
\newcolumntype{C}[1]{>{\hsize=#1\hsize\centering\arraybackslash}X}%

\usepackage{comment}
\usepackage{longtable}
\usepackage[export]{adjustbox}
\usepackage{utfsym}
\usepackage{hyperref}

\newcommand{\newcheckmark}{\usym{2713}}
\newcommand{\newcrossmark}{\usym{2717}}

\DeclareMathAlphabet{\mathbbold}{U}{bbold}{m}{n}

\newcommand{\tsup}[1]{\textsuperscript{#1}}

\renewcommand{\vec}[1]{\boldsymbol{#1}}

\newcommand{\mat}[1]{\vec{#1}}
\newcommand{\trp}[1]{{#1}^{\intercal}}
\newcommand{\vhat}[1]{\vec{\hat{#1}}}

\newcommand{\uone}{{\rm U}(1)}

\newcommand{\othree}{{\rm O}(3)}
\newcommand{\sotwo}{{\rm SO}(2)}
\newcommand{\sothree}{{\rm SO}(3)}
\newcommand{\het}{\tsup{3}He}

\definecolor{clink}{RGB}{0,70,255}
\hypersetup{colorlinks=true,linkcolor=clink,citecolor=clink,urlcolor=clink}

\definecolor{cgray}{RGB}{150,150,150}

\definecolor{HColor}{rgb}{0,0.6,0.6}
\definecolor{JGRColor}{rgb}{0.6,0.2,0.0}
\definecolor{PMColor}{rgb}{0.0,0.2,0.6}
\definecolor{newCyan}{rgb}{0, 0.858824, 0.909804}

\begin{document}

\title{Landau Theory for Non-relativistic Magnetism Beyond Altermagnetism}

\author{Hana Schiff}
\email{hschiff@uci.edu}
\affiliation{Department of Physics and Astronomy, University of California, Irvine, California 92697, USA}

\author{Judit Romhányi}
\email{jromhany@uci.edu}
\affiliation{Department of Physics and Astronomy, University of California, Irvine, California 92697, USA}

\author{Paul McClarty}
\email{paul.mc-clarty@cnrs.fr}
\affiliation{Laboratoire Léon Brillouin, CEA, CNRS, Université Paris-Saclay, CEA Saclay, 91191 Gif-sur-Yvette, France.}

\author{Jeffrey G. Rau}
\email{jrau@uwindsor.ca}
\affiliation{Department of Physics, University of Windsor, 401 Sunset Avenue, Windsor, Ontario, N9B 3P4, Canada}

\begin{abstract}
In this work we study phenomenological Landau theories of non-relativistic magnetic systems, generalizing work on the collinear case to coplanar and non-coplanar orders. Focusing on zero-wavevector ($\vec{k}=0$) magnetic orders
described by a single irreducible representation of the crystallographic point group, we construct the symmetry-allowed free energies to quartic order for all
possible multi-dimensional irreducible ordering channels. We find distinct
Landau theories for tetragonal, trigonal/hexagonal, and cubic symmetry. Similar
to the case of spin-nematic fluids, we find two natural outcomes: a collapse to
a lower-symmetry collinear phase or a genuinely coplanar or non-coplanar order
where the order parameters form a mutually orthogonal set in spin space --
reminiscent of the $A$ and $B$ phases in \het{}. We show in general that this
symmetry-broken phase realizes the spin space group associated with the spatial
representation of the order parameter sharing the same parent space group as the
paramagnetic phase. This provides a direct link between the Landau and spin
space group approaches. We calculate key observables using this Landau approach,
including net magnetization, spin conductivity, multipolar moments, and
piezomagnetism, connecting our results to constraints from the underlying spin
space group. Examples of coplanar and non-coplanar magnetic orders for each case
are worked out in detail. Finally, we discuss the relationship of the
lower-symmetry collinear phases to the usual collinear spin groups, as well as
connections to atomic altermagnetism, superfluid \het{} and electronic nematic
phases.
\end{abstract}

\date{\today}

\maketitle

\tableofcontents

\section{Introduction}
Magnetic order in solids exhibits a remarkable variety of structures, from
simple collinear ferromagnets and antiferromagnets, to coplanar spirals and
complex non-coplanar textures~\cite{coey2021magnetism}. Many of the properties
of these phases are dictated by their \emph{magnetic symmetry}: how the spatial
symmetries of the crystal can be combined with time-reversal symmetry to leave
the structure invariant. Just as crystalline symmetries have been codified into
the 230 space groups~\cite{bradley2009mathematical}, the classification of
magnetic symmetries has yielded 1,651 magnetic space
groups~\cite{litvin2008tables} that govern the possible symmetries of
magnetically ordered crystals.

Additional possibilities arise in the non-relativistic limit, where spin-orbit
and magnetostatic interactions are neglected.  In this limit, spin and spatial
degrees of freedom can transform separately, opening new possibilities for
operations that preserve the magnetic structure -- the so-called spin space
groups~\cite{Kitz1965,brinkman1966theory,Brinkman1966b,Litvin1974_SpinGroups,Litvin1977_SpinPoint,corticelli2022spin,Liu2022_SPG,Schiff2025_SPG,
xiao2024ssg,Chen2024_SSG_Reps,jiang2024ssg}. Classification of the spin space
groups was completed recently, yielding 1,421 collinear, 9,542 coplanar and
56,512 non-coplanar distinct symmetry
groups~\cite{xiao2024ssg,Chen2024_SSG_Reps,jiang2024ssg}. 

Though relativistic effects are always present to some degree in materials, the
non-relativistic limit has proven useful for understanding magnets with weak
spin-orbit coupling and without significant dipole-dipole interactions. An important class of these structures are the\emph{altermagnets}~\cite{smejkal2022emerging,smejkal2022beyond}. In the
non-relativistic limit, these correspond to magneto-crystalline structures whose
symmetries form a subset of the collinear spin spacegroups~\cite{smejkal2022beyond,Schiff2025_SPG,xiao2024ssg,Chen2024_SSG_Reps,jiang2024ssg}.
These spin space groups determine the symmetry-allowed observables and response
functions in the ordered phase, directly constraining microscopic descriptions.
In real altermagnetic materials, the leading-order magnetic energy scale has
these enhanced symmetries, which lead to a spin splitting of magnonic and
electronic band structures. This feature and other related properties have attracted attention as they have potential applications inspintronics~\cite{dal2024antiferromagnetic,song2025altermagnets,jungwirth2025altermagnetic}.

A complementary perspective to that of spin groups is furnished by Landau theory~\cite{landau1936theory,landau2013statistical,toledano1987landau}. Instead
of classifying the symmetries of the ordered phase directly, we consider
possible order parameters defined by an irreducible representation (irrep) of
the paramagnetic spin space group~\cite{toledano1987landau}. The symmetry of the
order parameter and its direction in spin and representation space then
determine the structure of the ordered phase.

Non-relativistic Landau theories that incorporate separate spin-rotation
symmetries have proven useful in classifying and understanding the properties of
collinear altermagnets~\cite{mcclarty2024landau,schiff2025collinear}, even
allowing for predictions of the weakly relativistic limit. They can directly
connect the web of characteristic features of altermagnetic materials --
including allowed response functions, transport coefficients, and secondary
multipolar order parameters -- to the symmetry of the order parameter, the
N\'eel vector, $\vec{N}$~\cite{mcclarty2024landau}. For this collinear case, one
can construct a direct map between the order parameter symmetry and the spin
space group of the ordered phase, tying the two approaches
together~\cite{schiff2025collinear}. However, for generalizations of
altermagnets, for example, to coplanar or non-coplanar structures, this remains
less well understood~\cite{hu2025spin}. While some specific cases relevant for
individual materials or for a broad class of odd-parity
magnets~\cite{hellenes2023p,brekke2024minimal} have been
explored~\cite{yu2025odd,luo2025spin}, it is also important to explore magnetic
structures in their entirety through the unifying perspective of Landau theory.
As explained below, this approach brings tremendous explanatory power.

In this paper, we develop phenomenological Landau theories of non-relativistic
magnetic systems, extending previous work on collinear
phases~\cite{mcclarty2024landau,schiff2025collinear} to encompass coplanar and
non-coplanar orders. Focusing on zero-wavevector ($\vec{k}=0$) structures, we
consider magnetic ordering driven by a single irreducible representation of the
crystallographic point group and construct the corresponding symmetry-allowed
Landau free energies to quartic order. For coplanar systems, the order parameter
is a pair of spin vectors $\vec{N}_1$, $\vec{N}_2$ that transform as a
two-dimensional irrep. For non-coplanar orders, three spin vectors $\vec{N}_1$,
$\vec{N}_2$, $\vec{N}_3$ transform as a three-dimensional irrep of the point
group. 

We find that there are two distinct quartic Landau theories corresponding to coplanar
orders with tetragonal symmetry or trigonal, hexagonal, or cubic symmetry and
one Landau theory corresponding to non-coplanar orders with cubic symmetry. No
other crystal systems admit multi-dimensional irreps, so within a single
irreducible ordering channel they cannot realize the coplanar or non-coplanar
structures considered here~\footnote{More general non-collinear structures can
arise from the simultaneous condensation of several irreducible representations.
We focus primarily on irreducible ordering channels, returning to reducible
order parameters and the additional spin groups they realize in
Appendix~\ref{sec:perp_reducible}.}. We map out the phase diagrams of these
Landau theories and show that there are two natural classes of order: those
where the $\vec{N}_a$ are all collinear, and those where the $\vec{N}_a$ are all
mutually orthogonal and of equal length.  
To illustrate these constructions, we provide detailed examples for each of the
three classes: two-dimensional irreps on the kagom\'e and square plaquette
lattices, and a three-dimensional irrep on the pyrochlore lattice. 

This phase structure mirrors that found in nematic and spin-nematicfluids~\cite{wu2004dynamic,wu2007fermi,fradkin2010nematic,hannappel2016}, with
the collinear phase being analogous to the $A$-phase of \het{} and the
orthogonal phase being analogous to the
$B$-phase~\cite{vollhardt2013superfluid}. We further show that these orthogonal
phases do not lower the symmetry of the parent space group, and that their
corresponding spin space group can be directly constructed from the
representation of the order parameter~\cite{xiao2024ssg}.  These Landau theories
generalize those of nematic and spin-nematic orders and provide a concrete link
between the Landau theory formulated from the paramagnetic phase and the
underlying spin space group of the ordered phase. In addition, these orthogonal
phases are directly related to so-called {\it regular magnetic
orders} (RMOs)~\cite{Messio2011_RMO}. It was observed in Ref.~\cite{Messio2011_RMO}
that these structures encompass many of the complex magnetic orders that appear
classically on lattices exhibiting geometric frustration and that they
correspond to extrema in the energy landscape for all couplings.

It is remarkable that so few Landau theories are sufficient to describe
magnetism in the non-relativistic limit. However, this simplification raises an
important question: how does this approach recover the very large number of spin
space groups? The answer to this question lies in the so-called {\it epikernels}
of the high-symmetry group. 

The paper is structured as follows. We begin, in Section~\ref{sec:LT}, with a
summary of the direct and inverse Landau problems and an introduction to
epikernels. Then, in turn, we introduce Landau theories for collinear, coplanar
and non-coplanar order in Section~\ref{sec:landau}, where we present the phase
diagrams of these Landau theories revealing collinear and orthogonal phases. In
Section~\ref{sec:ortho-proof}, we show that orthogonal phases with N\'{e}el
vectors of equal length correspond to magnetic structures that preserve the
space group of the paramagnetic phase up to spin rotations. These are the
regular magnetic orders. Section~\ref{sec:examples} gives detailed illustrations
of Landau theory methods as applied to magnetic structures on the kagom\'e
lattice, the square plaquette lattice and the pyrochlore lattice. For each
example, we provide explicit structures corresponding to the Landau theory
phases, parameterized by some set of angles.  For these examples, we compute the
epikernels associated with the principal collinear and orthogonal
order-parameter families, including configurations beyond the generic minima of
the quartic Landau theories. This illustrates how many distinct spin space
groups can emerge from a small number of Landau theories.

Having characterized magnetic structures from Landau theory, we examine their
observable signatures from a symmetry perspective
(Section~\ref{sec:observables}). We consider magnetization, spin transport, the
magnetoelectric effect, piezomagnetism, and multipolar order. Extension to other
observables is straightforward using the methods discussed here. Finally, in
Section~\ref{sec:discussion} we connect the Landau theory methods to other
topics. For example, we show that the insights of this approach map naturally to
cases where magnetic sites are associated with a complex pattern of spin
density, including atomic altermagnetism. We also elaborate on various
interesting connections to superfluid \het{} and electronic spin nematic phases.

There are extensive appendices where we explain, in detail, why we can formulate
Landau theory based merely on irreps of the parent space group rather than
having to worry about time reversal and spin rotation symmetries also present in
the paramagnetic phase. We give tables of epikernels on the pyrochlore
lattice. We explain how to build spin point groups unique up to conjugation. An
appendix is dedicated to explaining the symmetries of spin conductivity.
Appendix~\ref{app:rep_decomps} is a reference for identifying symmetry-allowed
responses in non-relativistic magnets. Appendix~\ref{app:schur} goes into more
detail about the connections between orthogonal phases, regular magnetic orders
and spin groups. 

\section{Direct and inverse Landau problems}\label{sec:LT}

Landau theory is a general phenomenological framework for understanding phase
transitions accompanied by a change in symmetry \cite{landau1936theory}. The
transition is characterized by an \emph{order parameter} that is zero in the
symmetric phase and nonzero in the ordered phase, where only a subset of the
original symmetries survive. More precisely, a Landau theory is defined by a
high-symmetry group $\mathcal{G}$, a low-symmetry group $\mathcal{H}$, and an
order parameter transforming as a non-trivial irrep $\Phi$ of $\mathcal{G}$ but
trivially under $\mathcal{H}$~\footnote{For our purposes, where we are
interested in intrinsically real quantities (physical observables), we can
formulate all our representation theory in terms of real numbers. For a more
detailed discussion of this decomplexification, see
Appendix~\ref{subapp:decomplex}.}.

It is useful to distinguish two cases of interest: the \emph{direct} and
\emph{inverse} Landau problems. The direct Landau problem asks which
$\mathcal{H}$ are possible given $\mathcal{G}$ and an order parameter
transforming as $\Phi$. The inverse Landau problem asks which irreps $\Phi$ (and
corresponding order parameters) could facilitate a transition from $\mathcal{G}$
to $\mathcal{H}$~\cite{Ascher1977_LProb,
Toledano1976_OParamSymm1,StokesHatch1989_230SG,
landau2013statistical,Birman1966_LT,Goldrich1968_LT,Jaric1977_LT,Hatch1981_LT,Hatch1984_LTIso}.

When $\Phi$ is one-dimensional the two problems are identical: $\mathcal{H}$ is
determined by $\mathcal{G}$ and $\Phi$, and similarly, $\Phi$ is fixed by
$\mathcal{G}$ and $\mathcal{H}$. However, when $\Phi$ is multi-dimensional the
direct and inverse problems do not necessarily have unique solutions. In the
direct Landau problem several $\mathcal{H}$ can be consistent with a given
$\mathcal{G}$ and $\Phi$, depending on the orientation of the order parameter in
this higher-dimensional space. Similarly, the inverse problem does not
necessarily yield a unique irrep $\Phi$ for a transition from $\mathcal{G}$ to
$\mathcal{H}$. This implies that in the case of higher-dimensional irreps more
information than simply $\Phi$, $\mathcal{G}$ and $\mathcal{H}$ is needed to
determine properties of the ordered phase. The direct and inverse problems are
nonetheless equivalent: when all possibilities are considered, both yield the
same set of Landau theories.

As we are interested in magnetism, we start from the paramagnetic phase and our
first goal is to carry out the \emph{direct} Landau approach to understand all
dipolar magnetic orders that arise in the non-relativistic limit. The
high-symmetry group, $\mathcal{S}$, of the paramagnetic phase includes all elements
of the crystallographic space group, $S$, as well as full spin-rotation symmetry
$\sothree$, and time-reversal,
$T$~\cite{Litvin1974_SpinGroups,Litvin1977_SpinPoint,
Liu2022_SPG,Schiff2025_SPG,jiang2024ssg,xiao2024ssg},
\begin{equation}\label{eq:PMGroup}
\mathcal{S} = \othree \times S,
\end{equation}
where $\othree = \sothree\times\{1,T\}$ identifies the improper rotations as
including time-reversal. The irreps of $\mathcal{S}$ determine the possible
order parameters of the low-symmetry phase.  Each can be written in terms of an
irrep of the space group part and the spin part (which naturally carries an
angular momentum label). Schematically, we write such a representation as
$$
\Phi = \Gamma^{(\ell)}_{S} \otimes \Gamma,
$$
where $\Gamma^{(\ell)}_{S}$ is the rank-$\ell$ angular momentum irrep of
$\othree$ and $\Gamma$ is some irrep of the space group $S$.

For ordering of spins into a conventional magnetic order we have $\ell=1$ and
only need to consider the vector representation $\Gamma^{(1)}_{S}$ of $\othree$.
Thus the Landau theory is fully defined through the space group irrep $\Gamma$;
in what follows we will leave the $\othree$ implicit. We will further consider
magnetic orders that do not enlarge the unit cell, so we seek irreps of the
space group $S$ with $\vec{k}=0$, which correspond to irreps of the point group
$G$ associated with the space group. 

For a point group $G$, the irrep $\Gamma$ can be one-, two-, or
three-dimensional. We treat each case in turn. 

For a one-dimensional irrep, we can write a single vector order parameter
$\vec{N}$ transforming as $\Phi$. Under spin rotations $\vec{N}$ rotates like a
vector, while under point group operations it acquires a sign fixed by the
one-dimensional irrep $\Gamma$. Explicitly, for a spin space operation
$[\mat{R}\Vert g]$ where $g \in G$ we have
$$
\vec{N} \xrightarrow{[\mat{R}\Vert g]} \Gamma(g) \mat{R}\vec{N},
$$
with $\Gamma(g) = \pm 1$. Here, knowing $\Gamma$ determines the spin space group
of the ordered phase, as shown in Ref.~\cite{schiff2025collinear}, independent
of the direction $\vhat{N}$.

For a two-dimensional irrep $\Gamma$, we write a pair of vector order parameters
$\vec{N}_1$, $\vec{N}_2$ that transform as $\Phi$. Under spin rotations, each of
$\vec{N}_1$, $\vec{N}_2$ rotates as a vector, while under point group operations
they form linear combinations. For a spin space operation $[\mat{R}\Vert g]$
where $g \in G$ this is
$$
\begin{pmatrix} \vec{N}_1 \\ \vec{N}_2 \end{pmatrix} \xrightarrow{[\mat{R}\Vert g]} 
\begin{pmatrix}
    \Gamma_{11}(g) & \Gamma_{12}(g) \\
    \Gamma_{21}(g) & \Gamma_{22}(g)
\end{pmatrix}
\begin{pmatrix} \mat{R}\vec{N}_1 \\  \mat{R}\vec{N}_2 \end{pmatrix},
$$
with $\mat{\Gamma}(g)$ being a two-by-two matrix.
Similarly, for a three-dimensional irrep $\Gamma$, we write a triad
of vector order parameters $\vec{N}_1$, $\vec{N}_2$, $\vec{N}_3$ that transform
as $\Phi$. Under spin rotations, each of $\vec{N}_1$, $\vec{N}_2$, $\vec{N}_3$ rotates as a vector, while under point group operations they form linear combinations.
For a spin space operation $[\mat{R}\Vert g]$ where $g \in G$ 
this is
$$
\begin{pmatrix} \vec{N}_1 \\ \vec{N}_2 \\ \vec{N}_3 \end{pmatrix} \xrightarrow{[\mat{R}\Vert g]} 
\mat{\Gamma}(g)
\begin{pmatrix} \mat{R}\vec{N}_1 \\  \mat{R}\vec{N}_2 \\ \mat{R}\vec{N}_3 \end{pmatrix},
$$
with $\mat{\Gamma}(g)$ being a three-by-three matrix that does not
act on the spin indices. Unlike the one-dimensional case, in the two- and three-dimensional cases, $\Gamma$ does not determine the spin space group uniquely. The spin group depends on the relative orientations of the vectors $\vec{N}_a$.

\subsection{Epikernels}\label{subsec:epikernels}
For higher-dimensional irreps, the spin space group of the ordered phase depends on the relative orientation of the different $\vec{N}_a$ vectors in the order parameter space. Our goal here is to determine all relative orientations that yield
a non-trivial spin space group in the ordered phase.
Historically, the question of finding such possible low-symmetry groups is referred to as searching for the \emph{isotropy subgroups} \cite{Toledano1976_OParamSymm1, Toledano1977_OParSymm2, Toledano1980_FerroElIso, Toledano1982_NonFerroIso, toledano1987landau, StokesHatch1989_230SG,StokesHatch2011_SGIso, Stokes2017_IsoSuperspace, Petricek2010_SGRepFriends} or the \emph{epikernels} \cite{Ascher1977_LProb,Ascher1977_Epikernel,Jaric1982_GT&PT,DamnjanovicSymm} of the high symmetry group.

A given orientation of the $\vec{N}_a$ may still preserve
some symmetries of the paramagnetic group, forming a subgroup $\mathcal{H}$ of $\mathcal{G} \equiv \othree \times G$.
The orientations that are compatible with a given subgroup satisfy
\begin{equation}
\label{eq:epikernel}
\vec{N}_a \xrightarrow{[\mat{R}_h\Vert h]} \sum_{a'} \Gamma_{aa'}(h) \mat{R}_h\vec{N}_{a'} = \vec{N}_a,
\end{equation}
for each $[\mat{R}_h\Vert h] \in \mathcal{H}$. The problem is thus to
determine the stabilizing subgroup for each possible $\vec{N}_a$.
Practically, we solve Eq.~(\ref{eq:epikernel}) for each element
of the paramagnetic group $\mathcal{G}$ and determine orientations
of the $\vec{N}_a$ consistent with that element. Note that the allowed spin rotations $\mat{R}_h$ are highly constrained by the structure of Eq.~(\ref{eq:epikernel}). Subgroups can then be identified via common orientations of the $\vec{N}_a$ in each solution, and we implement this program for the examples with two- and three-dimensional irreps in Sec.~\ref{sec:examples}. Alternatively, the same subgroups can be determined
by finding subgroups $\mathcal{H}$ where the restriction of $\Phi$ to the elements of $\mathcal{H}$ contains the trivial irrep, which we demonstrate on the kagom\'e example.

\section{Multi-dimensional Landau theories}
\label{sec:landau}
Having introduced the Landau framework, we now discuss the possible Landau theories for the higher-dimensional representations. We find that there are only four distinct symmetry-allowed forms: one for one-dimensional irreps, two for two-dimensional irreps, and one for three-dimensional irreps. We focus on the free energy, an expansion in low-order polynomials of $\vec{N}_a$, as is appropriate near a continuous phase transition.

\subsection{One-dimensional irreps}
The case of a one-dimensional irrep is particularly simple, as spin-rotation and time-reversal symmetry force the free energy, $F$, to be a function of only $\vec{N}\cdot\vec{N}$. Explicitly,
$$
F(\vec{N}) \equiv f(|\vec{N}|^2) = a_2 |\vec{N}|^2 + a_4 |\vec{N}|^4 +\dots.
$$
For this to be bounded from below at quartic order we require $a_4>0$. When $a_2>0$ there is only the trivial solution $\vec{N}=0$, while for $a_2<0$ we have a continuous family of
non-trivial solutions with $|\vec{N}| = \sqrt{|a_2|/(2a_4)}$. These describe collinear phases, with a continuous spin rotation symmetry about the order parameter direction $\vhat{N}$.

\subsection{A remark on crystal systems}
\label{sec:crystalsystems}

Crystals can be separated into seven crystal systems associated with their point group symmetry. These are triclinic, monoclinic, orthorhombic, trigonal, hexagonal, tetragonal, and cubic. One-dimensional irreps $\Gamma$ appear in the point groups belonging to all of these crystal systems. However, 2D and 3D irreps appear only in crystal types of higher symmetry. In particular, one finds 2D irreps may arise only in trigonal, tetragonal, hexagonal and cubic crystals while 3D irreps appear only in cubic crystals. 

\subsection{Two-dimensional irreps}
\label{sec:landau2}
The case of a two-dimensional irrep is more interesting, as there are multiple
distinct terms allowed at fourth order in $\vec{N}_1$, $\vec{N}_2$. As in the collinear case, time-reversal and spin-rotation impose strong constraints on the
free energy, forcing it to be a function of only $|\vec{N}_1|^2$, $|\vec{N}_2|^2$
and $\vec{N}_1\cdot \vec{N}_2$. Explicitly,
$$
F(\vec{N}_1,\vec{N}_2) \equiv f(|\vec{N}_1|^2,|\vec{N}_2|^2,\vec{N}_1\cdot\vec{N}_2).
$$
Within this class, we distinguish two cases: when the point group is tetragonal and when the point group is trigonal, hexagonal or cubic, as the allowed quartic terms differ.

Begin with the tetragonal case. The allowed terms can be classified using the
transformation properties of the three $\othree$ invariant quantities
$|\vec{N}_1|^2$, $|\vec{N}_2|^2$ and $\vec{N}_1\cdot \vec{N}_2$ which transform
as the symmetrized tensor square of $\Gamma$, $[\Gamma^2]$. 

For the non-cyclic tetragonal point groups, the symmetric square of an
$E$-type irrep contains one trivial irrep and two inequivalent non-trivial
one-dimensional irreps. For example, in $422$ ($D_4$), $[E^2]=A_1+B_1+B_2$. This implies a single allowed quadratic term, as there is only one scalar $A_{1}$. Then,
$\big([\Gamma^2]\big)^2$ yields three allowed quartic terms, one from the square of each one-dimensional irrep. Explicitly, the free energy can be written as
\begin{align}
\nonumber
F_{4}(\vec{N}_1,\vec{N}_2) &= 
a_2(|\vec{N}_1|^2 +|\vec{N}_2|^2)+a_4(|\vec{N}_1|^2 +|\vec{N}_2|^2)^2\\
&+b_4|\vec{N}_1|^2|\vec{N}_2|^2 + c_4 \left(\vec{N}_1\cdot\vec{N}_2\right)^2.
\label{eq:free-tet}
\end{align}

The cyclic tetragonal groups $4$, $\overline{4}$, and $4/m$ require a
qualification as in these groups $[E^2] \sim A + 2B$. The two $B$ components $|\vec{N}_{1}|^{2} - |\vec{N}_{2}|^{2}$ and $\vec{N}_{1}\cdot\vec{N}_{2}$ thus transform identically, and so their cross-term is symmetry-allowed. However, this term can be removed by a rotated choice of basis~\footnote{This is always possible because a rotation commutes with all two-dimensional $\Gamma(g)$ in these groups, as they are themselves rotations.}.
Thus, for any fixed set of coefficients,
the free energy can be brought to the canonical form of
Eq.~(\ref{eq:free-tet}). This does not reduce the number of independent quartic invariants in a fixed basis, but uses the freedom to choose the irrep basis after the coefficients are specified.

We will focus on the case of negative $a_2$, positive $a_4$ where there are non-trivial solutions for the $\vec{N}_1$, $\vec{N}_2$ vectors.
Note that when $b_4=c_4=0$, this free energy enjoys an enlarged ${\rm O}(6)$ symmetry as it only depends on $|\vec{N}_1|^2 +|\vec{N}_2|^2$. In this limit $|\vec{N}_1|^2 +|\vec{N}_2|^2=|a_2|/(2a_4)$, as in the one-dimensional case, and the direction in the higher-dimensional space (spin and representation index) is undetermined.

The qualitative structure of the phase diagram can be understood by perturbing
away from this limit, taking $b_4, c_4 \ll |a_2|, a_4$. Positive $b_4$ favors
one of $\vec{N}_1$ or $\vec{N}_2$ being zero, while negative $b_4$ prefers
finite $\vec{N}_1$ and $\vec{N}_2$ with equal magnitude. Positive $c_4$ favors
$\vec{N}_1 \perp \vec{N}_2$, while negative $c_4$ selects $\vec{N}_1 \|
\vec{N}_2$~\footnote{An antiparallel configuration is degenerate with the
parallel; however, these states are related by the tetragonal symmetries and
thus do not yield a new phase.}. 

For the non-cyclic tetragonal groups, we thus have three natural phases: a phase where one of $\vec{N}_1$ or $\vec{N}_2$ vanishes (we call this the ``single-$N$'' phase), and two phases where both are non-zero and of equal magnitude, one with $\vec{N}_1\|\vec{N}_2$ (the ``parallel'' phase) and one with $\vec{N}_1 \perp \vec{N}_2$ (the ``orthogonal'' phase). The single-$N$ and parallel phases are both collinear (they have a continuous $\uone$ symmetry about their common $\vhat{N}$ direction), while the orthogonal phase is coplanar (it only has a $2_{\perp} T$ spin-only symmetry, where the $2_{\perp}$ axis is along $\vhat{N}_1\times\vhat{N}_2$). 

For $4$, $\overline{4}$, and $4/m$, the same canonical free-energy form
may be used after a basis rotation, but the distinction between the
single-$N$ and parallel states is basis dependent. These should therefore
not be regarded as two symmetry-distinct collinear phases in the cyclic
tetragonal groups. Instead, the generic collinear minimum has an
orientation in the two-dimensional representation space that can vary
continuously with the quartic coefficients~\footnote{
Indeed, writing a collinear configuration as
$\vec{N}_1=N\cos\alpha \vhat{N}$, $\vec{N}_2=N\sin\alpha \vhat{N}$,
the additional invariant $\delta_4(|\vec{N}_1|^2-|\vec{N}_2|^2)(\vec{N}_1\cdot\vec{N}_2)$ gives
$$
F_4(\alpha)
=
{\rm const.}+
\frac{N^4}{8}\left[-(b_4+c_4) \cos4\alpha
+2\delta_4 \sin4\alpha\right],
$$
so that the preferred angle $\alpha$ is generically coefficient dependent.}.

\begin{figure}[tb!]
    \centering
    \includegraphics[width=\columnwidth]{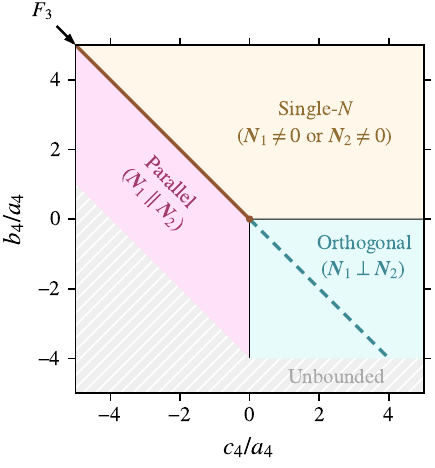}
    \caption{
    Phase diagram for the Landau theories with an order parameter transforming as a two-dimensional irrep with
    $a_2<0$ and $a_4>0$. 
    For the non-cyclic tetragonal point groups there are two independent
    parameters $b_4$ and $c_4$ and three distinct phases: the single-$N$ phase (only one of $\vec{N}_1$, $\vec{N}_2$ non-zero), the parallel phase ($\vec{N}_1\|\vec{N}_2$) and the orthogonal phase ($\vec{N}_1\perp \vec{N}_2$).
    For the cyclic tetragonal groups $4$, $\overline{4}$, and $4/m$, the
    general quartic free energy contains an additional mixed invariant (see text).
    In the parallel and orthogonal phases $|\vec{N}_1|=|\vec{N}_2|$. For the trigonal, hexagonal or cubic point groups, symmetry forces $b_4=-c_4$, indicated by the diagonal line labeled $F_3$. There are two phases: a degenerate collinear phase and the orthogonal phase.
    }
    \label{fig:phase-diagram-2d}
\end{figure}

The phase diagram is shown in Fig.~\ref{fig:phase-diagram-2d}. The theory is stable
when $4a_4+b_4\geq 0$ and $4a_4+c_4+b_4\geq 0$. The three phases expected from
the limiting analysis appear; their energies are 
\begin{align*}
    F_{\text{single-}N} &\!=\! -\frac{a_2^2}{4a_4}, &
    F_{\|} &\!=\! -\frac{a_2^2}{4a_4+b_4+c_4}, &
    F_{\perp} &\!=\! -\frac{a_2^2}{4a_4+b_4},
\end{align*}
where $F_{\text{single-}N}$ is the free energy of the single-$N$ phase, $F_{\|}$ is the
free energy of the parallel phase
and $F_{\perp}$ is the
free energy of the orthogonal phase.
The phase boundaries can be determined when these free energies become equal.
We find that the orthogonal phase appears for $b_4<0$ and $c_4>0$, the parallel
phase when $c_4<0$ and $b_4<-c_4$ and the single-$N$ phase when $b_4>0$ and $b_4>-c_4$.

Next we consider point groups with trigonal, hexagonal or cubic symmetry. As in the
tetragonal case, we can classify the allowed terms using the $\othree$
invariant $|\vec{N}_1|^2$, $|\vec{N}_2|^2$, $\vec{N}_1\cdot \vec{N}_2$ variables.
For an $E$-type irrep in such a group the symmetric square
contains the trivial irrep and a two-dimensional irrep (e.g., in the point group $D_6$, $[E^2_2] = A_1 + E_2$). This implies a single allowed quadratic 
term and only two allowed quartic terms (formed from the trivial square and the square of the two-dimensional irrep). The free energy can then be written as
\begin{align}
\nonumber
F_{3}(\vec{N}_1,\vec{N}_2) &= 
a_2(|\vec{N}_1|^2 +|\vec{N}_2|^2)+a_4(|\vec{N}_1|^2 +|\vec{N}_2|^2)^2\\
&+b_4\left(|\vec{N}_1|^2|\vec{N}_2|^2-(\vec{N}_1\cdot\vec{N}_2)^2\right).
\label{eq:f3}
\end{align}
This is the same form as in the tetragonal case but with $c_4=-b_4$. The phase
diagram is thus simply a cut of that derived for Eq.~(\ref{eq:free-tet}) as
indicated by the diagonal line in Fig.~\ref{fig:phase-diagram-2d}.

This cut runs along the phase boundary of the single-$N$ and
parallel phases, rendering these degenerate when $b_4>0$. Along this line, the configurations
which minimize the free energy are of the form $\vec{N}_1  = (N \cos{\alpha}) \vhat{N}$ and $\vec{N}_2 = (N \sin{\alpha}) \vhat{N}$ with the angle $\alpha$ arbitrary. To break this accidental symmetry, terms at sixth order in the order parameter are required. The orthogonal phase that appears for $b_4<0$ does not qualitatively change with respect to the tetragonal case.

\subsection{Three-dimensional irreps}
\label{sec:landau3}
The case of three-dimensional irreps is similar to the two-dimensional case.
Time-reversal and spin-rotation impose the same constraints, forcing the free
energy to be a function of only the six $\othree$ invariant quantities
$|\vec{N}_1|^2$, $|\vec{N}_2|^2$, $|\vec{N}_3|^2$, $\vec{N}_1\!\cdot\!\vec{N}_2$, $\vec{N}_2\!\cdot\!\vec{N}_3$, and $\vec{N}_3\!\cdot\!\vec{N}_1$. The free energy can then be written
$$
F(\vec{N}_1,\vec{N}_2,\vec{N}_3) \!\equiv\! f(|\vec{N}_1|^2\!,|\vec{N}_2|^2\!, |\vec{N}_3|^2,\! \vec{N}_1\cdot\vec{N}_2, \vec{N}_2\cdot\vec{N}_3,\vec{N}_3\cdot\vec{N}_1).
$$
These invariants transform as $[\Gamma^2]$ which for all $T$-type irreps
contains the trivial irrep and a two- and three-dimensional irrep (e.g., in $O$ we have $[T_2^2] = A_1 + E +T_2$). This implies a single quadratic invariant (from the trivial irrep) and three quartic invariants formed by the squares of the one-, two- and three-dimensional irreps. This yields
\begin{align}\label{eq:3D_cub}
F &\!=\! a_2 \left( |\vec{N}_{1}|^{2} + |\vec{N}_{2}|^{2} + |\vec{N}_{3}|^{2}\right)
+a_4 \left(|\vec{N}_{1}|^{2} + |\vec{N}_{2}|^{2}  + |\vec{N}_{3}|^{2}\right)^{2} \notag\\
&+ b_4 \left(|\vec{N}_{1}|^{2}|\vec{N}_{2}|^{2} +|\vec{N}_{2}|^{2}|\vec{N}_{3}|^{2} + |\vec{N}_{3}|^{2}|\vec{N}_{1}|^{2}\right)\notag  \\
&+c_4 \,\left( (\vec{N}_{1}\cdot\vec{N}_{2})^{2} + (\vec{N}_{2}\cdot\vec{N}_{3})^{2}+ (\vec{N}_{3}\cdot\vec{N}_{1})^{2} \right).
\end{align}
Following the approach discussed for the two-dimensional tetragonal case, we find
a nearly identical set of phases: a single-$N$ phase where only one of $\vec{N}_1$, $\vec{N}_2$, $\vec{N}_3$ is non-zero, a parallel phase where $\vec{N}_1 \| \vec{N}_2 \| \vec{N}_3$, all of equal magnitude, and an orthogonal phase where $\vec{N}_1$, $\vec{N}_2$, $\vec{N}_3$ are mutually orthogonal and of equal length.
The single-$N$ and parallel phases are collinear while the orthogonal phase is non-coplanar with no remnant spin-only symmetries.

\begin{figure}[tb!]
    \centering
    \includegraphics[width=\columnwidth]{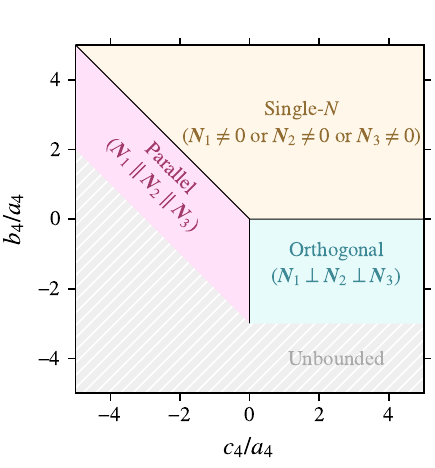}
    \caption{
    Phase diagram for the Landau theories with an order parameter transforming as a three-dimensional irrep with
    $a_2<0$ and $a_4>0$. 
    There are two independent
    parameters $b_4$ and $c_4$ and three distinct phases: the single-$N$ phase (only one of $\vec{N}_1$, $\vec{N}_2$, $\vec{N}_3$ non-zero), the parallel phase ($\vec{N}_1\|\vec{N}_2\|\vec{N}_3$) and the orthogonal phase ($\vec{N}_1\perp \vec{N}_2 \perp \vec{N}_3$). In the parallel and orthogonal phases the magnitudes of each vector are equal,  $|\vec{N}_1|=|\vec{N}_2|=|\vec{N}_3|$.
    }
    \label{fig:phase-diagram-3d}
\end{figure}

The phase diagram is shown in Fig.~\ref{fig:phase-diagram-3d}. The theory is stable
when $3a_4+b_4\geq 0$ and $3a_4+c_4+b_4\geq 0$. The three phases expected from
the limiting analysis appear; their energies are 
\begin{align*}
    F_{\text{single-}N} &\!=\! -\frac{a_2^2}{4a_4}, \\
    F_{\|} &\!=\! -\frac{3a_2^2}{4(3a_4+b_4+c_4)}, &
    F_{\perp} &\!=\! -\frac{3a_2^2}{4(3a_4+b_4)},
\end{align*}
where $F_{\text{single-}N}$ is the free energy of the single-$N$ phase, $F_{\|}$ is the
free energy of the parallel phase
and $F_{\perp}$ is the
free energy of the orthogonal phase.
Identical to the two-dimensional tetragonal case,
we find that the orthogonal phase appears for $b_4<0$ and $c_4>0$, the parallel
phase when $c_4<0$ and $b_4<-c_4$ and the single-$N$ phase when $b_4>0$ and $b_4>-c_4$.

\section{Orthogonal orders and maximal spin space symmetry}
\label{sec:ortho-proof}
A common feature of the Landau theories derived above is the appearance of a phase in which the vector order parameters have equal magnitude and
are mutually orthogonal,
\begin{equation}
    \vec{N}_a\cdot\vec{N}_b=N^2\delta_{ab}.
    \label{eq:orthogonal-condition}
\end{equation}
For a two-dimensional irrep this describes a coplanar state, while for a
three-dimensional irrep it describes a non-coplanar state. We now show
that orthogonal phases with all $\vert\vec{N}_a\vert$ equal preserve the full parent point group modulo spin rotations.

Consider an order parameter transforming under a real $d$-dimensional irrep
$\Gamma$ of the point group $G$, with $d\leq 3$. We collect the $d$ vector
order parameters into the $3\times d$ matrix
\begin{equation}
    \mat{\mathsf{N}}
    =
    \begin{pmatrix}
        | & & | \\
        \vec{N}_1 & \cdots & \vec{N}_d \\
        | & & |
    \end{pmatrix}.
    \label{eq:N-matrix}
\end{equation}
Under a spin space operation $[\mat{R}\Vert g]$, with $\mat{R}\in \othree$ and
$g\in G$, the order parameter transforms as
\begin{equation}
    \mat{\mathsf{N}}
    \longrightarrow
    \mat{R} \mat{\mathsf{N}} \trp{\mat{\Gamma}(g)}.
    \label{eq:N-matrix-transform}
\end{equation}
A spatial operation $g$ is therefore preserved by the ordered state, modulo
a spin rotation $\mat{R}_g$, when
\begin{equation}
    \mat{R}_g\mat{\mathsf{N}}
    =
    \mat{\mathsf{N}}\mat{\Gamma}(g).
    \label{eq:intertwiner}
\end{equation}
The consequences of Eq.~(\ref{eq:intertwiner}) are most easily seen by
introducing the Gram matrix $Q_{ab}\equiv \vec{N}_a\cdot\vec{N}_b$. In matrix form
\begin{equation}
    \mat{Q}
    \equiv
    \trp{\mat{\mathsf{N}}}\mat{\mathsf{N}}.
    \label{eq:gram-matrix}
\end{equation}
Since $\mat{R}_g$ is orthogonal, Eq.~(\ref{eq:intertwiner}) implies
\begin{align}\label{eq:Q_transformation}
    \mat{Q}
    = \trp{\mat{\mathsf{N}}}\trp{\mat{R}_g}\mat{R}_g\mat{\mathsf{N}}
    = \trp{\mat{\Gamma}(g)}\mat{Q}\mat{\Gamma}(g),
\end{align}
for every $g\in G$. This is equivalent to
\begin{equation}
    [\mat{Q},\mat{\Gamma}(g)]=0,
    \label{eq:Q-commutes}
\end{equation}
since $\mat{\Gamma}(g)$ is orthogonal.
Since the matrix $\mat{Q}$ is real and symmetric it can be diagonalized. From 
Eq.~(\ref{eq:Q-commutes}), each eigenspace of $\mat{Q}$ is
invariant under $\mat{\Gamma}$. Since $\mat{\Gamma}$ is irreducible, $\mat{Q}$ can have only
a single eigenvalue~\footnote{An irreducible representation does not preserve any non-zero proper
subspace, otherwise it would be reducible.}. Hence $\mat{Q}=N^2\mat{1}$ or equivalently
\begin{equation}
    \vec{N}_a\cdot\vec{N}_b=N^2\delta_{ab}.
    \label{eq:orthogonal-from-symmetry}
\end{equation}
Thus preservation of the full point group modulo spin rotations requires the
vector order parameters to have equal magnitude and to be mutually
orthogonal. This result can equivalently be viewed as an application of
Schur's lemma (see Appendix~\ref{app:schur}).

The converse also holds. Suppose the order parameter is in the orthogonal
phase of Eq.~(\ref{eq:orthogonal-condition}).
Collecting these vectors as the columns of a matrix gives
\begin{equation}
    \mat{\mathsf{N}}=N\mat{U},
\end{equation}
where $\trp{\mat{U}}\mat{U}=\mat{1}$.
For $d=2$, the columns of $\mat{U}$ span the plane of the two order
parameters, while for $d=3$ they form an orthonormal basis of spin space.

Now consider a spatial symmetry $g\in G$. The matrix
$\mat{\Gamma}(g)$ mixes the order parameters while preserving their inner
products. Since the vectors $\vhat{N}_a$ form an orthonormal set, the same
transformation can be realized as a spin rotation acting on the subspace
spanned by them. We therefore define $\mat{R}_g$ through
\begin{equation}
    \mat{R}_g\mat{U}
    =
    \mat{U}\mat{\Gamma}(g).
    \label{eq:R-def}
\end{equation}
For $d=3$, this determines the spin transformation uniquely,
giving $\mat{R}_g=\mat{U}\mat{\Gamma}(g)\trp{\mat{U}}$.
For $d=2$, Eq.~(\ref{eq:R-def}) determines the action of $\mat{R}_g$
within the plane spanned by $\vec{N}_1$ and $\vec{N}_2$; its action on the
normal direction can then be chosen so that $\mat{R}_g$ is an element of
$\othree$. Using that $\mat{\mathsf{N}}=N\mat{U}$, Eq.~(\ref{eq:R-def}) recovers
Eq.~(\ref{eq:intertwiner}).
Thus every spatial operation $g\in G$ can be compensated by a spin transformation $\mat{R}_g$, and the orthogonal phase preserves $G$ modulo spin rotations.

We therefore arrive at a direct equivalence: for an order parameter
transforming under a real irreducible representation, the full parent point group is preserved modulo spin rotations if and only if the order parameter vectors form an equal-length orthogonal set. These orthogonal Landau minima are natural realizations of {\it regular magnetic orders}~\cite{Messio2011_RMO} described in the introduction. They were found to be a useful class of variational states for classical Heisenberg models as they are extrema of the energy {\it and} frequently appear as ground states. 

Importantly, if the order parameters are not in this equal-length, orthogonal configuration, the symmetry of the resulting
phase must be lower than that of the crystal, even modulo spin rotations. Phases such as the collinear
parallel and single-$N$ phases have this character: the parent space group of their spin space group is only
a subgroup of the crystal's symmetry.

The spin space groups corresponding to these states can be constructed directly from the spatial representation $\Gamma$, following the framework of \citet{xiao2024ssg}, providing a natural generalization of the
same result for collinear orders~\cite{schiff2025collinear}. The lattice-specific examples considered below provide explicit realizations of this general result.

\section{Examples}
\label{sec:examples}
To illustrate this framework, we explicitly derive the relevant
Landau theories for three cases. Two involve two-dimensional irreps: a magnetic order on the kagom\'e
lattice where the symmetry is hexagonal and an order on a square plaquette lattice where the symmetry is tetragonal. 
Finally, we consider a case with a three-dimensional irrep on the pyrochlore lattice.

\subsection{Kagom\'e lattice}
\label{sec:kagome}
\begin{figure}
    \centering
    \includegraphics[width=\columnwidth]{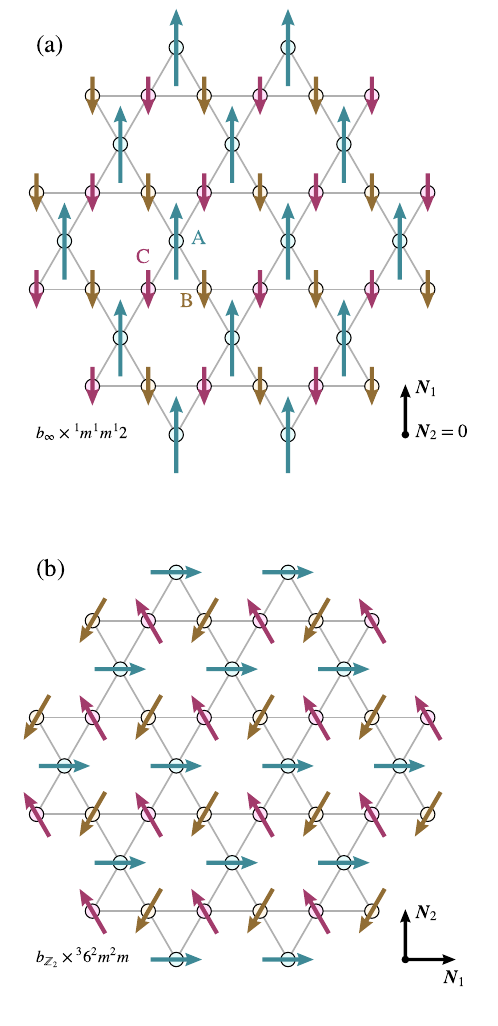}
    \caption{Illustration of two magnetic orders in the phase diagram of Fig.~\ref{fig:phase-diagram-2d} (on the $F_3$ line) on the kagom\'e lattice. (a) Collinear phase, corresponding to the single-$N$ phase with $\vec{N}_1=N\vhat{y}$ and $\vec{N}_2=0$. (b) Coplanar 120$^\circ$ phase, corresponding to the orthogonal phase
    with $\vec{N}_1=\sqrt{3/2}N\vhat{x}$, $\vec{N}_2=\sqrt{3/2}N\vhat{y}$.
    }
    \label{fig:kagome-orders}
    \vspace{-0.9cm}
\end{figure}

The unit cell of the kagom\'e lattice contains three atoms,
as illustrated in Fig.~\ref{fig:kagome-orders}. The point
group of the lattice is the hexagonal group $6mm$ ($C_{6v}$). We consider
a $\vec{k}=0$ magnetic order with moments $\vec{m}_A$, $\vec{m}_B$, $\vec{m}_C$ on these sublattices. 
Under the action of this group, the atoms permute, transforming as~\footnote{We can identify the two-dimensional part as $E_2$ due to the fact that the six-fold rotation fixes no atoms and thus has character zero.}
\begin{equation}\label{eq:kagome_decomposition}
\Gamma_{\rm perm} = A_1 + E_2.
\end{equation}
In terms of the magnetic moments, the $A_1$ component corresponds to a uniform ferromagnetic order, with order parameter
$\vec{M} = \vec{m}_A+\vec{m}_B+\vec{m}_C$.
The $E_2$ component corresponds to a coplanar compensated 
order parameter with
$$
\begin{pmatrix}
    \vec{N}_1 \\[0.5em]
    \vec{N}_2 
\end{pmatrix} = 
\begin{pmatrix}
    \frac{1}{\sqrt{6}}\left(2\vec{m}_A - \vec{m}_B - \vec{m}_C\right) \\[0.5em]
    \frac{1}{\sqrt{2}}\left(\vec{m}_C-\vec{m}_B\right)
\end{pmatrix}.
$$
Inverting this transformation, a finite $\vec{M}$ gives equal
$\vec{m}_A$, $\vec{m}_B$, $\vec{m}_C$. A finite $E_2$ order parameter instead yields
\begin{subequations}
\label{eq:n-to-m-kagome}
\begin{align}
    \vec{m}_A &= \sqrt{\frac{2}{3}} \vec{N}_1, \\
    \vec{m}_B &= -\frac{1}{\sqrt{6}} \vec{N}_1 - \frac{1}{\sqrt{2}}\vec{N}_2, \\
    \vec{m}_C &= -\frac{1}{\sqrt{6}} \vec{N}_1 + \frac{1}{\sqrt{2}}\vec{N}_2,
\end{align}
\end{subequations}
with zero total magnetization $\vec{m}_A+\vec{m}_B+\vec{m}_C=0$.

The Landau theory for such $E_2$ orders is described
in Sec.~\ref{sec:landau2} by the $F_3$ free energy given in Eq.~(\ref{eq:f3}). There are two natural phases, one degenerate phase where $\vec{N}_1$ and $\vec{N}_2$ share a direction, but not necessarily the same magnitude, and one orthogonal phase with $\vec{N}_1 \perp \vec{N}_2$. 
We illustrate the magnetic moment configurations for two example states from these phases in Fig.~\ref{fig:kagome-orders}, using the mapping given in Eq.~(\ref{eq:n-to-m-kagome}).

For the orthogonal phase we choose $\vec{N}_1=\sqrt{3/2}N\vhat{x}$, $\vec{N}_2=~\sqrt{3/2}N\vhat{y}$; the state is coplanar, with moments
\begin{subequations}
\begin{align}
    \vec{m}_A &= N \vhat{x}, \\
    \vec{m}_B &= N\left(-\frac{1}{2}\vhat{x}-\frac{\sqrt{3}}{2}\vhat{y}\right), \\
    \vec{m}_C &= N\left(-\frac{1}{2}\vhat{x}+\frac{\sqrt{3}}{2}\vhat{y}\right),
\end{align}
\end{subequations}
arranged in a 120$^{\circ}$ pattern.

In this hexagonal case, the single-$N$ and parallel phases are degenerate, allowing for arbitrary mixing between them. This family can be parameterized by an angle $\alpha$.
Explicitly,
\begin{align}
\label{eq:degen}
    \vec{N}_1 &= (N\cos{\alpha})\vhat{N}, &
    \vec{N}_2 &= (N\sin{\alpha})\vhat{N}.
\end{align}
A particular value of $\alpha$ is only selected by sextic terms in the order parameter, breaking the 
accidental $\uone$ symmetry.
We illustrate one of these states, choosing $\alpha=0$, that has $\vec{N}_1 = \vec{N}$, $\vec{N}_2=0$ in Fig.~\ref{fig:kagome-orders}. This gives
\begin{align}
    \vec{m}_A &= \sqrt{\frac{2}{3}}  \vec{N}, &
    \vec{m}_B &= \vec{m}_C =  -\frac{1}{\sqrt{6}}  \vec{N}, 
\end{align}
with unequal moments, $|\vec{m}_A| = 2|\vec{m}_B|=2|\vec{m}_C|$. 

The epikernels for this two-dimensional order parameter
can be worked out explicitly. Recall, to find the epikernel
we solve for each $g \in 6mm$
the equation
\begin{equation}
\label{eq:d6-epi}
\begin{pmatrix} \vec{N}_1 \\ \vec{N}_2 \end{pmatrix} =
\begin{pmatrix}
    \Gamma_{11}(g) & \Gamma_{12}(g) \\
    \Gamma_{21}(g) & \Gamma_{22}(g)
\end{pmatrix}
\begin{pmatrix} \mat{R}\vec{N}_1 \\  \mat{R}\vec{N}_2 \end{pmatrix}.
\end{equation}
When the solution spaces for each $g$ are known, we can
identify the relevant subgroup $\mathcal{H} \leq \mathcal{G}$ via the elements that leave a common subspace
invariant. 

We begin the analysis with the collinear degenerate phase where we can write (without loss of generality) $\vec{N}_1 =N_1\vhat{N}$, $\vec{N}_2=N_2\vhat{N}$. 
First note that these order parameters are preserved by any $\othree$ matrix $\mat{R}$ satisfying $\mat{R}\vhat{N} = \pm \vhat{N}$. 
The elements that preserve $\vhat{N}$ are the elements of the collinear spin-only group, which contains $\sotwo$ rotations about the collinear axis as well as a two-fold $2_{\perp}$ spin rotation perpendicular to this axis combined with time-reversal. 
We denote this group as $b_{\infty} \equiv \sotwo \rtimes \{1, 2_{\perp}T\}$.
The elements that map $\vhat{N} \to -\vhat{N}$ are those of the spin-only group combined with time-reversal, which we denote as $b_{\infty} T$. 
Solutions for $\mat{R}$ in Eq.~(\ref{eq:d6-epi}) can thus be parameterized entirely in terms of a sign $\gamma = \pm 1$ encoding whether $\mat{R}$ belongs to $b_{\infty}$ ($\gamma = +1$) or $b_{\infty} T$ ($\gamma = -1$). 

To determine these signs, consider one representative from each conjugacy class of $6mm$ ($C_{6v}$): $1$, $2_{z}$, $6_{z}$, $3_{z}$, $m_{x}$, and $m_y$.
The identity $1$ and the $2_{z}$ operation are trivial in the $E_2$ irrep, and thus $\mat{R}$ can be any element in $b_{\infty}$ and so $\gamma=+1$. Next consider the $6_{z}$ operation which acts as
$$
\begin{pmatrix} {N}_1 \\ {N}_2 \end{pmatrix} 
= \gamma
\begin{pmatrix}
    -\tfrac{1}{2} & -\tfrac{\sqrt{3}}{2} \\[0.25em]
    +\tfrac{\sqrt{3}}{2} & -\tfrac{1}{2}
\end{pmatrix}
\begin{pmatrix} {N}_1 \\ {N}_2 \end{pmatrix}.
$$
For $\gamma=+1$ this is a rotation by $2\pi/3$, and for $\gamma=-1$ a rotation by $-\pi/3$, in the $(N_1,N_2)$ plane; neither has a fixed vector, so there is no non-zero solution.
Similarly, the defining equation for $3_{z}$ also has no non-trivial solutions. Finally, for the mirror $m_x$, following the same logic, we have
$$
\begin{pmatrix} {N}_1 \\ {N}_2 \end{pmatrix} 
= \gamma
\begin{pmatrix}
    +1  & 0 \\
    0 & -1
\end{pmatrix}
\begin{pmatrix} {N}_1 \\ {N}_2 \end{pmatrix}.
$$
There are two solutions: if $\gamma=+1$ ($\mat{R} \in b_{\infty}$) we have $N_1 \neq 0$, $N_2=0$, while if $\gamma=-1$ ($\mat{R} \in b_{\infty}T$) then $N_1=0$, $N_2 \neq 0$. An identical analysis follows for $m_y$ since $\mat{\Gamma}(m_x) = \mat{\Gamma}(m_y)$ for $E_2$.

\begin{figure}
    \centering
    \includegraphics[width=\columnwidth]{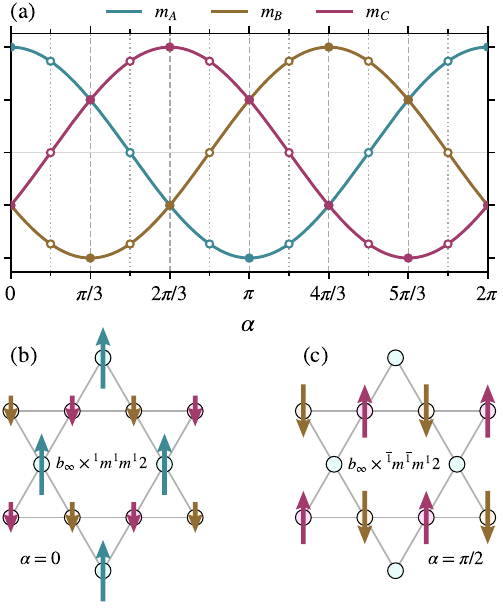}
    \caption{Illustration of the degenerate collinear phases of the $E_2$ order on the kagom\'e lattice, 
    parameterized by the angle $\alpha$ [Eq.~(\ref{eq:degen})].
    (a) The magnetic moments are compensated (at linear order) and collinear, and we show the modulation of the signed moments $m_A \equiv \vec{m}_A\cdot\vhat{N}$, $m_B \equiv \vec{m}_B\cdot\vhat{N}$, $m_C \equiv \vec{m}_C\cdot\vhat{N}$. (b),(c)
    We show two high-symmetry states at $\alpha=0$ ($\vec{N}_1 \neq 0$, $\vec{N}_2=0$) and $\alpha=\pi/2$ ($\vec{N}_1 = 0$, $\vec{N}_2 \neq 0$) with epikernels $b_{\infty} \times {}^1 m {}^1 m {}^1 2$ and $b_{\infty} \times {}^{\bar{1}} m {}^{\bar{1}} m {}^1 2$ respectively.
    Generic values of $\alpha$ have epikernel
    ${b_{\infty}\times{^{1}2}}$. We indicate 
    $\alpha$ values equivalent to $\alpha=0$ in (a) by filled circles, and equivalent to $\alpha=\pi/2$ by open circles.
    }
    \label{fig:kagome-collinear-orders}
\end{figure}

From these solutions, we can identify the relevant epikernels. For a generic choice of $N_1$, $N_2$ only
the equations for $1$ and $2_{z}$ can be satisfied, and thus the epikernel is $b_{\infty} \times {}^1 2$. For $N_1 \neq 0$ and $N_2=0$ additionally $m_x$, $m_y$ can be satisfied, without inclusion of time-reversal and so the epikernel is $b_{\infty} \times {}^1 m {}^1 m {}^1 2$.
This is $\alpha=0$ in the language of Eq.~(\ref{eq:degen}).
Lastly, when $N_1=0$ and $N_2 \neq 0$ the $m_x$, $m_y$ operations are accompanied by time-reversal and so the epikernel is $b_{\infty} \times {}^{\bar{1}} m {}^{\bar{1}} m {}^1 2$, corresponding to $\alpha=\pi/2$. Additional epikernels can be generated by considering mirrors generated by conjugating $m_x$ and $m_y$ with the $6_{z}$ or $3_{z}$ rotations, corresponding to choices of $\alpha=(2n)\pi/6$ or $\alpha=(2n+1)\pi/6$. 
We show illustrations of these two states, along with the variation of the moment lengths, in Fig.~\ref{fig:kagome-collinear-orders}.

Note that the $\alpha=0$ and $\alpha=\pi/2$ states can be viewed as collinear structures in $mm2 \leq 6mm$, with the sign structures defining one-dimensional irreps $A_1$ ($\alpha=0$) or $A_2$ ($\alpha=\pi/2$). In this setting the $\alpha=0$ phase is a ferrimagnet (see Sec.~\ref{subsubsec:obs_odd_vectors_3})
and the $\alpha=\pi/2$ phase is an altermagnet ($d$-wave). There is a natural interpretation of these assignments
in terms of a Landau theory formulated directly for these collinear orders. Instead of proceeding via the two-dimensional $E_2$ order parameter, we could define a one-dimensional $A_1$ or $A_2$ order parameter in the subgroup $\othree \times mm2$ of the original paramagnetic group.

Next, we consider the orthogonal phase where
$\vec{N}_1$ and $\vec{N}_2$ are of equal magnitude and
orthogonal. Note that there is a coplanar spin-only
group that leaves these invariant: this is a $2_\perp T$ rotation about the axis $\vhat{z}$.
We denote this group $b_{\mathbb{Z}_2} \equiv \{1,2_\perp T\}$. In Eq.~(\ref{eq:d6-epi}) the matrix $\mat{R}$ must preserve the plane defined by $\vec{N}_1$ and $\vec{N}_2$. Such a transformation must take the form of a (proper or improper) rotation about the normal to the plane, $\vhat{N}_1 \times \vhat{N}_2$, or a two-fold rotation about an axis lying within the plane. Some of these elements act non-trivially on $\vhat{N}_1 \times \vhat{N}_2$; combining such elements with the $2_{\perp}T$ element of the spin-only group renders their action on this direction trivial. Thus we can consider only $\mat{R}$ acting within the $\vec{N}_1$, $\vec{N}_2$ plane.

Following the choice made for Fig.~\ref{fig:kagome-orders}(b), set $\vhat{N}_1=\vhat{x}$ and $\vhat{N}_2=\vhat{y}$. The defining equation for the epikernels [Eq.~(\ref{eq:d6-epi})] is then
\begin{subequations}
\begin{align}
    N \vhat{x} &= N\left(\Gamma_{11}(g) \mat{R}\vhat{x} + \Gamma_{12}(g) \mat{R}\vhat{y}\right), \\
    N \vhat{y} &= N\left(\Gamma_{21}(g) \mat{R}\vhat{x} + \Gamma_{22}(g) \mat{R}\vhat{y}\right).
\end{align}
\end{subequations}
Keeping in mind that the $\mat{R}$ rotation acts only on $\vhat{x}$, $\vhat{y}$, and expanding the matrix-vector products, e.g., $\mat{R}\vhat{x} = R_{11}\vhat{x} + R_{21}\vhat{y}$, we obtain
$$
\begin{pmatrix}
    N\vhat{x} \\ N\vhat{y}
\end{pmatrix}
= 
\begin{pmatrix}
    [\mat{\Gamma}(g)\trp{\mat{R}}]_{11} &  [\mat{\Gamma}(g)\trp{\mat{R}}]_{12} \\
    [\mat{\Gamma}(g)\trp{\mat{R}}]_{21} &  [\mat{\Gamma}(g)\trp{\mat{R}}]_{22}
\end{pmatrix}
\begin{pmatrix}
    N\vhat{x} \\ N\vhat{y}
\end{pmatrix} \equiv
\mat{\Gamma}(g) \trp{\mat{R}} \begin{pmatrix}
    N\vhat{x} \\ N\vhat{y}
\end{pmatrix}.
$$
This then immediately implies that to satisfy this equation we must have $
\mat{R} = \mat{\Gamma}(g)$.

We thus see that for this phase each element $g$ of
$6mm$ can be paired with a spin rotation corresponding to $\mat{\Gamma}(g)$. 
The epikernel is then isomorphic to $6mm$ (modulo the spin-only group), and is given by $b_{\mathbb{Z}_2} \times {}^3 6 {}^{2} m {}^{2} m$.
We list the specific matrices, $\mat{R}$, for each conjugacy class of $6mm$ in Table~\ref{tab:d6-epi}. We also list the determinant of
each of these matrices. Since each element that acts non-trivially on $\vhat{z}$ is paired with the spin-only $2_\perp T$, the determinant of $\mat{R}$ specifies whether the operation is unitary ($+1$) or antiunitary ($-1$). In Table~\ref{tab:kagome_subduction}, we explicitly verify the epikernels by subduction by identifying the spin point subgroups of $\othree\times6mm$ for which the restriction of $\Gamma_{S}^{(1)}\otimes \Gamma$ contains the trivial irrep.

\begin{table}[tb]
    \centering
    \begin{tabular}{ccc@{\hspace{2em}}ccc}
        \toprule
        \multicolumn{3}{c}{Degenerate}
        & \multicolumn{3}{c}{Orthogonal} \\
        \cmidrule(lr){1-3} \cmidrule(lr){4-6}
        $g$ & $\gamma$ & Subspace
        & $g$ & $\mat{R}$ & $\det \mat{R}$ \\[0.5em]
        \midrule

        $1$
        & $+1$
        & $\begin{pmatrix} N_1 \\ N_2 \end{pmatrix}$
        & $1$
        & $\begin{pmatrix} 1 & 0 \\ 0 & 1 \end{pmatrix}$
        & $+1$ \\[1em]

        $2_{z}$
        & $+1$
        & $\begin{pmatrix} N_1 \\ N_2 \end{pmatrix}$
        & $2_{z}$
        & $\begin{pmatrix} 1 & 0 \\ 0 & 1 \end{pmatrix}$
        & $+1$ \\[1em]

        $3_{z}$
        &
        & $\begin{pmatrix} 0 \\ 0 \end{pmatrix}$
        & $3_{z}$
        & $\begin{pmatrix}
            -\frac{1}{2} & \frac{\sqrt{3}}{2} \\
            -\frac{\sqrt{3}}{2} & -\frac{1}{2}
          \end{pmatrix}$
        & $+1$ \\[1em]

        $6_{z}$
        &
        & $\begin{pmatrix} 0 \\ 0 \end{pmatrix}$
        & $6_{z}$
        & $\begin{pmatrix}
            -\frac{1}{2} & -\frac{\sqrt{3}}{2} \\
            \frac{\sqrt{3}}{2} & -\frac{1}{2}
          \end{pmatrix}$
        & $+1$ \\[1em]

        \multirow{2}{*}{$m_x$}
        & $+1$
        & $\begin{pmatrix} N \\ 0 \end{pmatrix}$
        & \multirow{2}{*}{$m_x$}
        & \multirow{2}{*}{$\begin{pmatrix} 1 & 0 \\ 0 & -1 \end{pmatrix}$}
        & \multirow{2}{*}{$-1$} \\

        & $-1$
        & $\begin{pmatrix} 0 \\ N \end{pmatrix}$
        &
        &
        & \\[1em]

        \multirow{2}{*}{$m_y$}
        & $+1$
        & $\begin{pmatrix} N \\ 0 \end{pmatrix}$
        & \multirow{2}{*}{$m_y$}
        & \multirow{2}{*}{$\begin{pmatrix} 1 & 0 \\ 0 & -1 \end{pmatrix}$}
        & \multirow{2}{*}{$-1$} \\

        & $-1$
        & $\begin{pmatrix} 0 \\ N \end{pmatrix}$
        &
        &
        & \\[1em]

        \bottomrule
    \end{tabular}
    \caption{Solutions of Eq.~(\ref{eq:d6-epi}) for the two phases of $F_3$ [Eq.~(\ref{eq:f3})] of the $E_2$ order
    parameter on the kagom\'e lattice. For a representative $g$ of each conjugacy class of
    $6mm$ we specify the associated rotation $\mat{R}$ together with the subspace of $(N_1,N_2)$ that is then
    left invariant (a subspace $(0,0)$ indicates that Eq.~(\ref{eq:d6-epi}) has no non-trivial
    solution). For the degenerate collinear phase only the sign $\gamma$ is listed, fixing
    $\mat{R}\vhat{N} = \gamma\vhat{N}$, modulo elements of the spin-only group, $b_{\infty}$. For the orthogonal phase, the spin rotation $\mat{R}$ associated with $g$ is $\mat{\Gamma}(g)$, modulo elements of $b_{\mathbb{Z}_2}$, with $\det\mat{R} = +1$ or $-1$ indicating whether the operation is
    unitary or antiunitary.}
    \label{tab:d6-epi}
\end{table}

\subsection{Square plaquette lattice}
\label{sec:tetragonal}
The kagom\'e example is constrained by its hexagonal point group to the $F_3$ line of
Fig.~\ref{fig:phase-diagram-2d};  to illustrate the generic
non-cyclic tetragonal case, with $b_4$ and $c_4$ independent, we now
consider a square-plaquette example with the non-centrosymmetric point group $4mm$.

\begin{figure*}[ht!]
    \centering
    \includegraphics[width=\textwidth]{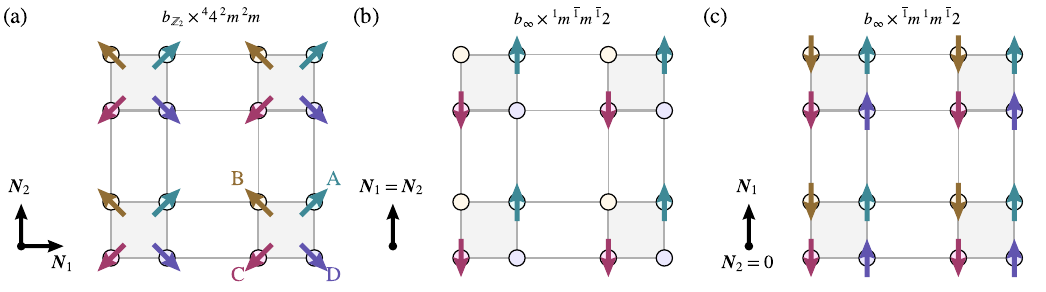}
    \caption{Three magnetic orders in the phase diagram of
    Fig.~\ref{fig:phase-diagram-2d} for the square plaquette lattice, with $A$, $B$,
    $C$, $D$ sublattices. 
    (a) Orthogonal phase, $\vec{N}_1 = 2\sqrt{2}N\vhat{x}$, $\vec{N}_2=2\sqrt{2}N\vhat{y}$: every
    moment has length $N$ and points radially out of its plaquette, the planar analogue of the
    ``all-in/all-out'' state.
    (b) Parallel phase, $\vec{N}_1=\vec{N}_2=2N\vhat{y}$: collinear, with zero moment on $B$ and $D$.
    (c) Single-$N$ phase, $\vec{N}_1 = 4N\vhat{y}$, $\vec{N}_2=\vec{0}$: collinear stripes running along
    $\vhat{y}$, with all four moments of equal length.
    Each panel gives the state's epikernel, and shows the direction of
    the order parameters. Circles denote the positions of magnetic ions, while ligands (not shown) explicitly break inversion to realize a $4mm$ point group.
    }
    \label{fig:tetragonal-orders}
\end{figure*}

We take a lattice with four magnetic ions in the unit cell located at the corners of a
square plaquette, which we label $A$, $B$, $C$ and $D$, as illustrated in
Fig.~\ref{fig:tetragonal-orders}. The point group of the lattice is 
$4mm$ ($C_{4v}$). We consider $\vec{k}=0$ order within this unit cell with moments $\vec{m}_A$, $\vec{m}_B$, $\vec{m}_C$, $\vec{m}_D$ on these four sublattices.
Under the action of the point group the four sites permute, transforming as~\footnote{The character of the permutation representation on each class is the number of sites left fixed, giving $(4,0,0,0,2)$ on the classes $(1, 2C_{4z}, C_{2z}, 2m_{v}, 2m_{d})$; only the diagonal mirrors $m_{x\pm y}$, which pass through a pair of sites, fix any site.}
\begin{equation}\label{eq:sq_plaquette_decomp}
\Gamma_{\rm perm} = A_1 + B_2 + E.
\end{equation}
The $A_1$ component is the uniform ferromagnetic order parameter
$\vec{M} = \vec{m}_A + \vec{m}_B + \vec{m}_C + \vec{m}_D$. Unlike the kagom\'e case,
$\Gamma_{\mathrm{perm}}$ also contains a non-trivial one-dimensional irrep, $B_{2}$,
$$
\vec{L} = \vec{m}_A - \vec{m}_B + \vec{m}_C - \vec{m}_D,
$$ describing conventional N\'eel order. Being one-dimensional, its Landau theory is the collinear one discussed in Sec.~\ref{sec:landau}, so we do not consider it further. The interesting case is the $E$ component, with a two-component order parameter
\begin{subequations}
\label{eq:tet_N_moments}
\begin{align}
\vec{N}_1 &= \vec{m}_A - \vec{m}_B - \vec{m}_C + \vec{m}_D, \\
\vec{N}_2 &= \vec{m}_A + \vec{m}_B - \vec{m}_C - \vec{m}_D.
\end{align}
\end{subequations}
Inverting Eq.~(\ref{eq:tet_N_moments}) with the constraint $\vec{M}=\vec{L}=\vec{0}$ gives 
\begin{subequations}
\label{eq:tet_moments_ITO_N}
\begin{align}
\vec{m}_A &= \tfrac{1}{4}\left(\phantom{-}\vec{N}_1 + \vec{N}_2\right), &
\vec{m}_B &= \tfrac{1}{4}\left(-\vec{N}_1 + \vec{N}_2\right), \\
\vec{m}_C &= \tfrac{1}{4}\left(-\vec{N}_1 - \vec{N}_2\right), &
\vec{m}_D &= \tfrac{1}{4}\left(\phantom{-}\vec{N}_1 - \vec{N}_2\right).
\end{align}
\end{subequations}
The matrices $\mat{\Gamma}(g)$ representing $4mm$ on $(\vec{N}_1,\vec{N}_2)$ follow directly from the
site permutations and are listed in Table~\ref{tab:d4-epi}.

Since the point group is tetragonal, the Landau theory for this order is defined by the $F_4$ free energy of
Eq.~(\ref{eq:free-tet}) with two independent quartic terms $b_4$ and $c_4$. As discussed in Sec.~\ref{sec:landau2}
there are three phases: the single-$N$ phase where one of $\vec{N}_1$, $\vec{N}_2$ vanishes, the
parallel phase with $\vec{N}_1 \| \vec{N}_2$ of equal magnitude, and the orthogonal phase with
$\vec{N}_1 \perp \vec{N}_2$ of equal magnitude. In contrast to the kagom\'e case, where the single-$N$ and parallel phases are degenerate, here
these are two distinct phases. We illustrate one example
state from each phase in Fig.~\ref{fig:tetragonal-orders}, using the mapping given in
Eq.~(\ref{eq:tet_moments_ITO_N}).

For the orthogonal phase we choose $\vec{N}_1 = 2\sqrt{2}N\vhat{x}$ and
$\vec{N}_2 = 2\sqrt{2}N\vhat{y}$, which yields
\begin{align}
    \vec{m}_A &= \frac{N}{\sqrt{2}}\left(\phantom{-}\vhat{x}+\vhat{y}\right), &
    \vec{m}_B &= \frac{N}{\sqrt{2}}\left(-\vhat{x}+\vhat{y}\right), \notag \\
    \vec{m}_C &= \frac{N}{\sqrt{2}}\left(-\vhat{x}-\vhat{y}\right), &
    \vec{m}_D &= \frac{N}{\sqrt{2}}\left(\phantom{-}\vhat{x}-\vhat{y}\right).
\end{align}
Each moment points radially outward from the center of the plaquette; this is the planar analogue of the ``all-in/all-out'' state found in the pyrochlore lattice. As in the kagom\'e example, this coplanar state has only a $2_\perp T$ spin-only
symmetry, with the $2_\perp$ axis along $\vhat{N}_1\times\vhat{N}_2$.

For the single-$N$ phase we take $\vec{N}_1 = 4N\vhat{y}$ and $\vec{N}_2=\vec{0}$, so that the moments become
\begin{align}
    \vec{m}_A &=  N\vhat{y}, &
    \vec{m}_B &= -N\vhat{y}, &
    \vec{m}_C &= -N\vhat{y}, &
    \vec{m}_D &=  N\vhat{y},
\end{align}
a collinear stripe order with all moments of equal length, running along $\vhat{y}$ and alternating
along $\vhat{x}$. Finally, for the parallel phase we take $\vec{N}_1 = \vec{N}_2 = 2N\vhat{y}$, which
gives
\begin{align}
    \vec{m}_A &= N\vhat{y}, &
    \vec{m}_B &= \vec{0}, &
    \vec{m}_C &= -N\vhat{y}, &
    \vec{m}_D &= \vec{0}.
\end{align}
As in the degenerate phase of the kagom\'e lattice with $\alpha=\pi/2$, this state is collinear and compensated at leading order but has unequal moments, with the order residing entirely
on the $A$--$C$ diagonal while the $B$ and $D$ sites carry no moment at all.

The epikernels follow from solving, for each $g \in 4mm$,
\begin{equation}
\label{eq:d4-epi}
\begin{pmatrix} \vec{N}_1 \\ \vec{N}_2 \end{pmatrix} =
\mat{\Gamma}(g)
\begin{pmatrix} \mat{R}\vec{N}_1 \\  \mat{R}\vec{N}_2 \end{pmatrix},
\end{equation}
exactly as in Eq.~(\ref{eq:d6-epi}), and identifying the subgroup $\mathcal{H}\leq\mathcal{G}$
whose elements leave a common subspace invariant.

For the orthogonal phase, the analysis of Sec.~\ref{sec:kagome} carries over verbatim. The coplanar
spin-only group $b_{\mathbb{Z}_2} = \{1, 2_\perp T\}$ allows us to restrict $\mat{R}$ to the plane spanned by $\vec{N}_1$ and
$\vec{N}_2$. Setting $\vhat{N}_1 = \vhat{x}$ and $\vhat{N}_2=\vhat{y}$, Eq.~(\ref{eq:d4-epi}) reduces
to
$$
\begin{pmatrix} N\vhat{x} \\ N\vhat{y} \end{pmatrix} =
\mat{\Gamma}(g)\trp{\mat{R}} \begin{pmatrix} N\vhat{x} \\ N\vhat{y} \end{pmatrix},
$$
and is solved by $\mat{R} = \mat{\Gamma}(g)$ for every $g$, identically to the kagom\'e case. Each element of $4mm$ is paired with a definite spin rotation, and the epikernel is isomorphic to $4mm$ itself. The fourfold rotation acts as a $90^{\circ}$ rotation in the $(\vec{N}_1,\vec{N}_2)$ plane and so is paired with a fourfold spin rotation, in contrast to the kagom\'e case where the $E_2$ irrep rotates at twice the
spatial rate and $6_{z}$ is paired with a spin three-fold rotation. The mirrors are paired with improper $\mat{R}$ and are therefore antiunitary, while the rotations are unitary. The epikernel is
$b_{\mathbb{Z}_2}\times {}^{4}4\, {}^{2}m\, {}^{2}m$. A representative orthogonal state is shown in Fig.~\ref{fig:tetragonal-orders}(a), and the explicit $\mat{R}$ are given in Table~\ref{tab:d4-epi}. 

\begin{table}[tb]
    \centering
    \begin{tabular}{cccc@{\hspace{1.5em}}cc}
        \toprule
        & & \multicolumn{2}{c}{Collinear}
        & \multicolumn{2}{c}{Orthogonal} \\
        \cmidrule(lr){3-4} \cmidrule(lr){5-6}
        $g$ & $\mat{\Gamma}(g)$ & single-$N$ & parallel
        & $\mat{R}$ & $\det \mat{R}$ \\[0.5em]
        \midrule

        $1$
        & $\begin{pmatrix} 1 & 0 \\ 0 & 1 \end{pmatrix}$
        & $+1$ & $+1$
        & $\begin{pmatrix} 1 & 0 \\ 0 & 1 \end{pmatrix}$
        & $+1$ \\[1em]

        $2_{z}$
        & $\begin{pmatrix} -1 & 0 \\ 0 & -1 \end{pmatrix}$
        & $-1$ & $-1$
        & $\begin{pmatrix} -1 & 0 \\ 0 & -1 \end{pmatrix}$
        & $+1$ \\[1em]

        $4_{z}$
        & $\begin{pmatrix} 0 & 1 \\ -1 & 0 \end{pmatrix}$
        & --- & ---
        & $\begin{pmatrix} 0 & 1 \\ -1 & 0 \end{pmatrix}$
        & $+1$ \\[1em]

        $4_{z}^{-1}$
        & $\begin{pmatrix} 0 & -1 \\ 1 & 0 \end{pmatrix}$
        & --- & ---
        & $\begin{pmatrix} 0 & -1 \\ 1 & 0 \end{pmatrix}$
        & $+1$ \\[1em]

        $m_x$
        & $\begin{pmatrix} -1 & 0 \\ 0 & 1 \end{pmatrix}$
        & $-1$ & ---
        & $\begin{pmatrix} -1 & 0 \\ 0 & 1 \end{pmatrix}$
        & $-1$ \\[1em]

        $m_y$
        & $\begin{pmatrix} 1 & 0 \\ 0 & -1 \end{pmatrix}$
        & $+1$ & ---
        & $\begin{pmatrix} 1 & 0 \\ 0 & -1 \end{pmatrix}$
        & $-1$ \\[1em]

        $m_{x-y}$
        & $\begin{pmatrix} 0 & 1 \\ 1 & 0 \end{pmatrix}$
        & --- & $+1$
        & $\begin{pmatrix} 0 & 1 \\ 1 & 0 \end{pmatrix}$
        & $-1$ \\[1em]

        $m_{x+y}$
        & $\begin{pmatrix} 0 & -1 \\ -1 & 0 \end{pmatrix}$
        & --- & $-1$
        & $\begin{pmatrix} 0 & -1 \\ -1 & 0 \end{pmatrix}$
        & $-1$ \\[1em]

        \bottomrule
    \end{tabular}
    \caption{Solutions of Eq.~(\ref{eq:d4-epi}) for the three tetragonal $E$ phases appearing in $F_4$ [Eq.~(\ref{eq:free-tet})] (single-$N$, parallel and orthogonal) on the square plaquette lattice. For each element $g$ of $4mm$ we give the matrix $\mat{\Gamma}(g)$ of the
    $E$ irrep, the sign $\gamma$ fixing $\mat{R}\vhat{N} = \gamma\vhat{N}$ in the two collinear
    phases (a dash indicates Eq.~(\ref{eq:d4-epi}) has no non-trivial solution), and the spin
    rotation $\mat{R}$ for the orthogonal phase, where $\det\mat{R} = +1$ or $-1$
    indicates whether the operation is unitary or antiunitary.}
    \label{tab:d4-epi}
    \vspace{-0.6cm}
\end{table}

For the two collinear phases, the structure of the spin-only group is identical to that discussed in
Sec.~\ref{sec:kagome}: the solutions for $\mat{R}$ are parameterized by a sign $\gamma=\pm 1$
depending on whether $\mat{R}$ lies in $b_{\infty}$ ($\gamma=+1$) or $b_{\infty}T$ ($\gamma=-1$), so
that $\mat{R}\vhat{N} = \gamma \vhat{N}$. The sign structures $\gamma$ of the two collinear phases are collected in Table~\ref{tab:d4-epi}.

In the parallel phase, the equation we need to solve is
$$
\begin{pmatrix} N \\ N \end{pmatrix} = \gamma \mat{\Gamma}(g) \begin{pmatrix} N \\ N \end{pmatrix},
$$
which requires $\mat{\Gamma}(g)$ to preserve or reverse the vector $(1,1)$. The identity element $1$ and the
diagonal mirror $m_{x-y}$ preserve $(1,1)$, yielding $\gamma=+1$, while $2_{z}$ and $m_{x+y}$ reverse it, requiring 
$\gamma=-1$. The fourfold rotations and the axis mirrors have no solution. The four relevant elements form an $mm2$ subgroup generated by the \emph{diagonal}
mirrors, and the epikernel is $b_{\infty}\times {}^{1}m\, {}^{\bar{1}}m\, {}^{\bar{1}}2$ with the
mirror planes normal to $[1\bar{1}0]$ and $[110]$. These symmetries can be easily checked using the example state of $\vec{N}_1 = \vec{N}_2 = 2N\vhat{y}$ shown in Fig.~\ref{fig:tetragonal-orders}(b).  

For the single-$N$ phase, for each $g \in 4mm$ we must solve 
$$
\begin{pmatrix} N \\ 0 \end{pmatrix} = \gamma \mat{\Gamma}(g) \begin{pmatrix} N \\ 0 \end{pmatrix}.
$$
This has a solution only when $\mat{\Gamma}(g)$ preserves or reverses the
vector $(1,0)$.  The fourfold rotations and the diagonal mirrors provide no solution. The elements $1$ and $m_y$ preserve $(1,0)$, with $\gamma=+1$, while $2_{z}$ and $m_x$
reverse it, requiring $\gamma=-1$.
These four elements form a different $mm2$ subgroup of $4mm$ than the parallel phase, generated by $m_x$ and $m_y$.
The epikernel is therefore
$b_{\infty}\times {}^{\bar{1}}m\, {}^{1}m\, {}^{\bar{1}}2$. This is readily checked against the
state shown in Fig.~\ref{fig:tetragonal-orders}(c): $m_y$ exchanges $A \leftrightarrow D$ and $B\leftrightarrow C$ and so is a
symmetry on its own, while $m_x$ and $2_{z}$ exchange up and down moments and must be accompanied by
time reversal.

The single-$N$ and parallel collinear phases thus have isomorphic epikernels, which are distinct subgroups of $\mathcal{G}$: they are built from the two inequivalent mirror classes of $4mm$ and are not related by conjugation, but instead by a change of axes. 
These are the tetragonal
counterparts of the $\alpha=0$ and $\alpha=\pi/2$ states of the kagom\'e example, except here they do not belong to a common degenerate manifold of states. In this case both the single-$N$ and parallel phases are altermagnets, with order
parameters transforming as $B_1$ or $B_2$ in $mm2$, depending on the choice of basis.

\subsection{Pyrochlore lattice}
\label{sec:pyrochlore}
\begin{figure*}
    \centering
    \includegraphics[width=\textwidth]{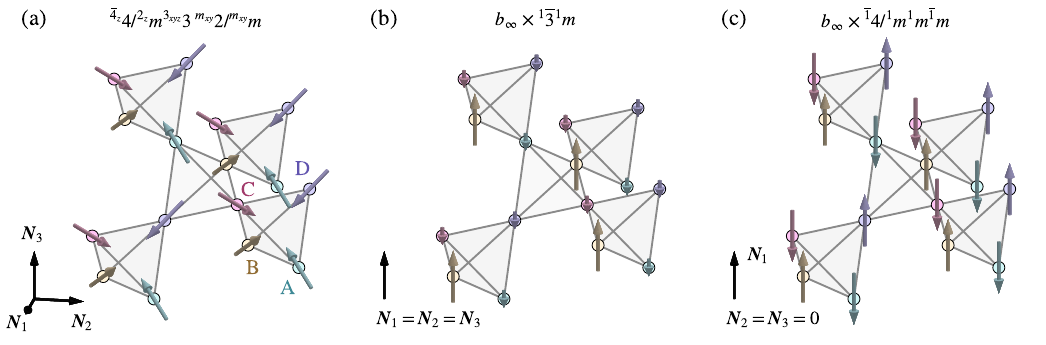}
    \caption{Illustrations of three magnetic orders in the phase diagram of
    Fig.~\ref{fig:phase-diagram-3d} for the pyrochlore lattice.
    (a) Orthogonal phase, $\vec{N}_1 = 4N\vhat{x}/\sqrt{3}$, $\vec{N}_2 = 4N\vhat{y}/\sqrt{3}$,
    $\vec{N}_3 = 4N\vhat{z}/\sqrt{3}$: every moment has length $N$ and lies along the local
    $\langle 111\rangle$ axis of its own tetrahedron, the ``all-in/all-out'' state.
    (b) Parallel phase, $\vec{N}_1 = \vec{N}_2 = \vec{N}_3 = 4N\vhat{z}/3$: collinear and
    compensated (at leading order), but with unequal moments. Sublattice $B$ carries moment $N$ and the other three carry $N/3$ oppositely.
    (c) Single-$N$ phase, $\vec{N}_1 = 4N\vhat{z}$, $\vec{N}_2=\vec{N}_3=\vec{0}$: collinear with all moments of length $N$, $A$ and $C$ opposite $B$ and $D$, reminiscent of the ``two-in/two-out''
    arrangement of spin ice.
     Each panel gives the state's epikernel and the order parameter directions.
    }
    \label{fig:pyrochlore}
\end{figure*}

The point groups of the kagom\'e and square plaquette lattices do not support any three-dimensional irreducible representations. As an example of an irreducible non-coplanar phase, we consider $\vec{k}=0$ orders on the pyrochlore lattice, illustrated in Fig.~\ref{fig:pyrochlore}. This three-dimensional cubic lattice has a point group $m\overline{3}m$ ($O_h$), and the permutation representation of the four atoms in the primitive cell is
\begin{equation}\label{eq:pyrochlore_decomposition}
    \Gamma_{\rm perm} = A_{1g} +T_{2g}.
\end{equation}
In terms of the magnetic moments of the four sites, which we label $\vec{m}_A$, $\vec{m}_B$, $\vec{m}_C$, and $\vec{m}_D$, the $A_{1g}$ part is the total magnetization
$\vec{M} = \vec{m}_{A} + \vec{m}_{B} +\vec{m}_{C} + \vec{m}_{D}$ and the $T_{2g}$ part is
\begin{subequations}
    \label{eq:Pyro_N_moments}
\begin{align}
\vec{N}_{1} &=  -\vec{m}_{A} + \vec{m}_{B} -\vec{m}_{C} + \vec{m}_{D}, \\
\vec{N}_{2} &=  -\vec{m}_{A} + \vec{m}_{B} + \vec{m}_{C} -\vec{m}_{D}, \\
\vec{N}_{3} &= \phantom{-}\vec{m}_{A} + \vec{m}_{B} - \vec{m}_{C} - \vec{m}_{D},
\end{align}
\end{subequations}
which may act as the order parameters on the pyrochlore lattice. Inverting Eq.~(\ref{eq:Pyro_N_moments}) with the constraint that $\vec{M}=\vec{0}$, we can express the moments on each sublattice in terms of these order parameters, 
\begin{subequations}
\label{eq:pyro_moments_ITO_N}
\begin{align}
\vec{m}_{A} &= \frac{1}{4} \left(  - \vec{N}_{1} - \vec{N}_{2} + \vec{N}_{3}\right),\\
\vec{m}_{B} &=\frac{1}{4} \left( \phantom{-} \vec{N}_{1} + \vec{N}_{2} +\vec{N}_{3} \right),\\
\vec{m}_{C} &= \frac{1}{4}\left( - \vec{N}_{1} + \vec{N}_{2} -\vec{N}_{3}\right),  \\
\vec{m}_{D} &= \frac{1}{4} \left(\phantom{-} \vec{N}_{1} - \vec{N}_{2} - \vec{N}_{3} \right).
\end{align}
\end{subequations}

The Landau theory for such $T_{2g}$ orders is described
in Sec.~\ref{sec:landau3} by the free energy given in Eq.~(\ref{eq:3D_cub}). There are three natural phases, 
the single-$N$ phase where all but one of $\vec{N}_1$, $\vec{N}_2$ or $\vec{N}_3$ vanish, the parallel phase where $\vec{N}_1 \| \vec{N}_2 \| \vec{N}_3 $ with equal magnitude and an orthogonal phase where $\vec{N}_1 \perp \vec{N}_2 \perp \vec{N}_3$ with equal magnitude. 
We illustrate the magnetic moment configurations for three example states of these phases in Fig.~\ref{fig:pyrochlore}, using the mapping given in Eq.~(\ref{eq:pyro_moments_ITO_N}).

For the orthogonal phase we choose $\vec{N}_1 = 4N \vhat{x}/\sqrt{3}$, $\vec{N}_2 = 4N\vhat{y}/\sqrt{3}$, $\vec{N}_3 = 4N\vhat{z}/\sqrt{3}$. This yields
\begin{equation}
    \begin{aligned}
        \vec{m}_A &= \frac{N}{\sqrt{3}}\left(-\vhat{x}-\vhat{y}+\vhat{z}\right), &
        \vec{m}_B &= \frac{N}{\sqrt{3}}\left(\phantom{-}\vhat{x}+\vhat{y}+\vhat{z}\right), \\
        \vec{m}_C &= \frac{N}{\sqrt{3}}\left(-\vhat{x}+\vhat{y}-\vhat{z}\right), &
        \vec{m}_D &= \frac{N}{\sqrt{3}}\left(\phantom{-}\vhat{x}-\vhat{y}-\vhat{z}\right),
    \end{aligned}
\end{equation}
where the spins are arranged in an ``all-in/all-out'' pattern. 

For the single-$N$ phase, we consider $\vec{N}_1=4N\vhat{z}$ and $\vec{N}_2=\vec{N}_3=\vec{0}$, which gives
\begin{equation}
    \begin{aligned}
        \vec{m}_A &= -N\vhat{z},\quad &
        \vec{m}_B &= +N\vhat{z}, \\
        \vec{m}_C &= -N\vhat{z},\quad &
        \vec{m}_D &= +N\vhat{z},
    \end{aligned}
\end{equation}
which is reminiscent of a spin-ice ``two-in/two-out'' configuration. Lastly, we have the parallel phase
where we consider $\vec{N}_1 = \vec{N}_2 = \vec{N}_3=4N\vhat{z}/3$. This gives
\begin{equation}
    \begin{aligned}
        \vec{m}_A &= -\frac{N}{3}\vhat{z}, &\quad
        \vec{m}_B &= N\vhat{z}, \\
        \vec{m}_C &= -\frac{N}{3}\vhat{z}, &\quad
        \vec{m}_D &= -\frac{N}{3}\vhat{z}.
    \end{aligned}
\end{equation}
Like the single-$N$ case ($\alpha=0$) in the kagom\'e example, this phase is compensated at leading order but has unequal moments.

The epikernels can be worked out explicitly by solving for each $g \in O_h$ the equation
\begin{equation}
\label{eq:t2g-epi}
\begin{pmatrix} \vec{N}_1 \\ \vec{N}_2 \\ \vec{N}_3 \end{pmatrix} =
\mat{\Gamma}(g)
\begin{pmatrix} \mat{R}\vec{N}_1 \\  \mat{R}\vec{N}_2 \\ \mat{R}\vec{N}_3 \end{pmatrix},
\end{equation}
where $\mat{\Gamma}(g)$ is a three-by-three matrix, not acting on the spin indices.
As in the kagom\'e case, once the solution spaces for each $g$ are known we can identify the relevant subgroup $\mathcal{H} \leq \mathcal{G}$ via the elements leaving a common subspace invariant. 

For the collinear phases the structure of the spin-only
group is identical to that of Sec.~\ref{sec:kagome}. For the single-$N$ or parallel phases we thus require a choice of sign $\gamma$ where $\mat{R}\vhat{z} = \gamma\vhat{z}$.

For the single-$N$ phase, we try to solve for each $g$ in $m\overline{3}m$
$$
\begin{pmatrix}
    N \\ 0 \\ 0
\end{pmatrix}=
\gamma 
\mat{\Gamma}(g)
\begin{pmatrix}
    N \\ 0 \\ 0
\end{pmatrix},
$$
where $\mat{\Gamma}(g)$ is the $T_{2g}$ representation.
This equation has solution only when this matrix preserves or reverses the $(1,0,0)$ vector. The elements $1$, $2_{x}$, $2_{[011]}$, $2_{[01\bar{1}]}$ have $\gamma=1$ while $4_{x}$, $4_{x}^{-1}$, $2_{y}$ and $2_{z}$ have $\gamma=-1$. We have only listed proper rotations since $T_{2g}$ is inversion even. With inversion, these generate a $4/mmm$ ($D_{4h}$) subgroup of $m\overline{3}m$. By restoring the improper elements, the epikernel is then $b_{\infty} \times {}^{\bar{1}} 4 / {}^{1}m {}^{1}m {}^{\bar{1}} m$.

For the parallel phase we need to solve a similar equation
$$
\begin{pmatrix}
    N \\ N \\ N
\end{pmatrix}=
\gamma 
\mat{\Gamma}(g)
\begin{pmatrix}
    N \\ N \\ N
\end{pmatrix}.
$$
This requires that the element preserves or reverses the $(1,1,1)$ vector. No elements reverse this vector, and those that preserve form a $\overline{3}m$ ($D_{3d}$) subgroup of $m\overline{3}m$ with the signs, $\gamma$, corresponding to the $A_{1g}$ irrep. The epikernel is then $b_{\infty} \times {}^{1}\bar{3}{}^{1}m$.

For the orthogonal phase, unlike the other examples considered so far, the spin-only group is trivial.
If we set $\vhat{N}_1=\vhat{x}$, $\vhat{N}_2=\vhat{y}$, $\vhat{N}_3=\vhat{z}$, the defining equation for the epikernels is then
\begin{subequations}
\begin{align}
    N \vhat{x} &= N\left(\Gamma_{11}(g) \mat{R}\vhat{x} + \Gamma_{12}(g) \mat{R}\vhat{y}+\Gamma_{13}(g) \mat{R}\vhat{z}\right), \\
    N \vhat{y} &= N\left(\Gamma_{21}(g) \mat{R}\vhat{x} + \Gamma_{22}(g)\mat{R}\vhat{y} +\Gamma_{23}(g) \mat{R}\vhat{z}\right),\\
    N \vhat{z} &= N\left(\Gamma_{31}(g) \mat{R}\vhat{x} + \Gamma_{32}(g) \mat{R}\vhat{y}+\Gamma_{33}(g) \mat{R}\vhat{z}\right).
\end{align}
\end{subequations}
Expanding the matrix-vector products yields
$$
\begin{pmatrix}
    N\vhat{x} \\ N\vhat{y} \\ N\vhat{z}
\end{pmatrix}
= 
\mat{\Gamma}(g) \trp{\mat{R}} \begin{pmatrix}
    N\vhat{x} \\ N\vhat{y} \\ N\vhat{z}
\end{pmatrix}.
$$
To satisfy this equation we find the same condition as in the two-dimensional
case, $ \mat{R} = \mat{\Gamma}(g)$. Whether this rotation is proper or improper
specifies the pairing with time-reversal. Thus, the fourfold rotations and
diagonal twofold rotations are paired with time-reversal. The epikernel is
isomorphic to $m\overline{3}m$ and is given explicitly by
${^{\overline{4}_{z}}4/^{2_{z}}m^{3_{xyz}}3 ^{m_{xy}}2/^{m_{xy}}m}$.

Only a small subset of possible phases is realized in the phase diagram Fig.~\ref{fig:phase-diagram-3d} of Eq.~(\ref{eq:3D_cub}), because the free energy is truncated at quartic order. Including higher-order terms may realize phases other than these three. The epikernel analysis described here also reveals symmetry-distinct states beyond the generic quartic minima. In Appendix~\ref{app:pyro_tables}, we give a complete enumeration of the collinear and mutually orthogonal epikernels for the pyrochlore $T_{2g}$ order parameter.

\section{Observables}\label{sec:observables}
Having laid out the framework for understanding the possible non-relativistic symmetry-broken phases via Landau theory, we turn our attention to how observable quantities depend on these order parameters. Observable quantities can be viewed as functions of the equilibrium state, characterized by the order parameters $\vec{N}_{a}$. In particular, near the phase transition where the $\vec{N}_{a}$ are small, these observables may be expanded into powers of our order parameters. At each order in this expansion, the symmetry-allowed terms are combinations of the $\vec{N}_{a}$ transforming the same way as the observable, as we will describe throughout this section.

Because the order parameters are generically higher-dimensional objects, carrying with them spin and spatial structure, it is desirable to first classify the kinds of products of $\vec{N}_{a}$ that may appear at each order of an expansion. In doing so, we will restrict ourselves to the linear, quadratic and cubic order in the Landau theory. This is sufficient for a wide variety of observables; the extension to higher order proceeds analogously, if tediously.

As the $\vec{N}_{a}$ are themselves time-reversal odd, observables even (odd) under time-reversal must exclusively consist of even (odd) powers of the $\vec{N}_{a}$: linear and cubic functions are time-reversal odd, whereas quadratic functions are even under time-reversal. The inequivalence of the time-reversal even and odd co-representations of $\mathcal{G}$ is discussed in Appendix~\ref{subapp:reality}. 

The spin structure of the observable further restricts the form of the products; all of the quantities we consider here are either scalars or vectors in the spin space. Thus, an observable's spin space transformation as a scalar or vector, in combination with parity under time-reversal ($T$), leads to four types of products of the $\vec{N}_{a}$ up to cubic order, summarized in Table~\ref{tab:products_observables}. These can be grouped into familiar scalar and pseudo-scalar quantities like $1$, $\vec{N}_a\cdot \vec{N}_b$ and $\vec{N}_a \cdot (\vec{N}_b \times \vec{N}_c)$ as well as 
vectors and pseudo-vectors $\vec{N}_a$, $\vec{N}_a\times \vec{N}_b$ and
$(\vec{N}_a \cdot \vec{N}_b)\vec{N}_c$.

\begin{table}[t]
\setlength{\tabcolsep}{5pt}
\renewcommand{\arraystretch}{1.3}
    \centering
\begin{tabular}{ccccc}
    \hline\hline
     & \multicolumn{2}{c}{$T$ odd}  & \multicolumn{2}{c}{$T$ even}  \\[0em]
    & Product & Rep & Product & Rep\\
    \hline
    \multirow{2}{*}{Spin scalar}
        & $\vec{N}_{a}\cdot (\vec{N}_{b}\times \vec{N}_{c})$
        & $\{\Gamma^{3}\}$
        & $1$
        & $\Gamma_{0}$ \\
    & & & $\vec{N}_{a}\cdot\vec{N}_{b}$ & $[\Gamma^{2}]$ \\
    \hline
    \multirow{2}{*}{Spin vector}
        & $\vec{N}_{a}$
        & $\Gamma$
        & \multirow{2}{*}{$\vec{N}_{a}\times\vec{N}_{b}$}
        & \multirow{2}{*}{$\{\Gamma^{2}\}$} \\
        & $(\vec{N}_{a}\cdot\vec{N}_{b})\vec{N}_{c}$
        & $[\Gamma^{2}]\otimes \Gamma$
        & & \\
    \hline\hline
\end{tabular}
    \caption{Four classes of product terms that may appear in the expansion of an observable up to cubic order, in terms of the order parameters $\vec{N}_{a}$, based on its parity under time-reversal, $T$, and its behavior in spin-space as a scalar or a vector.}
    \label{tab:products_observables}
    \vspace{-0.8cm}
\end{table}

Having classified the spin  transformations of each product, we now specify the
real space transformation law of the identified product terms. Since the
$\vec{N}_{a}$ transform under the irrep $\Gamma$ in real space,
$\vec{N}_{a}\cdot \vec{N}_{b}$ and $\vec{N}_{a}\times \vec{N}_{b}$ transform as
the symmetric and antisymmetric square representations of $\Gamma$, denoted by
$[\Gamma^{2}]$ and $\{\Gamma^{2}\}$, respectively. The third order product
$\vec{N}_{a}\cdot(\vec{N}_{b}\times\vec{N}_{c})$ is antisymmetric under any
permutation of the constituent vectors, and thus it transforms as the fully
antisymmetric cube $\{\Gamma^{3}\}$. On the other hand,
$(\vec{N}_{a}\cdot\vec{N}_{b})\vec{N}_{c}$ transforms as $[\Gamma^{2}]\otimes
\Gamma$. These transformation laws are listed in
Table~\ref{tab:products_observables}.

We now consider observables of known time-reversal parity and spin structure,
whose real space transformation law is described by a representation $D$ (which
may be irreducible or reducible) of the crystal point group, $G$. We aim to
construct the lowest-order polynomial in the $\vec{N}_{a}$ that reproduces the
observable's time-reversal parity and spin structure, while also transforming as
(a component of) $D$. Practically, this means we pick product terms from
Table~\ref{tab:products_observables} to satisfy the spin structure and
time-reversal parity, and we then verify whether the representation of these
products shares any irreps with $D$. We carry out this program for the time-odd
vectors in Sec.~\ref{subsec:obs_odd_vectors}, for time-even vectors in
Sec.~\ref{subsec:obs_even_vectors}, and for time-even and time-odd scalars in
Sec.~\ref{subsec:obs_even_scalars} and Sec.~\ref{subsec:obs_odd_scalars},
respectively.

In real space, the observables we will consider transform as tensors,
constructed from polar and axial vectors, and their symmetrized or
antisymmetrized tensor powers. To simplify our discussion, we utilize the Jahn
notation to specify tensor symmetriesthroughout~\cite{Jahn1949,Etxebarria2019_mtensor,Etxebarria2025_STENSOR}. In the
Jahn notation, $V$ denotes a polar vector and $e$ a pseudoscalar (odd under
improper operations), so $eV$ is an axial vector, while $[V^{n}]$ and
$\{V^{n}\}$ denote rank-$n$ tensors symmetric or antisymmetric under index
transpositions. We provide the decompositions of some of these representations
into irreps for all point groups in Table~\ref{tab:decomps}.

For each observable property, we establish whether mutual collinearity,
 coplanarity or non-coplanarity restricts the effect, perhaps
 depending on the crystal system. Constraints from spin group symmetry have been
 recently explored \cite{Etxebarria2025_STENSOR,Watanabe2024_Disentangling,Xiao2026_TensorSymm};
 we provide a complementary perspective that immediately follows from the order
 parameter. We then apply our analysis to the kagom\'e, square plaquette, and
 pyrochlore examples developed in Sec.~\ref{sec:examples}~\footnote{ When both
 orbital and spin contributions are possible we take only the spin contribution,
 unless otherwise stated. The symmetry properties of the orbital contributions
 to various equilibrium and transport effects under spin point group symmetry
 have been outlined in Ref.~\cite{Etxebarria2025_STENSOR}.}.

\subsection{Time-odd spin vectors}\label{subsec:obs_odd_vectors}
As we can see from Table~\ref{tab:products_observables}, time-reversal odd vectors may appear at linear or at cubic order. We will explore many observables appearing linearly in Sec.~\ref{subsubsec:obs_odd_vectors_1}, while for the cubic expansion we restrict our focus to the net magnetization in Sec.~\ref{subsubsec:obs_odd_vectors_3}. At the linear order, we will analyze the inverse magnetoelectric effect, piezomagnetism, the symmetric and antisymmetric time-odd spin conductivities, and multipolar spin density distributions. Further, we will see that for any non-trivial irrep $\Gamma$, a net magnetization cannot appear at the linear order -- the lowest order at which it may appear is cubic, whereas the RMOs remain compensated at all orders. 

\subsubsection{Linear order time-odd spin vectors}\label{subsubsec:obs_odd_vectors_1}
Here we begin with the inverse magnetoelectric effect, where the magnetoelectric tensor transforms as $D=V$, and we work our way up in spatial complexity to the multipolar spin density distributions where $D = [V^{n}]$. From Table~\ref{tab:products_observables}, a linear coupling will be allowed when $\Gamma$ is contained in $D$.\\

\noindent \textit{Inverse magnetoelectric effect.} The inverse magnetoelectric effect generates a magnetic moment $\vec{M}$ by the application of an electric field $\vec{E}$ \cite{Etxebarria2019_mtensor,Etxebarria2025_STENSOR}, and at zero spin-orbit coupling (SOC), it is governed by 
$$
\vec{M} = \sum_i \vec{\alpha}_{i}E_{i}.
$$ 
Bold vector notation specifies spin space components while lower indices denote real space components. Whether $\Gamma$ is contained in $V$ can be discerned by the decomposition of $V$ into point group irreps, provided in Table~\ref{tab:decomps}.

The cubic point groups forbid $\vec{\alpha}_{i}$ in their collinear orders, because these groups do not have one-dimensional irreps in the decomposition of $V$.  It is also forbidden for coplanar cubic orders as $V$ is irreducible and three-dimensional.  Likewise, for the irreducible coplanar RMOs we see that systems with triclinic, monoclinic, or orthorhombic point groups also forbid the inverse magnetoelectric effect, as these groups do not possess two-dimensional irreps, and so none will appear in the decomposition of $V$. For irreducible non-coplanar RMOs, only the cubic point groups admit an inverse magnetoelectric effect, as these are the only point groups for which $V$ contains a three-dimensional irrep. 

We can illustrate this for the examples of Sec.~\ref{sec:examples}. On the kagom\'e lattice, $E_2$ is not contained in $V$, and so the corresponding coplanar RMO does not exhibit the inverse magnetoelectric effect. For the square plaquette lattice $E$ does appear in $V$ and gives non-zero $\vec{\alpha}_{x}$ and $\vec{\alpha}_{y}$ within the spin plane. Indeed, this is expected for time-odd vector observables in coplanar systems: spin components perpendicular to the plane are forbidden, while components in the plane are allowed. This is consistent with the findings of Ref.~\cite{Etxebarria2025_STENSOR}. On the pyrochlore lattice, $V =  T_{1u}$ does not overlap with $T_{2g}$, so the magnetoelectric effect is forbidden.  \\

\noindent \textit{Piezomagnetism \& symmetric time-odd spin conductivity.} The piezomagnetic effect generates a magnetic moment $\vec{M}$ by the application of strain $\epsilon_{ij}$, governed by \cite{Etxebarria2019_mtensor,Etxebarria2025_STENSOR} $$
\vec{M} = \sum_{ij}\vec{\Lambda}_{ij}\epsilon_{ij}.
$$
The piezomagnetic tensor $\vec{\Lambda}_{ij}$ is a symmetric rank-2 tensor in real space, $D=[V^{2}]$. 

The spin conductivity tensor $\vec{\sigma}_{ij}$, whose $\sigma^{a}_{ij}$ component characterizes $a$-polarized spin currents in the $i$ direction generated by applied electric fields along $j$, is governed by 
$$
\vec{J}_{i} = \sum_j \vec{\sigma}_{ij} E_{j}.
$$
From Eq.~(\ref{eq:sc_-+}) we see that the time-odd index-symmetric component of the spin conductivity, $\vec{\sigma}_{[ij]}^{-}$, and the piezomagnetic tensor $\vec{\Lambda}_{ij}$ transform identically. The symmetry properties of the spin conductivity in relativistic systems have been explored previously \cite{ GonzalezHernandez2021_SpinCond,Xiao2026_TensorSymm}, and recently in the context of non-relativistic spin group symmetries \cite{Etxebarria2025_STENSOR, Watanabe2024_Disentangling}. 

\begin{table}[t]
\centering
\caption{Allowed $\vec{\Lambda}_{ij}$ (or equivalently, $\vec{\sigma}_{[ij]}^{-}$) for the irreducible collinear (upper), coplanar (middle) and non-coplanar (lower) RMOs. Kagom\'e and square plaquette cannot host non-coplanar RMOs as their point groups have no three-dimensional irreps, while the pyrochlore lattice does not host coplanar irreducible RMOs. We have omitted the ferromagnetic phase on the square plaquette lattice. }
\label{tab:rmo_piezo}
\small
\setlength{\tabcolsep}{4pt}
\renewcommand{\arraystretch}{1.3}
\begin{tabular}{ccc}
\hline\hline
Kagom\'e & Square plaquette & Pyrochlore \\
\hline
$A_{1}$ & $B_{2}$ & $A_{1g}$ \\
\begin{tabular}[c]{@{}c@{}}$\vec{\Lambda}_{xx}+\vec{\Lambda}_{yy}\sim\vec{N}$\\ $\vec{\Lambda}_{zz}\sim\vec{N}$\end{tabular}
 & $\vec{\Lambda}_{xy}\sim\vec{N}$
 & \begin{tabular}[c]{@{}c@{}}$\vec{\Lambda}_{xx}+\vec{\Lambda}_{yy}+\,\vec{\Lambda}_{zz}\sim\vec{N}$\end{tabular} \\
\hline
$E_{2}$ & $E$ & 
\\
\begin{tabular}[c]{@{}c@{}}$\vec{\Lambda}_{xx}-\vec{\Lambda}_{yy}\sim\vec{N}_{1}$\\ $\vec{\Lambda}_{xy}\sim\vec{N}_{2}$\end{tabular}
 & \begin{tabular}[c]{@{}c@{}}$\vec{\Lambda}_{xz}\sim\vec{N}_{1}$\\ $\vec{\Lambda}_{yz}\sim\vec{N}_{2}$\end{tabular}
 &
 \\ \hline
 & & $T_{2g}$ \\
 & & \begin{tabular}[c]{@{}c@{}}$\vec{\Lambda}_{yz}\sim\vec{N}_{1}$, \, $\vec{\Lambda}_{xz}\sim\vec{N}_{2}$\\ $\vec{\Lambda}_{xy}\sim\vec{N}_{3}$\end{tabular} \\
\hline\hline
\vspace{-0.5cm}
\end{tabular}
\end{table}

In every point group, $[V^{2}]$ contains the trivial irrep. Non-trivial one-dimensional irreps, corresponding to compensated collinear phases, appear in $[V^{2}]$ only for the monoclinic, orthorhombic and tetragonal point groups. For irreducible coplanar RMOs, $[V^{2}]$ contains (real) two-dimensional  irreps in all tetragonal, trigonal, hexagonal and cubic point groups. Only the cubic groups allow non-zero $\vec{\Lambda}_{ij}$ or $\vec{\sigma}_{[ij]}^{-}$ in irreducible non-coplanar RMOs, as these are the only point groups with three-dimensional irreps; in all cases one is contained in $[V^{2}]$. 

In Table~\ref{tab:rmo_piezo}, we express the components of $\vec{\Lambda}_{ij}$ (or equivalently $\vec{\sigma}^{-}_{[ij]}$) transforming as the $\vec{N}_{a}$ for the kagom\'e, square plaquette and pyrochlore lattices of Sec.~\ref{sec:examples}. These components of $\vec{\Lambda}_{ij}$ thus have contributions linear in the $\vec{N}_a$. From the expansions, it is clear that a time-odd spin vector in collinear systems must be aligned with the collinear axis, whereas it must lie in the plane for coplanar systems, consistent with Ref.~\cite{Etxebarria2025_STENSOR}.
\\

\noindent \textit{Antisymmetric time-odd spin conductivity.} From Eq.~(\ref{eq:sc_def}), the spin conductivity also has a time-odd component antisymmetric in its indices, $\vec{\sigma}_{\{ij\}}^{-}$, identified in Eq.~(\ref{eq:sc_--}) transforming as $D = \{V^{2}\}$. Only the cubic point groups forbid $\vec{\sigma}_{\{ij\}}^{-}$ in irreducible collinear RMOs, as their decompositions do not contain any one-dimensional irreps. Simultaneously, only the cubic point groups permit $\vec{\sigma}_{\{ij\}}^{-}$ in irreducible non-coplanar RMOs, as only they contain three-dimensional irreps in $\{V^{2}\}$. The irreducible coplanar RMOs forbid $\vec{\sigma}_{\{ij\}}^{-}$ in the triclinic, monoclinic, orthorhombic and cubic point groups.

In the kagom\'e and pyrochlore examples, the decompositions of Eqs.~(\ref{eq:kagome_decomposition}) and (\ref{eq:pyrochlore_decomposition}) do not share irreps with $\{V^{2}\}$, so RMOs on these lattices forbid linear $\vec{\sigma}_{\{ij\}}^{-}$. In the $E$ phase of the square plaquette lattice, $\vec{\sigma}_{\{xz\}}^{-}\sim\vec{N}_{1}$ and $\vec{\sigma}_{\{yz\}}^{-}\sim\vec{N}_{2}$.\\

\noindent \textit{Multipolar spin density distribution.} Ferroically ordered multipoles within each unit cell can also be examined within this framework. Multipoles of the form $$\int d^{3}\vec{r}\, [r_{\mu_{1}}...r_{\mu_{n}}] \vec{m}(\vec{r}),$$ where $[...]$ indicates symmetrization and the $r_{\mu_{i}}$ indicate the Cartesian coordinates, have been discussed in the context of altermagnetism and unconventional antiferromagnetism \cite{Bhowal2024_dwave,Bhowal2024_triakontadipolar,HayamiMultipole1,HayamiMultipole2,HayamiMultipole3,clusterMultipoles2017,schiff2025collinear,Radaelli2024_tensorial_approach,mcclarty2024landau,jaeschke2025atomic,Buiarelli2025_multipolesAM}. They transform as time-odd spin vectors, and as $D = [V^{n}]$ in the real space. Decompositions for $V$, $[V^{2}]$, $[V^{3}]$ and $[V^{4}]$ are provided in Table~\ref{tab:decomps}.

Irreducible orders with $\Gamma$ one-dimensional, non-trivial and inversion even
correspond to altermagnets~\cite{mcclarty2024landau,schiff2025collinear},
possessing multipoles with $n=2$, 4 or 6 ($d$-, $g$- or $i$-wave) reflecting
non-relativistic alternating spin (chirality) splitting of electronic (magnonic)
bands, in addition to odd order $n$ multipoles in the non-centrosymmetric altermagnets.

The triclinic, monoclinic and orthorhombic systems always only contain
one-dimensional irreps, corresponding to multipoles in irreducibly collinear
orders. For even $n$, all remaining point groups allow multipole components in
irreducible coplanar RMOs, while for $n=1$ and $n=3$ the coplanar RMOs
have no multipoles in the cubic point groups. The irreducible non-coplanar RMOs couple to
multipoles only in the cubic groups.

In the kagom\'e example, the $E_{2}$ coplanar RMO will produce $n\geq 2$
multipoles, i.e., octupoles or higher, while the ferromagnetically ordered
trivial $A_{1}$ irrep will produce multipoles at all orders. In the pyrochlore
example, the $T_{2g}$ non-coplanar RMO produces only even multipoles (here
$n=2,4$). Ferromagnetic ordering corresponding to the trivial irrep will also
produce non-trivial multipoles for $n=2,4$. The square-plaquette $E$ order instead permits
$n=1$ magnetic quadrupoles, while a ferromagnetic order has an $n=0$ magnetic
dipole.

\subsubsection{The cubic contribution to the net magnetization}\label{subsubsec:obs_odd_vectors_3}
At zero spin-orbit coupling the net magnetization, $\vec{M}$, transforms as a vector in spin space, and is odd under time-reversal. In real space, it is uniform and transforms trivially. Thus, $D$ is the trivial irreducible representation. From Table~\ref{tab:products_observables}, then, we see $\vec{M}$ can only appear at the linear order if our $\vec{N}_{a}$ have $\Gamma$ equal to the trivial irrep. Whenever $\Gamma$ is non-trivial, the magnetization is at least cubic in the $\vec{N}_{a}$. We may generically express the magnetization as
\begin{equation}\label{eq:generic_M}
\vec{M} = \sum_{a} f_{a}(\mat{Q}) \vec{N}_{a}
\end{equation}
where $a$ runs from 1 to the dimension of $\Gamma$, $|\Gamma|$, and $\mat{Q}$ is the Gram matrix of Eq.~(\ref{eq:gram-matrix}). For the cubic order expansion of $\vec{M}$, the coefficients $f_{a}(\mat{Q})$ are linear functions of $\mat{Q}$, and the scalar products comprising $\mat{Q}$ make Eq.~(\ref{eq:generic_M}) contain the terms identified in Table~\ref{tab:products_observables}. Higher order contributions to $\vec{M}$ are obtained by more general coefficient functions. We first show that the orthogonal phase for non-trivial $\Gamma$ is compensated at all orders.

For $\vec{M}$ to transform trivially under $G$, from Eqs.~(\ref{eq:N-matrix-transform}) and (\ref{eq:gram-matrix}), it follows that for every $g$ 
\begin{equation} \mat{f}\left(\Gamma(g)\mat{Q}\Gamma(g)^{\mathsf{T}}\right) = \Gamma(g)\mat{f}(\mat{Q}),
\end{equation} where $\mat{f} = \begin{pmatrix} f_{1}, \dots, f_{|\Gamma|} \end{pmatrix}^{\mathsf{T}}$ is the vector of coefficients in Eq.~(\ref{eq:generic_M}). This means that $\mat{f}$ transforms as $\Gamma$. In the orthogonal phase, we have seen in Sec.~\ref{sec:ortho-proof} that $\mat{Q} = N^{2}\mat{1}$, in which case 
\begin{equation}
    \mat{f}(N^{2}\mat{1}) = \Gamma(g) \mat{f}(N^{2}\mat{1}),
\end{equation} and for non-trivial $\Gamma$ the only fixed vector is $\mat{f} = 0$. Thus, for the orthogonal phase of $\Gamma$, $\vec{M}$ is zero at all orders. 

Lower-symmetry phases descending from non-trivial irreps, however, need not be compensated. A moment cubic in the $\vec{N}_{a}$ will be allowed by symmetry whenever the trivial irrep is contained in $[\Gamma^{2}]\otimes \Gamma$, following Table~\ref{tab:products_observables}. Because our $\Gamma$ are real, another way to phrase this criterion is that cubic $\vec{M}$ is allowed whenever $[\Gamma^{2}]$ contains $\Gamma$. 

When $\Gamma$ is one-dimensional and non-trivial, $[\Gamma^{2}]$ is trivial while $\Gamma$ is not. This eliminates the triclinic, monoclinic and orthorhombic groups, as they only contain one-dimensional irreps. For the two- and three-dimensional irreps, as a general rule: whenever $\Gamma(g) = - \mat{1}$ for some element $g$, $\Gamma$ cannot be contained in $[\Gamma^{2}]$ or in any even power of $\Gamma$, for that matter, and hence will be compensated at all orders. This eliminates the tetragonal groups. In the remaining trigonal, hexagonal and cubic groups, this eliminates the $E_{u}$, $E_{1}$, $E''$, $E_{1g}$, $E_{1u}$ and $E_{2u}$, as well as $T_{u}$, $T_{1u}$ and $T_{2u}$. Additionally, the $T_{1}$ irrep of $432$ ($O$) and $\overline{4}3m$ ($T_{d}$) and the $T_{1g}$ irrep of $m\overline{3}m$ ($O_{h}$) fail to contain themselves in the symmetric square (see Table~\ref{tab:decomps}).

To make this analysis concrete, consider the cubic contribution to the net magnetization on the kagom\'e, the square plaquette, and the pyrochlore lattices at zero SOC, for phases with non-trivial $\Gamma$.

In the $E_{2}$ phase of the kagom\'e lattice, $[E_{2}^{2}]$ contains $E_{2}$ and yields a unique third-order coupling
\begin{equation}\label{eq:kagome_M3}   \vec{M}\sim\vec{N}_{1}\left(|\vec{N}_{1}|^{2}-|\vec{N}_{2}|^{2}\right)-2\vec{N}_{2}\left(\vec{N}_{1}\cdot\vec{N}_{2}\right).
\end{equation}
In the degenerate collinear phase of Eq.~(\ref{eq:degen}), this gives $\vec{M}\propto N^{3}\cos3\alpha\,\vhat{N}$. The moment vanishes for $\alpha=(2n+1)\pi/6$, where the mirrors are paired with time reversal (see Fig.~\ref{fig:kagome-collinear-orders}), but is otherwise finite. When $\vec{M}$ is finite, a ferrimagnet forms. In the orthogonal phase Eq.~(\ref{eq:kagome_M3}) vanishes, as required.

On the square plaquette lattice, $[E^{2}]$ never contains $E$, and more generally no even power of $E$ contains itself, which is easily seen from the fact that $E(C_{2z})=-\mat{1}$. Thus, every resulting phase is compensated to all orders.

In the $T_{2g}$ phase of the pyrochlore lattice, $[T_{2g}^{2}]$ contains $T_{2g}$ once, leading to
\begin{equation}\label{eq:pyrochlore_M3}
    \vec{M}\sim\vec{N}_{1}(\vec{N}_{2}\cdot\vec{N}_{3})+\vec{N}_{2}(\vec{N}_{3}\cdot\vec{N}_{1})+\vec{N}_{3}(\vec{N}_{1}\cdot\vec{N}_{2}),
\end{equation}
which vanishes in the single-$N$ and orthogonal phases but is finite in the parallel phase, $\vec{M}\propto N^{3}\vhat{N}$. Here, the unequal sublattice moments depicted in Fig.~\ref{fig:pyrochlore}(b) are not forbidden from developing a net moment.

\subsection{Time-even spin vectors}\label{subsec:obs_even_vectors}

From Table~\ref{tab:products_observables}, time-even spin vectors of the form $\vec{N}_a \times \vec{N}_b$ are only non-zero for non-collinear order. Here, we will use the time-even channel of the spin conductivity (see Appendix~\ref{app:spin_conductivity}) as our example observable.

For our three examples, none of the higher-dimensional $\Gamma$ have $\{\Gamma^{2}\}$ sharing an irrep with $[V^{2}]$. Thus, $\vec{\sigma}^{+}_{[ij]}$ is forbidden at second order. By contrast, $\{\Gamma^{2}\}$ is contained in $\{V^{2}\}$ for all three. In the $E_{2}$ phase of the kagom\'e lattice and the $E$ phase of the square plaquette lattice, $\{\Gamma^{2}\}=A_{2}$, giving $\vec{\sigma}^{+}_{\{xy\}}\sim\vec{N}_{1}\times\vec{N}_{2}$,
while in $T_{2g}$ of the pyrochlore lattice $\{T_{2g}^{2}\}=T_{1g}$, resulting in
$
\vec{\sigma}^{+}_{\{yz\}}\sim\vec{N}_{2}\times\vec{N}_{3}$, $\vec{\sigma}^{+}_{\{zx\}}\sim\vec{N}_{3}\times\vec{N}_{1}$, and $
\vec{\sigma}^{+}_{\{xy\}}\sim\vec{N}_{1}\times\vec{N}_{2}$. 

In fact, we can make some general statements. Collinear RMOs always forbid $\vec{\sigma}^{+}_{\{ij\}}$, because $\{\Gamma^{2}\}$ for a one-dimensional irrep is null. For the irreducible coplanar RMOs in non-cubic point groups, $\{\Gamma^{2}\} = \det\Gamma$ is a one-dimensional irrep, transforming as the axial vector component aligned with the principal rotation axis; as such, it is always contained in $\{V^{2}\} \cong eV$, and in these cases $\vec{\sigma}^{+}_{\{ij\}}$ will be allowed. In the cubic groups, there is no invariant axis and $eV$ is three-dimensional, so $\vec{\sigma}^{+}_{\{ij\}}$ is forbidden for coplanar orders in the cubic groups. Finally, for the non-coplanar irreducible RMOs, $\{ \Gamma^{2} \} \cong \{V^{2}\}$ identically, and so $\vec{\sigma}^{+}_{\{ij\}}$ is allowed. By contrast, $\vec{\sigma}^{+}_{[ij]}$ requires $\{\Gamma^{2}\}$ to be contained in $[V^{2}]$, which occurs only for coplanar orders in $3$, $\bar{3}$, $4$, $\bar{4}$, $4/m$, $6$, $\bar{6}$ and $6/m$, and for irreducible non-coplanar orders in $23$ and $m\bar{3}$.

\subsection{Time-even spin scalars}\label{subsec:obs_even_scalars}

\begin{table}[t!]
\centering
\caption{In each example lattice of Sec.~\ref{sec:examples} with higher-dimensional $\Gamma$, listed in the first column, the second column lists the overlap of $[\Gamma^{2}]$ with $[V^{2}]$. The third column lists second-order contributions to the symmetric conductivity $\sigma_{[ij]}$.}
\label{tab:sigma_second_order}
\small
\setlength{\tabcolsep}{4pt}
\renewcommand{\arraystretch}{1.2}
\begin{tabular}{ccl}
\hline\hline
Phase & Irrep & $\sigma_{[ij]}$ \\
\hline
\multirow{3}{*}{\begin{tabular}{@{}c@{}}Kagom\'e\\ $E_{2}$\end{tabular}}
 & $A_{1}$ & $\sigma_{[xx]}+\sigma_{[yy]},\,\sigma_{[zz]}\sim|\vec{N}_{1}|^{2}+|\vec{N}_{2}|^{2}$ \\
 & \multirow{2}{*}{$E_{2}$} & $\sigma_{[xx]}-\sigma_{[yy]}\sim |\vec{N}_1|^2-|\vec{N}_2|^2$ \\
 & & $\sigma_{[xy]}\sim-\vec{N}_{1}\cdot\vec{N}_{2}$ \\
\hline
\multirow{3}{*}{\begin{tabular}{@{}c@{}}Square\\ plaquette $E$\end{tabular}}
 & $A_{1}$ & $\sigma_{[xx]}+\sigma_{[yy]},\,\sigma_{[zz]}\sim|\vec{N}_{1}|^{2}+|\vec{N}_{2}|^{2}$ \\
 & $B_{1}$ & $\sigma_{[xx]}-\sigma_{[yy]}\sim  |\vec{N}_1|^2-|\vec{N}_2|^2$ \\
 & $B_{2}$ & $\sigma_{[xy]}\sim\vec{N}_{1}\cdot\vec{N}_{2}$ \\
\hline
\multirow{6}{*}{\begin{tabular}{@{}c@{}}Pyrochlore\\ $T_{2g}$\end{tabular}}
 & $A_{1g}$ & $\sigma_{[xx]}+\sigma_{[yy]}+\sigma_{[zz]}\sim|\vec{N}_{1}|^{2}+|\vec{N}_{2}|^{2}+|\vec{N}_{3}|^{2}$ \\
 & \multirow{2}{*}{$E_{g}$} & $\sigma_{[xx]}-\sigma_{[yy]}\sim  |\vec{N}_1|^2-|\vec{N}_2|^2$ \\
 & & $\sigma_{[yy]}-\sigma_{[zz]}\sim  |\vec{N}_2|^2-|\vec{N}_3|^2$ \\
 & \multirow{3}{*}{$T_{2g}$} & $\sigma_{[xy]}\sim\vec{N}_{1}\cdot\vec{N}_{2}$ \\
 & & $\sigma_{[xz]}\sim\vec{N}_{1}\cdot\vec{N}_{3}$ \\
 & & $\sigma_{[yz]}\sim\vec{N}_{2}\cdot\vec{N}_{3}$ \\
\hline\hline
\end{tabular}
\end{table}

As an example observable for the time-even spin scalars, we choose the symmetric conductivity tensor, $\sigma_{[ij]}$. It is a rank-two tensor in real space, symmetrized in both of its indices, so that $D = [V^{2}]$. A contribution is generally present already in the paramagnetic phase.
From Table~\ref{tab:products_observables}, the leading magnetic correction must be constructed from scalar products $\vec{N}_{a}\cdot\vec{N}_b$. Indeed, it appears at second order in all three example lattices. Focusing on the higher-dimensional $\Gamma$, we decompose the symmetric squares of $E_{2}$ on kagom\'e, $E$ on the square plaquette lattice, and $T_{2g}$ in pyrochlore. Their overlap with $[V^{2}]$ (found in Table~\ref{tab:decomps}) is given in Table~\ref{tab:sigma_second_order}, with the resulting quadratic dependence of the $\sigma_{[ij]}$ on the $\vec{N}_{a}$.

From Table~\ref{tab:sigma_second_order} we see that the off-diagonal components of the symmetric conductivity are proportional to the scalar products between different $\vec{N}_{a}$. In the orthogonal phases, these terms vanish and the purely longitudinal conductivities may survive.

\subsection{Time-odd spin scalars}\label{subsec:obs_odd_scalars}

The time-odd scalar terms in Table~\ref{tab:products_observables} vanish whenever the $\vec{N}_{a}$ lie in the same plane. Thus, only non-coplanar orders may host time-odd scalars at the third order. For the irreducible orders, then, $\Gamma$ must be three-dimensional, and in this case $\{\Gamma^{3}\}=\det\Gamma$ is a one-dimensional irrep. The orbital magnetization and geometric Hall effect serve as practical examples, both transforming as an axial vector $eV$ in real space, for which it is known that non-coplanar order is required \cite{Watanabe2024_Disentangling,Etxebarria2025_STENSOR}. The orbital part of the pyrotoroidic effect, a polar vector $V$ in real space, and the spontaneous Faraday effect, transforming as $\{V^{2}\}$ under the point group, also fall within this class.

Irreducible, three-dimensional $\Gamma$ exist only in the cubic point groups. There, $V$, $eV$ and $\{V^{2}\}$ are all irreducible and three-dimensional, so none of them can contain the one-dimensional $\det\Gamma$. Non-coplanarity is thus necessary but not sufficient: for any order parameter arising from three-dimensional $\Gamma$, the orbital magnetization, geometric Hall, pyrotoroidic and spontaneous Faraday effects are forbidden at third order. 

\section{Discussion}
\label{sec:discussion}
The non-relativistic theories explored in the previous sections have been motivated by dipolar orders with zero propagation vector. However, related symmetry analyses may apply beyond this scenario. Examples include recently introduced atomic altermagnets \cite{jaeschke2025atomic} discussed in Sec.~\ref{subsec:atomicAM}, superfluid \het{}, and nematic and spin-nematic phases, discussed in Sec.~\ref{subsec:other_weak_soc}.

\subsection{Descendants of higher-dimensional irreps}\label{subsec:collinear_descendants}
The dimensionality of the order parameter’s irrep determines whether the magnetic order can be collinear, coplanar, or non-coplanar. In the case of one-dimensional irreps, the order parameter corresponds to a single $\vec{N}$ vector, whose direction fixes the magnetic dipole moments to be either parallel or antiparallel to it. Therefore, one-dimensional irreps can only produce collinear phases. Two-dimensional irreps contain two $\vec{N}$ vectors, $\vec{N}_1$ and $\vec{N}_2$, that can span a plane, resulting in a coplanar phase, or if they are parallel or antiparallel, define a collinear phase. Similarly, the three components, $\vec{N}_1$, $\vec{N}_2$, and $\vec{N}_3$ of a three-dimensional irrep can realize collinear, coplanar, and non-coplanar orders, too, depending on the dimension of the span of the three $\vec{N}_a$ vectors.  

As shown in Sec.~\ref{sec:ortho-proof}, among the higher-dimensional order parameters, only those whose components are mutually orthogonal and of equal length preserve the full symmetry of the parent point group (modulo spin rotations), corresponding to a regular magnetic order.  Any other solution, such as orthogonal components of unequal length, collinear $\vec{N}_a$, or a generic solution, necessarily lowers the point group symmetry. 

This property can help in interpreting the appearance of some of the collinear phases in the worked examples. 
As an illustration, in the square plaquette case, we obtain two collinear phases, the single-$N$ and parallel phases, that are not realizable while maintaining the full point group $4mm$ of the crystal. These two phases instead lower the real space symmetry to $mm2$. For example, the single-$N$ phase can be understood as a one-dimensional irrep of that point group. In this case it is either altermagnetic $B_1$ or $B_2$, depending on the choice of axis. Collinear phases of this type can then be understood using the framework of Ref.~\cite{schiff2025collinear} once the parent group and the relevant
Wyckoff positions in the lower-symmetry setting are identified.

This does not always have to correspond to an irrep of the lower-symmetry group. 
In the kagom\'e $\alpha=0$ phase, the reduced $mm2$ symmetry allows two independent collinear amplitudes, both transforming as $A_1$.
Symmetry in this subgroup therefore does not fix their relative magnitude. There are thus two inequivalent Wyckoff positions, each with an independent collinear ferromagnetic $A_1$ order parameter described by a reducible representation $A_1 \oplus A_1$, rather than the irrep found in the square plaquette case. The primary combination (a linear combination of the two $A_1$ irreps) inherited from the parent $E_2$ order and the secondary uniform magnetization both transform as $A_1$ in this lower-symmetry description.

The dimensional reduction described above, going from a two- or three-dimensional irrep to a lower-dimensional one, is not limited to collinear order. Indeed, a three-dimensional irrep can reduce to a two-dimensional one describing a coplanar order.
These kinds of order are not natural at the level of the quartic free energies discussed in Sec.~\ref{sec:landau}. Stabilization of these phases requires additional terms at higher order in the $\vec{N}_a$. However, we can still explore this systematically, by considering the solutions for the epikernels [Eq.~(\ref{eq:epikernel})] beyond these minima. We list
a subset of such epikernels, those with mutually orthogonal $\vec{N}_a$, in Table~\ref{tab:pyro_epi_perp_groups} of Appendix~\ref{app:pyro_tables}. For example, for the $T_{2g}$ irrep on the pyrochlore lattice the choice $\vec{N}_1 \perp \vec{N}_2$ with $|\vec{N}_1|=|\vec{N}_2|$ and $\vec{N}_3=0$ represents a coplanar order with epikernel $b^z_{\mathbb{Z}_2}\times{}^{\bar{4}_z}4/{}^{2_z}m ^{2_x}m ^{m_{xy}}m$ which is a coplanar spin group associated with the $4/mmm$ subgroup of $m\bar{3}m$. In addition, there are a variety of collinear phases beyond the single-$N$ and parallel phases discussed in Sec.~\ref{sec:pyrochlore}; we list all such collinear epikernels for the $T_{2g}$ pyrochlore irrep in Table~\ref{tab:Pyrochlore_groups_(i)} of Appendix~\ref{app:pyro_tables}.

\subsection{Atomic altermagnetism}\label{subsec:atomicAM}

\begin{figure*}[t!]
    \centering
    \includegraphics[width=\textwidth]{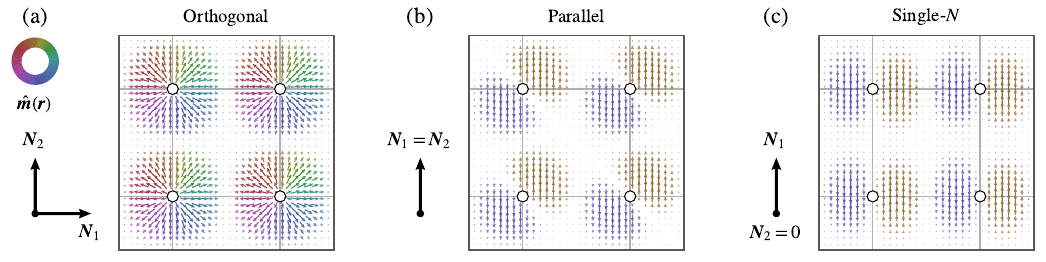}
\caption{
Magnetization density $\vec{m}(\vec{r})$ for multipolar order on the square lattice transforming spatially as $E$ of the point group $4mm$. The direction of $\vec{m}(\vec{r})$ is indicated by the color wheel, and lattice sites are marked by open circles. The three panels are the intra-atomic analogues of the three phases of Fig.~\ref{fig:phase-diagram-2d}, with the order parameters $\vec{N}_1$ and $\vec{N}_2$ describing components of the local (multipolar) magnetization density rather than distinct sublattice moments. (a) Orthogonal phase, $\vec{N}_1=N\vhat{x}$ and $\vec{N}_2=N\vhat{y}$: a coplanar magnetic quadrupolar order. (b) Parallel phase, $\vec{N}_1=\vec{N}_2=N\vhat{y}$: a collinear state with lobes along $\vhat{x}+\vhat{y}$.
(c) Single-$N$ phase, $\vec{N}_1=N\vhat{y}$ and $\vec{N}_2=0$: a collinear state with lobes along $\vhat{x}$.
}
    \label{fig:square-eu}
\end{figure*}

The examples considered thus far are microscopically realized through magnetic dipoles located on multiple sublattices within the crystallographic unit cell. However, the same Landau theories can arise even when there is only a \emph{single} magnetic site per unit cell, provided the magnetization density on this site has a non-trivial spatial structure.  This naturally describes the recently introduced atomic altermagnets~\cite{d2025altermagnetism,jaeschke2025atomic}, which can be seen as continuous, intra-atomic analogues of the multi-sublattice dipolar structures.

As a simple example, consider a square lattice with space group $P4mm$ (No.~99). Its point group is the same $4mm$ symmetry used in the square-plaquette example of Sec.~\ref{sec:tetragonal}, with the relevant order parameter in both cases transforming as the two-dimensional $E$ irrep. The magnetic structures that can be realized depend on the degrees of freedom at the occupied Wyckoff positions. At the $1a$ position, a conventional dipole can furnish only the trivial representation at zero wave vector, which must be a ferromagnet. To realize non-ferromagnetic spin point groups at the $1a$ position, we allow the local magnetization density to carry non-trivial spatial multipoles. Consider a pair of $p$-orbital-like form factors furnishing a basis of the $E$ irrep of $4mm$,
\begin{align*}
f_1(\vec{r}) &\equiv f(r) r_x, &
f_2(\vec{r}) &\equiv f(r) r_y,
\end{align*}
where $f(r)$ is a radial form factor. We write the magnetization density $\vec{m}(\vec{r})$ in terms of these states as
$$
\vec{m}(\vec{r}) = \sum_i \sum_{a} \vec{N}_a f_a(\vec{r}-\vec{r}_i),
$$
where $\vec{N}_1$ and $\vec{N}_2$ are the two components of the $E$ order parameter and $\vec{r}_i$ is the position of lattice site $i$. Since $(\vec{N}_1,\vec{N}_2)$ transforms as an $E$ doublet, the resulting Landau theory is precisely the two-dimensional tetragonal theory of Sec.~\ref{sec:landau2}. It admits the same three phases as the $E$-irrep order on the square plaquette lattice: a single-$N$ phase, a parallel phase, and an orthogonal phase.  Thus the distinction between the two realizations is microscopic rather than symmetry-theoretic.  

The orthogonal phase depicted in Fig.~\ref{fig:square-eu}(a) can be represented by the choice $\vec{N}_1 = N\vhat{x}$ and $\vec{N}_2 = N\vhat{y}$, which gives
$$
\vec{m}(\vec{r}) = N \sum_i\left( f_1(\vec{r}-\vec{r}_i)\vhat{x}+f_2(\vec{r}-\vec{r}_i)\vhat{y}\right).
$$
This is a coplanar ferroic magnetic quadrupolar texture.  Like the orthogonal phase in Sec.~\ref{sec:tetragonal}, the equal-length and orthogonal order parameters preserve the full $4mm$ point group modulo spin rotations.  This state is the smooth analogue of the radial configuration (Fig.~\ref{fig:tetragonal-orders}(a)) on the square plaquette: instead of four discrete moments pointing along different directions, the magnetization direction points radially outward and varies continuously in the region around a single magnetic site.

The two collinear phases also have direct analogues. For the single-$N$ phase (Fig.~\ref{fig:square-eu}(c)), $\vec{N}_1=N\vhat{y}$ and $\vec{N}_2=\vec{0}$ yields
$$
\vec{m}(\vec{r}) = N \vhat{y} \sum_i f_1(\vec{r}-\vec{r}_i),
$$
so that the magnetization is everywhere collinear while its amplitude changes sign across each site. Likewise, the parallel phase defined by $\vec{N}_1 = \vec{N}_2 = N\vhat{y}$ (Fig.~\ref{fig:square-eu}(b)) gives
$$
\vec{m}(\vec{r}) = N \vhat{y} \sum_i \left(f_1(\vec{r}-\vec{r}_i)+f_2(\vec{r}-\vec{r}_i)\right),
$$
corresponding to a collinear texture with lobes oriented along $\vhat{x}+\vhat{y}$. The fourfold spatial symmetry is reduced to an $mm2$ subgroup, as in Sec.~\ref{sec:tetragonal}. Thus the same single-$N$, parallel, and orthogonal phases need not be realized through discrete sublattice moments: they can equally well occur as smooth magnetization textures within a single crystallographic site. 

Conversely, the same $E$-irrep structures can be realized using ordinary magnetic dipoles once a Wyckoff position with sufficiently large multiplicity is available. For example, on the $4d$ Wyckoff site [$(xxz)$ and equivalents] of $P4mm$, the four magnetic sites furnish the same type of multi-sublattice realization considered in the square-plaquette example of Sec.~\ref{sec:tetragonal}. Suitable collinear arrangements~\cite{schiff2025collinear} realize conventional altermagnetic structures, while a pattern of four moments pointing radially outward, as in Fig.~\ref{fig:tetragonal-orders}(a), can realize the analogue of the orthogonal $E$ phase.  Thus, the atomic and multi-sublattice constructions are distinct microscopic realizations of the same $E$ order parameter, with the same Landau theory and spin space group structure.

\subsection{Connection to other weakly spin-orbit coupled phases}\label{subsec:other_weak_soc}

Our discussion of zero spin-orbit coupled magnetic phases has certain parallels with the physics of \het{} and nematic-spin-nematics. In this section we draw out some of the connections between the Landau theories for these systems.

\subsubsection{Superfluid \texorpdfstring{\het{}}{3He}}

\het{} is endowed with a nuclear spin one-half and, though the electronic configuration has zero orbital (and spin) angular momentum, Cooper pairing through a spin-fluctuation mediated interaction leads to a net internal angular momentum too. Relativistic effects coming from the nuclear dipolar interaction are weak so the \het{} superfluid enjoys an order parameter space composed of approximately independent
spin and spatial rotational symmetries $\sothree{}_{L} \times \sothree{}_{S}$ in addition to an overall phase $\uone_N$ due to number conservation. 

At millikelvin temperatures \het{} condenses into a spin-triplet $p$-wave superfluid phase. 
The pairing order parameter of a triplet superfluid can be written in terms of a two-by-two matrix
in spin space
$$
\Delta_{\alpha\beta}(\vec{k}) \equiv \left[(\vec{d}(\vec{k})\cdot\vec{\sigma}) i\mat{\sigma}_y\right]_{\alpha\beta}
$$
where $\vec{d}(\vec{k})$ is a complex vector encoding the orbital component and $\vec{\sigma}$ are the Pauli matrices. 
For a $p$-wave orbital state the wave-vector dependence is linear with
$$
\vec{d}(\vec{k}) \equiv \sum_{a=1}^3 \vec{d}_a \hat{k}_a
$$
where $a=1,2,3$ are three orthogonal spatial directions. This set of three
vectors $\vec{d}_a$ can be interpreted as a three-by-three complex matrix $\textbf{\textsf{d}}$ with
$\left[\textbf{\textsf{d}}\right]_{a,\mu} \equiv d^{\mu}_a$ where the $\vec{d}_a$ are the rows of $\textbf{\textsf{d}}$. This complex order parameter has $18$ real parameters. Under
a spin space rotation $[\mat{U}\|\mat{R}]$ this matrix transforms as
$$
\textbf{\textsf{d}} \rightarrow \mat{R} \textbf{\textsf{d}}\trp{\mat{U}}
$$
and rotational invariance in spin and orbital space can reduce the number of free parameters. 

The connection to our magnetic problem is immediate. The vectors $\vec{d}_a$ play essentially the same role as the N\'eel vectors $\vec{N}_a$, with the important distinction that the orbital index $a$ transforms under the continuous rotation group $\sothree_L$ rather than a discrete three-dimensional crystallographic representation. In addition, the superfluid order parameter is complex, allowing a richer set of phases than the real magnetic order parameters considered here.

Preserving all the rotational symmetries restricts the Ginzburg-Landau theory for the superfluid to have the form
\begin{widetext}
\begin{align}
F(\vec{d}_a) & = v_2 ( \vert \vec{d}_1\vert^2 + \vert \vec{d}_2\vert^2  + \vert \vec{d}_3\vert^2) + v_{4,1} \left\vert  \sum_a \vec{d}^2_a \right\vert^2 + v_{4,2} \sum_{a,b} \left( \vec{d}_a^*\cdot \vec{d}^*_b \right)  \left( \vec{d}_a\cdot \vec{d}_b \right) +  v_{4,3} \left(   \sum_a \vert \vec{d}_a\vert ^2 \right)^2 \nonumber \\
& + v_{4,4} \sum_{a,b} \left( \vec{d}_a^*\cdot \vec{d}_b \right)  \left( \vec{d}_a^*\cdot \vec{d}_b \right) + v_{4,5} \sum_{a,b} \left( \vec{d}^*_a\cdot \vec{d}_b \right)  \left( \vec{d}_a\cdot \vec{d}^*_b \right)
\end{align}
\end{widetext}
with one quadratic and five quartic invariant polynomials. The larger number of quartic terms reflects the complex nature of the \het{} order parameter. Restricting it to be real leaves only two independent quartic invariants, corresponding to the $\sothree_L$-symmetric limit $c_4=-b_4$ of our three-dimensional Landau theory [Eq.~(\ref{eq:3D_cub})].
The observed superfluid phases of \het{} can be found among the solutions obeying the so-called {\it unitary condition} $\vec{d}(\vec{k})\times \vec{d}^*(\vec{k})=0$ on the order parameter. This implies either (i) that $\vec{d}(\vec{k})$ is real (up to an overall phase) or, if complex, that (ii) $\vec{d}(\vec{k})=c(\vec{k}) \hat{\vec{n}}$ for real unit vector $\hat{\vec{n}}$ and complex $c(\vec{k})$. 

Some of the phases of \het{} have direct analogues to our non-relativistic Landau theories. 
However, since the magnetic phases are constrained both to have real order parameters and only crystalline
spatial symmetries, they only form a subclass of those phases realizable in \het{}.
 
The closest analogue is the \emph{Balian--Werthamer (BW) state}~\cite{balian1963superconductivity}. The order parameter is
$$
\textbf{\textsf{d}}_{\textsf{BW}} \equiv d \mat{1} 
$$
where $\mat{1}$ is the identity matrix and $d$ is a real amplitude. This preserves an $\sothree{}$
subgroup, $[\mat{R}\|\mat{R}]$, of the original symmetry group, where spin and space transform
identically. The BW phase is realized in \het{} at low pressures. Bearing in mind the freedom to choose spin and orbital components and the invariance of the energy under multiplication by a complex phase, the order parameter may be written 
$$
\vec{d}(\vec{k})_{\textsf{BW}} = d \vec{O} \hat{\vec{k}}
$$
where $\vec{O}$ is an orthogonal matrix and $d$ is complex. This state is isotropic in spin and orbital space, having therefore zero total angular momentum.

In terms of the three vectors $\vec{d}_a$, the BW state therefore consists of three mutually orthogonal vectors of equal magnitude, directly analogous to the orthogonal magnetic phases discussed above. The residual diagonal $\sothree{}$ symmetry is the continuous analogue of preserving the full spatial symmetry modulo a compensating spin rotation. In this sense, the BW phase is the closest analogue in \het{} to the regular magnetic orders arising naturally in our Landau theories.

The $A$ phase, realized in \het{} at elevated pressure and temperature, has features in common with our collinear orders, but differs in important aspects. It carries a complex order parameter
$$
\textbf{\textsf{d}}_{\textsf{A}} \equiv d (\vhat{u}+i\vhat{v})
\trp{\vhat{d}}
$$
where $\vhat{u}$, $\vhat{v}$ are real and orthogonal, $\vhat{d}$ is a real unit vector, and $d$ can be chosen to be real, since its phase can be absorbed into the complex vector part. In terms of the vectors $\vec{d}_a$
we are saying
$$
\vec{d}_a = d (\hat{u}_a +i \hat{v}_a) \vhat{d} \equiv |\vec{d}_a| e^{i\phi_a} \vhat{d}.
$$
Thus the $\vec{d}_a$ are collinear vectors with (potentially different) complex amplitudes.
This has a remnant $\sotwo$ symmetry about the fixed spin axis, like our collinear phases, but also a $\uone$ symmetry combining a gauge transformation with a spatial rotation about $\vhat{u} \times \vhat{v}$.

Other superfluid phases of \het{} realize structures that do not have such a direct correspondence with the magnetic phases considered here. The planar and polar phases provide examples in which only two or one independent orbital components participate, respectively, and are therefore loosely analogous to coplanar and collinear magnetic orders. The $A$ phase instead has an intrinsically complex order parameter with collinear spin vectors carrying different complex orbital amplitudes, illustrating possibilities absent for the real magnetic order parameters studied here. Still more general phases, including other inert states and pairing in higher orbital angular-momentum channels, are possible in principle, but we will not consider them further.

\subsubsection{Nematic-spin-nematic phases}

Spin-triplet Pomeranchuk instabilities provide another close analogue of the
non-relativistic Landau theories considered in this work~\cite{wu2004dynamic,wu2007fermi,fradkin2010nematic,hannappel2016}.
In an isotropic Fermi liquid, a Pomeranchuk instability in an orbital angular momentum channel $\ell>0$ can develop in the spin-triplet particle-hole channel. An order parameter $n_{\mu a}$ can be defined where $\mu$ is a spin index and $a$ labels the orbital harmonics in the
$\ell$ channel. Unlike the triplet pairing order parameter of
superfluid \het{}, this particle-hole order parameter is real. ~\footnote{Under time reversal, this order parameter has parity $(-1)^{\ell+1}$.
Thus, for odd $\ell$, the correspondence discussed below concerns the
rotational structure of the uniform Landau theory rather than
time-reversal parity.}
In three dimensions, the spin and orbital indices transform independently under
$\mathrm{SO}(3)_S$ and $\mathrm{SO}(3)_L$, respectively.  Thus, writing
the columns of $n_{\mu a}$ as spin vectors $\vec{n}_a$, these vectors
play essentially the same role as our $\vec{N}_a$ vectors.  The microscopic
origin, however, is different: the index $a$ here labels angular harmonics of the Fermi surface rather than
components of a real crystallographic point-group representation.

This correspondence is particularly transparent in two dimensions.  For a
circular Fermi surface, every $\ell>0$ channel has two real orbital
components $\cos(\ell\theta)$ and $\sin(\ell\theta)$.
The order parameter therefore consists of two spin vectors
$\vec{n}_1$ and $\vec{n}_2$.  Up to quartic order, the uniform
Landau free energy takes the form
\begin{align}
F &=
a_2\left(
|\vec{n}_1|^2+|\vec{n}_2|^2
\right)
+a_4\left(
|\vec{n}_1|^2+|\vec{n}_2|^2
\right)^2
+b_4
|\vec{n}_1\times\vec{n}_2|^2 . \nonumber
\end{align}
This has precisely the same form as the Landau theory of Eq.~(\ref{eq:f3}).
For $b_4>0$, the two vectors are collinear, with their relative amplitudes
undetermined at quartic order.  This is the $\alpha$ phase of
Refs.~\cite{wu2004dynamic,wu2007fermi}.  For $b_4<0$, they are orthogonal
and of equal magnitude, giving the $\beta$ phase.  These are the direct
counterparts of the degenerate collinear and orthogonal phases of
Eq.~(\ref{eq:f3}), respectively.

The $\alpha$ phase with $\ell=2$ is the state commonly referred to as a
nematic-spin-nematic.  Each spin sector acquires quadrupolar Fermi-surface distortions
whose principal axes are perpendicular to one another, while the total
magnetization remains zero.  The corresponding $\beta$ phase with $\ell=2$
instead retains circular Fermi surfaces while developing a non-trivial
momentum-dependent spin texture. Thus the $\alpha$ and $\beta$ phases realize, respectively, the collinear
and orthogonal alternatives that recur throughout our Landau theories.

There is an analogous correspondence in three dimensions.  For an
$\ell=1$ instability, the orbital sector has three components and the order
parameter may be written as three spin vectors
$\vec{n}_1$, $\vec{n}_2$, and $\vec{n}_3$.  In the centrosymmetric case the uniform
Landau free energy is
\begin{align}
F &=
a_2\sum_{a=1}^{3}|\vec{n}_a|^2
+a_4\left(\sum_{a=1}^{3}|\vec{n}_a|^2\right)^2
+b_4\sum_{a<b}
|\vec{n}_a\times\vec{n}_b|^2 . \nonumber
\end{align}
This is the $\mathrm{SO}(3)_{L}$-symmetric limit $c_4=-b_4$ of our
non-coplanar Landau theory in Eq.~(\ref{eq:3D_cub}).
For $b_4>0$ the vectors are mutually collinear, giving the
$\alpha$ phase with $\ell=1$, whereas for $b_4<0$ they are equal-length and mutually orthogonal, giving the $\beta$ phase with $\ell=1$.  The latter has the same equal-length orthogonal structure as our
orthogonal phase and the BW state of \het{}. This correspondence concerns the
uniform Landau free energy; in the itinerant $\ell=1$ problem additional
gradient invariants are allowed and can drive a Lifshitz instability
\cite{wu2007fermi}.

In a crystalline metal, the continuous orbital rotation symmetry is reduced to the point group of the lattice, and an angular-momentum multiplet decomposes into crystallographic irreducible representations.
The resulting order parameters are then of the same representation-theoretic form as those considered throughout this work.  
When this crystal symmetry yields a multi-dimensional point-group irrep, these rotationally invariant theories reduce to the tetragonal, trigonal/hexagonal, or cubic Landau theories discussed in Sec.~\ref{sec:landau}. 
From this perspective, the magnetic phases studied here can be regarded as crystalline counterparts of the $\alpha$ and $\beta$ phases of spin-triplet Pomeranchuk order, with real-space magnetic configurations replacing spin-dependent deformations or textures of the Fermi surface.

\section{Summary and conclusions}

We have surveyed the landscape of magnetism in the non-relativistic limit using
the language of Landau theory, focusing primarily on phases arising from a
single irreducible representation $\Gamma$ of the parent point group. We have
seen that this approach cleanly unifies the very rich physics expected in this
limit. Within this setting, $\Gamma$ determines the allowed Landau invariants
and provides the organizing principle for the magnetic phases and their
responses. Remarkably, the large variety of spin-space-group symmetries is
organized by only a small number of quartic Landau theories.

Collinear phases arising from 2D or 3D irreps are often associated with detailed
magnetic structures with unequal moments and even with sites that have vanishing
moments. However, from the point of view of spin space groups these are
collinear spin groups of a lower-symmetry parent group, described by
one-dimensional order parameters (or direct sums of them) of that
subgroup~\cite{schiff2025collinear}.

We have seen that the orthogonal phases with N\'{e}el vectors of equal length
have a special status since the spin space groups associated with these phases are
those that preserve all the lattice symmetries up to spin rotations. This
puts into a different context earlier work noting that these phases are useful
variational states for classical frustrated systems. Landau theory has
confirmed that these are very common phases across all crystal
systems that admit multi-dimensional irreps. Our work also directly ties these explicitly to spin space groups and
points out similarities to superfluid \het{}. 

We have shown that the representation $\Gamma$, together with the direction of
the order parameter within its representation space, strongly constrains the
experimentally accessible responses of non-relativistic magnetic phases.

We have focused on uniform phases but the tools we have discussed can be applied
also to phases at finite wave vector (see, for example,
Refs.~\cite{belashchenko2026,yu2025odd}). A larger task is to address, for
non-collinear systems, the problem we solved for collinear systems in previous
work: are there matching conditions that inform us of the
nature of the relativistic response solely given information about the
non-relativistic limit \cite{mcclarty2024landau,schiff2025collinear}? This
question is of great interest when we consider real materials. While the results
of this paper provide crucial insight into the dominant physics in many magnetic
systems, it is desirable to bridge the divide between such materials and those,
for example, in which spin-orbit effects are strong.

\begin{acknowledgments}
H.S. and J.R. were supported by the NSF through Grant No. DMR-2142554. Work at
the University of Windsor (J.G.R.) was funded by the Natural Sciences and
Engineering Research Council of Canada (NSERC) (Funding Reference No.
RGPIN-2020-04970). P.M. acknowledges funding from the Centre National de la
Recherche Scientifique (CNRS) et le Commissariat à l’Énergie Atomique (CEA). 

The authors acknowledge the use of generative artificial intelligence tools
(Anthropic Claude) for language editing and for assistance in writing code used
to generate figures. All symmetry analysis, derivations, and physical
conclusions were carried out and verified by the authors, who take full
responsibility for the content of this work.
\end{acknowledgments}

\appendix

\section{(Co)irreps of \texorpdfstring{$\mathcal{G}$}{G}}\label{app:Fpm}

Here, we provide clarification on the irreducible co-representations of the high-symmetry paramagnetic groups $$\mathcal{G} = \othree\times G = \sothree\times G \cup T(\sothree\times G ),$$ and the irreps of the point groups $G$.

\subsection{Equivalence of decomplexification}\label{subapp:decomplex}

Let $\tilde{\mathcal{G}} = \sothree\times G$, let $\Gamma$ be an irrep of $G$, and let $\Gamma_{S}^{(\ell)}$ denote an irrep of $\sothree$. Then, the irreps of $\tilde{\mathcal{G}}$ are given by $\tilde{\gamma} \equiv \Gamma_{S}^{(\ell)}\otimes \Gamma$. Since $\tilde{\mathcal{G}}$ is the unitary halving subgroup of $\mathcal{G}$, its irreps determine the co-irreps $\gamma$ of $\mathcal{G}$. Throughout this paper we have worked directly with $\tilde{\gamma}$ and not with the co-irreps. In this appendix, we explain why this makes sense provided we work with the {\it decomplexified} analog of $\tilde{\gamma}$: the co-irreps induced from decomplexified $\tilde{\gamma}$ are equivalent to $\gamma$. 

There are three cases for the induction of $\gamma$ from $\tilde{\gamma}$, based on the Dimmock indicator of $\tilde{\gamma}$ \cite{bradley2009mathematical}. Because the antiunitary coset representative is pure time-reversal $T$, the Dimmock indicator reduces to the Frobenius-Schur indicator for the reality of the irrep $\tilde{\gamma}$. This is because the Dimmock indicator asks whether the $^{*}T$-conjugated irrep $\tilde{\gamma}^{(*T)} \equiv \{\tilde{\gamma}^{*}(Tg T^{-1}) \,|\, g \in\tilde{\mathcal{G}}\}$ is equivalent to $\tilde{\gamma}$, while the Frobenius-Schur indicator asks whether $\tilde{\gamma}^{*} \equiv \{\tilde{\gamma}^{*}(g)\,|\,g\in\tilde{\mathcal{G}}\}$ is equivalent to $\tilde{\gamma}$. Since $T$ commutes with all $g\in\tilde{\mathcal{G}}$, $\tilde{\gamma}^{*}(T g T^{-1}) = \tilde{\gamma}^{*}(g)$. Because the $\Gamma_{S}^{(\ell)}$ are all real, then, the reality of $\Gamma$ determines the reality of $\tilde{\gamma}$. The only point group irreps that are complex are one-dimensional, and belong to the groups $C_{n}$, $S_{n}$, $C_{nh}$ ($n\geq3$), $23$ and $m\overline{3}$.

Using the previous conclusion, in the induction algorithm for $\gamma$ from $\tilde{\gamma}$, $\gamma$ will always be real. If $\tilde{\gamma}$ is real, this feature is retained during induction, because the co-irrep $\gamma$ will have the form $$\gamma(g) = \tilde{\gamma}(g),\quad \gamma(T g) = \tilde{\gamma}(g)$$ for $g\in\tilde{\mathcal{G}}$ (assuming $\tilde{\gamma}$ is expressed in a real basis, so that $\tilde{\gamma}^{*}$ is not just equivalent to $\tilde{\gamma}$, but equal to it). If $\tilde{\gamma}$ is complex, it is paired with its complex conjugate $\tilde{\gamma}^{*}$ in $\gamma$, and so $\gamma$ takes the form $$\gamma(g) = \begin{bmatrix}\tilde{\gamma}(g) & 0\\ 0 & \tilde{\gamma}^{*}(g)\end{bmatrix}, \quad \gamma(T g) = \begin{bmatrix} 0 & \tilde{\gamma}^{*}(g)\\ \tilde{\gamma}(g) & 0\end{bmatrix},$$ since the matrix $\begin{bmatrix} 0 & \mathbb{I}_{|\tilde{\gamma}|} \\ \mathbb{I}_{|\tilde{\gamma}|} & 0\end{bmatrix}$ will represent $T$ in conjunction with complex conjugation, where $\mathbb{I}_{|\tilde{\gamma}|}$ is a $|\tilde{\gamma}| \times |\tilde{\gamma}|$ identity.

We will now show that the co-irreps $\gamma'$ induced from $\tilde{\gamma}_{\mathsf{R}} = \Gamma_{S}^{(\ell)} \otimes \Gamma_{\mathsf{R}}$ are equivalent to $\gamma$, given $\Gamma_{\mathsf{R}}$ are the \emph{decomplexified} irreps of $G$. When $\Gamma$ is real it is equivalent to its decomplexification. Thus, we need only consider complex $\Gamma$. The decomplexified irrep $\Gamma_{\mathsf{R}}$ may be expressed as \begin{equation}\label{eq:decomplexified_delta}
\Gamma_{\mathsf{R}} \cong \begin{bmatrix}\Gamma & 0 \\ 0 & \Gamma^{*} \end{bmatrix}.
\end{equation} Then, the irrep $\tilde{\gamma}_{\mathsf{R}}$ of $G$ is real. According to the induction algorithm, then, $\gamma'$ will take the form 
\begin{align*}
    \gamma'(g) &=  \Gamma_{S}^{(\ell)} \otimes \begin{bmatrix} \Gamma(\mathsf{g}) & 0 \\ 0 & \Gamma^{*}(\mathsf{g}) \end{bmatrix}, \\ \gamma'(T g) &= \left( \mathbb{I}_{\ell} \otimes \begin{bmatrix} 0 & \mathbb{I}_{|\Gamma|} \\ \mathbb{I}_{|\Gamma|} & 0\end{bmatrix}\right) \left(  \Gamma_{S}^{(\ell)} \otimes \begin{bmatrix} \Gamma(\mathsf{g}) & 0 \\ 0 & \Gamma^{*}(\mathsf{g})\end{bmatrix}\right) \\
    &= \Gamma_{S}^{(\ell)}\otimes \begin{bmatrix} 0 & \Gamma^{*}(\mathsf{g})\\\Gamma(\mathsf{g}) & 0\end{bmatrix}
\end{align*}
for $g\in\tilde{\mathcal{G}}$ of the form $g \equiv [R\,\|\, \mathsf{g}]$ with $R \in \sothree$ and $ \mathsf{g}\in G$. This is because the matrix $$Z = \begin{bmatrix} 0 & \mathbb{I}_{|\Gamma|} \\ \mathbb{I}_{|\Gamma|} & 0\end{bmatrix}$$ imposes equivalence between $\Gamma_{\mathsf{R}}$ and its complex conjugate $\Gamma_{\mathsf{R}}^{*}$, that is, $\Gamma_{\mathsf{R}} = Z^{-1}\Gamma_{\mathsf{R}}^{*} Z$. 

Two co-representations are equivalent if their characters agree on the unitary coset, and if the antiunitary characters differ by a phase \cite{wigner2012group}. Let us compare $\chi(\gamma)$ and $\chi(\gamma')$:
\begin{align}
    \chi(\gamma(g)) & = \mathrm{Tr(\tilde{\gamma})+Tr(\tilde{\gamma}^{*}) = Tr(\Gamma_{S}^{(\ell)})\left(Tr(\Gamma)+Tr(\Gamma^{*}) \right)}, \nonumber\\
    \chi(\gamma(T g )) &= 0, \nonumber\\
    \chi(\gamma'( g)) & =  \mathrm{Tr(\Gamma_{S}^{(\ell)})\left(Tr(\Gamma)+Tr(\Gamma^{*}) \right)},\\
    \chi(\gamma'(T g)) &= 0. \nonumber 
\end{align}
We can see that the characters coincide on $\tilde{\mathcal{G}}$ and on $T \tilde{\mathcal{G}}$, and so these co-irreps are equivalent, and $\gamma\sim\gamma'$.

\subsection{Inequivalence of time-even \& time-odd co-irreps over \texorpdfstring{$\mathbb{R}$}{R}}\label{subapp:reality}

The two co-irreps $\Gamma_{S}^{(1+)}\otimes \Gamma$ and $\Gamma_{S}^{(1-)}\otimes\Gamma$, characterized by equal representations on the unitary coset of $\mathcal{G}$ and differing only by a sign on the antiunitary coset, are \emph{equivalent} as complex co-irreps. Under a change of basis $B$, a co-representation transforms as $D(g)\to BD(g)B^{-1}$ on the unitary coset but as $D(T g)\to BD(T g)(B^{*})^{-1}$ on the antiunitary coset, so the choice $B=i\,\mathbb{I}$ leaves $D(g)$ unchanged while sending $D(T g)\to-D(T g)$. More generally, two co-representations whose antiunitary cosets differ by a phase are always equivalent over $\mathbb{C}$ \cite{wigner2012group, bradley2009mathematical}.

In a theory over $\mathbb{R}$, however, $B$ must be real, and the sign can no longer be removed. Since $T$ commutes with every element of $\mathcal{G}$ and $T^{2}=1$, in the real basis of Appendix~\ref{subapp:decomplex} the time-reversal element is represented by $\gamma(T)=\pm\mathbb{I}_{|\gamma|}$, and no real similarity transformation can change the sign of a multiple of the identity. This holds equally for real and for decomplexified $\Gamma$. The character $\chi(T)=\pm|\gamma|$ is an invariant of any real theory, and it distinguishes a magnetic order parameter from a non-magnetic one
with identical spatial symmetry. Because tabulated co-irreps are usually classified up to complex equivalence, they do not fix this sign, and it must be specified separately.

Thus, we must distinguish between the time-reversal even and odd co-irreps $\Gamma_{S}^{(1\pm)}\otimes\Gamma$ defining our order parameter. Because we wish for our order parameter to behave as a spin dipole in spin space, which is time-reversal odd, it transforms as $\Gamma_{S}^{(1-)}\otimes\Gamma$ of $\mathcal{G}$, with $\Gamma$ itself taken time-reversal even. It is this parity that fixes the time-reversal character of the physical responses in Sec.~\ref{sec:observables}: terms of odd order in the $\vec{N}_{a}$ are time-reversal odd, and terms of even order are time-reversal even.

\section{Relation between orthogonal phases, RMOs and spin groups}
\label{app:schur}

In Sec.~\ref{sec:ortho-proof}, we demonstrated that orthogonal phases with equal-length $\vec{N}_{a}$ correspond to phases that preserve the parent point group up to spin rotations (defining regular magnetic order). Here we provide another proof. In Sec.~\ref{sec:perp_spg}, we use a known correspondence \cite{xiao2024ssg} between $\Gamma$ and the spin point groups to establish a one-to-one relation between an orthogonal phase and its low-symmetry spin point group. Together, this means that every lattice possesses a unit-cell-preserving RMO, whose spin point group is uniquely determined.

With the procedure for determining spin group epikernels in Sec.~\ref{subsec:epikernels}, along with the following correspondence between an RMO and its spin point group, we establish a complete framework for understanding the low-symmetry phases of all $\vec{k}=0$ magnetic orders at zero SOC.

\subsection{Orthogonal phase and spin groups}\label{sec:perp_spg}
 
In general, there are many spin point groups with parent point group $G$. A natural question is: which spin point groups are realized by the orthogonal phase? 

Here, we demonstrate that in the Landau theory arising from an irrep $\Gamma$ of $G$, the spin point group of the orthogonal phase is directly determined by $\Gamma$ itself. This construction closely follows that of Ref.~\cite{xiao2024ssg}, and may be classified according to whether the order is collinear, coplanar, or non-coplanar. A formally different approach based on projective symmetries is utilized in Ref.~\cite{Messio2011_RMO}, and its utility for spin-orbit-free systems becomes apparent when recast into the language of spin groups. We first briefly summarize the salient features of spin point groups that we will utilize in our analysis.

The spin point groups are discrete subgroups of $\othree\times\mathrm{O(3)}$, unique up to conjugacy, which amounts to an orthogonal change of basis in spin and real space. The elements of the spin point groups are denoted by $[R\|\mathsf{g}]$, where $R \in \othree$ acts on the spin degrees of freedom, while $\mathsf{g} \in \mathrm{O(3)}$ acts on the spatial degrees of freedom. When $R$ is an improper rotation, it is of the form $R = T \tilde{R}$ where $\tilde{R}$ is a global, proper spin rotation and $T$ is the time-reversal element, acting as the inversion element of $\othree$. These elements $[R\|\mathsf{g}]$ act on an arrangement of magnetic moments $\vec{m}(\vec{r})$ as $$[R\|\mathsf{g}]\, m^{i}(\vec{r}) = \sum_{j} R_{ij}\,m^{j}\left(\mathsf{g}^{-1}\vec{r}\right).$$

Every spin point group $X$ is of the form $X = {b}\times S$, where ${b}$ contains elements acting \emph{only} on the spin degrees of freedom, of the form $[R\|E]$. Here, $E$ is the trivial group element. Accordingly, this subgroup ${b}$ is called the \emph{spin-only group}, and it depends only on the mutual collinearity, coplanarity or non-coplanarity of the spins. Respectively, in each of these cases, it is given by ${b}_{\infty} = \mathrm{SO(2)} +T 2_{\perp} \mathrm{SO(2)}$ (where $2_{\perp}$ is perpendicular to the collinear spin axis), ${b}_{\mathbb{Z}_{2}} = \{E, T 2_{\perp}\}$ (where $2_{\perp}$ is orthogonal to the plane of the spins), and ${b_{\mathrm{triv}}} = \{E\}$ for non-coplanar arrangements. 

The subgroup $S$ is called the \emph{non-trivial spin point group}, and it contains both elements of the form $[E\|\mathsf{g}]$ and $[R\|\mathsf{g}]$, but it contains no elements of the form $[R\|E]$ with $R\neq E$.  There are 598 such groups, from which all total spin point groups $X$ can be derived by pairing with compatible ${b}$ \cite{Litvin1974_SpinGroups,Litvin1977_SpinPoint,Liu2022_SPG,Schiff2025_SPG,xiao2024ssg,jiang2024ssg,Chen2024_SSG_Reps}.

For a parent point group $G$, the non-trivial spin point groups are formed by identifying all homomorphisms $\Phi: G\rightarrow \othree$. Such homomorphisms ensure that the elements $[\Phi(\mathsf{g})\|\mathsf{g}]$ form a group. These $\Phi$ are three-dimensional, real representations of $G$, and can be constructed from the decomplexified irreducible representations $\Gamma$ of $G$ \cite{Lazic2013_spinLine,Schiff2025_SPG,xiao2024ssg,jiang2024ssg,Chen2024_SSG_Reps}. 

In the following, we demonstrate that, uniquely up to equivalence in $\othree\times\mathrm{O(3)}$, each one-dimensional irrep can be identified with a collinear spin point group; each two-dimensional irrep can be identified with a coplanar spin point group; and each three-dimensional irrep can be identified with a non-coplanar spin point group \cite{xiao2024ssg}.

To proceed, we first notice that an explicit choice of basis for the matrices $\Phi(\mathsf{g})$ establishes a basis in spin space, and we require that this basis is chosen to be orthogonal; any other basis obtained by an orthogonal transformation yields an equivalent spin point group \cite{xiao2024ssg,Liu2022_SPG,Schiff2025_SPG, jiang2024ssg}. 

The orthogonal phase is characterized by $|\Gamma|$ vectors in spin space, $\vec{N}_{a}$, all orthogonal to one another. We will use $\Gamma$ to define the homomorphism $\Phi$ by aligning the $\vec{N}_{a}$ along $|\Gamma|$ basis axes. Then, the $|\Gamma|$-dimensional subspace spanned by the $\vec{N}_{a}$ is represented by $\Gamma$ and there is a choice we can make, introduced in Ref.~\cite{xiao2024ssg}, for the remaining $(3-|\Gamma|)$-dimensional subspace such that a unique spin point group is obtained.

\subsubsection{Spin groups for collinear RMOs}\label{subapp:SPG_collin_RMO}

When the irrep $\Gamma$ of $G$ is one-dimensional, there is only one order parameter vector $\vec{N}$, and the resulting orthogonal phase has a collinear magnetic order. Without loss of generality, let us define the spin $\hat{\vec{z}}$ axis as the one aligned with $\vec{N}$. Then, the matrices $\Phi(\mathsf{g})$ will be of the form 
\begin{equation}\label{eq:oxy_collinear}
    \Phi(\mathsf{g}) = \left[ 
\begin{array}{cc|c} 
  &    &  0   \\
\multicolumn{2}{c|}{\smash{\raisebox{.5\normalbaselineskip}{$O_{xy}$}}}
     &      0  \\
 \hline \\[-\normalbaselineskip]
 0 &  0  & \Gamma(\mathsf{g})
\end{array} 
\right], 
\end{equation}
where $O_{xy}$ is a two-dimensional orthogonal matrix. Following Ref.~\cite{xiao2024ssg}, we can choose $O_{xy} = \Gamma(\mathsf{g}) \, \mathbbold{1}_{2\times2}$. In this way, we obtain a non-trivial spin point group $S$ where $\Phi(\mathsf{g})= -\mathbbold{1}$ when $\Gamma(\mathsf{g}) = -1$ (corresponding to $R = T$) and $\Phi(\mathsf{g}) = \mathbbold{1}$ when $\Gamma(\mathsf{g}) = 1$ (corresponding to $R = E$). 

There are 105 real, one-dimensional irreps $\Gamma$ of the crystallographic point groups. When deriving a collinear spin point group according to this procedure, fifteen of the irreps lead to redundant groups that are conjugate to one another in $\othree\times\mathrm{O(3)}$ (see Appendix~\ref{app:redundant_spgs}). Removing these redundancies correctly reproduces the $90$ non-trivial collinear spin point groups (SPGs).

To obtain the total spin point group, we take the direct product of the non-trivial spin point group and the collinear spin-only group ${b}_{\infty}$. The direct product is possible because all elements of ${b}_{\infty}$ commute with the two possible choices of $\Phi(\mathsf{g})$ in Eq.~(\ref{eq:oxy_collinear}), and the intersection ${b}_{\infty}\cap S$ is trivial. 

\subsubsection{Spin groups for coplanar RMOs}\label{subapp:SPG_coplanar_RMO}

For two-dimensional irreps $\Gamma$ of $G$, there are two spin vectors $\vec{N}_{1}$ and $\vec{N}_{2}$ constituting our order parameter, and the orthogonal phase gives rise to a coplanar magnetic order. Without loss of generality, let $\vec{N}_{1}$ and $\vec{N}_{2}$ lie in the $xy$ plane. Then, the non-trivial spin point group $S$ is formed by selecting the matrices $\Phi(\mathsf{g})$ that are of the form
\begin{equation}\label{eq:oz_coplanar}
    \Phi(\mathsf{g}) = \left[ 
\begin{array}{cc|c} 
  &    &  0   \\
\multicolumn{2}{c|}{\smash{\raisebox{.5\normalbaselineskip}{$\Gamma(\mathsf{g})$}}}
     &      0  \\
 \hline \\[-\normalbaselineskip]
 0 &  0  & O_{z}
\end{array} 
\right], 
\end{equation}
where $O_{z}$ may be chosen as $O_{z} = \det(\Gamma(\mathsf{g})) $ \cite{xiao2024ssg}. This choice always leads to $\det(\Phi(\mathsf{g})) = 1$.

Notice that $\Phi(\mathsf{g})$ in Eq.~(\ref{eq:oz_coplanar}) commutes with $T 2_{\perp}\in{b}_{\mathbb{Z}_{2}}$, whose $\othree$ matrix is $\mathrm{diag}(1,1,-1)$. Combined with the fact that ${b}_{\mathbb{Z}_{2}}\cap S$ is trivial, we are now able to take the direct product of ${b}_{\mathbb{Z}_{2}}$ and $S$ to obtain the total coplanar spin point groups describing orthogonal phases with $|\Gamma|=2$.

There are 41 two-dimensional irreps $\Gamma$ of the crystallographic point groups (including decomplexifications), and none of them lead to redundancies in $\othree\times\mathrm{O(3)}$ (see Appendix~\ref{app:redundant_spgs}). Yet, there are 253 non-trivial SPGs corresponding to coplanar moments;  as we will see, the remaining 212 are described by reducible orthogonal orders, as described in Sec.~\ref{sec:perp_reducible}.

\subsubsection{Spin groups for non-coplanar RMOs}\label{subapp:SPG_noncoplanar_RMO}

When the irrep $\Gamma$ of $G$ is three-dimensional, three vectors $\vec{N}_{1}$, $\vec{N}_{2}$ and $\vec{N}_{3}$ form our order parameter, and the orthogonal phase gives rise to a non-coplanar magnetic order. Without loss of generality, let $\vec{N}_{1}$, $\vec{N}_{2}$ and $\vec{N}_{3}$ be aligned along the $\vec{\hat{x}}$, $\vec{\hat{y}}$, and $\vec{\hat{z}}$ axes, respectively. Then, the non-trivial spin point group is formed by selecting $\Phi(\mathsf{g})$ equal to $\Gamma(\mathsf{g})$, as this is already a three-dimensional, real representation. Since the spin-only group is trivial, we have found the entire spin point group.

There are 11 three-dimensional irreps of the crystallographic point groups, and none of them lead to redundant spin point groups in $\othree \times \mathrm{O(3)}$ (see Appendix~\ref{app:redundant_spgs}). However, there are 598 non-coplanar spin point groups. Similar to the coplanar case, we will see that the remaining 587 are described by reducible orthogonal orders, as described in Sec.~\ref{sec:perp_reducible}.

\subsubsection{Generic appearance of orthogonality}\label{subsubsec:genericOrtho}

The fact that every irrep can actualize an RMO hints that the appearance of orthogonality is generic, and promotes the spin groups from abstract constructions to the symmetries of energy-minimizing magnetic orders. Here we prove that the spin point groups constructed in Secs.~\ref{subapp:SPG_collin_RMO}--\ref{subapp:SPG_noncoplanar_RMO}, with elements $[\Phi(\mathsf{g})\|\mathsf{g}]$ built from $\Gamma(\mathsf{g})$, force all $\vec{N}_{a}$ to be orthogonal. This proof is also equally applicable to the reducible order parameters of the following section, Sec.~\ref{sec:perp_reducible}. 

Here, we repackage the order parameter into a $|\Gamma|\times3$ matrix $\mathfrak{N}$ where the matrix element $\mathfrak{N}_{ai}$ is the $i$th component of $\vec{N}_{a}$. Under the element $[\Phi(\mathsf{g})\|\mathsf{g}]$, then, invariance of $\mathfrak{N}$ implies 
\begin{equation}\label{eq:invarCondition}
    \mathfrak{N} = \Gamma(\mathsf{g})\, \mathfrak{N} \,\Phi^{\mathtt{T}}(\mathsf{g}).
\end{equation}
Let $\Phi_{0}(\mathsf{g})$ denote the matrices constructed above, namely Eq.~(\ref{eq:oz_coplanar}) for $|\Gamma|=2$ and $\Phi_{0}(\mathsf{g}) = \Gamma(\mathsf{g})$ for $|\Gamma|=3$; the collinear case $|\Gamma|=1$ is trivial, as there is a single vector $\vec{N}$. We will demonstrate that the choice 
\begin{equation}\label{eq:choicePhi}
    \Phi(\mathsf{g}) = R^{\mathtt{T}} \,\Phi_{0}(\mathsf{g})\, R
\end{equation}
satisfies this condition while enforcing the $\vec{N}_{a}$ are orthogonal. The matrix $R$ in Eq.~(\ref{eq:choicePhi}) is an orthogonal transformation in spin space, encoding freedom in the choice of spin space coordinates. With this choice, the invariance condition becomes
$$\mathfrak{N} = \Gamma(\mathsf{g}) \mathfrak{N} R^{\mathtt{T}} \Phi_{0}^{\mathtt{T}}(\mathsf{g}) R,$$
implying that
\begin{equation}\label{eq:commutingN}
(\mathfrak{N} R^{\mathtt{T}} )\,\Phi_{0}(\mathsf{g}) = \Gamma(\mathsf{g}) (\mathfrak{N}R^{\mathtt{T}}).
\end{equation}
For $|\Gamma|=3$, $\Phi_{0} = \Gamma$, so $X \equiv \mathfrak{N}R^{\mathtt{T}}$ commutes with $\Gamma(\mathsf{g})$ for all $\mathsf{g}$. For $|\Gamma|=2$, we write $\mathfrak{N}R^{\mathtt{T}} = [X\,|\,\vec{c}\,]$, with $X$ a $2\times2$ block and $\vec{c}$ a column; Eq.~(\ref{eq:commutingN}) then splits into $X\Gamma(\mathsf{g}) = \Gamma(\mathsf{g})X$ and $\det\Gamma(\mathsf{g})\,\vec{c} = \Gamma(\mathsf{g})\,\vec{c}$. The latter would make $\vec{c}$ span a one-dimensional invariant subspace of $\Gamma$, which is impossible for irreducible $\Gamma$, so $\vec{c} = 0$. In both cases, $X$ commutes with $\Gamma(\mathsf{g})$ for all $\mathsf{g}$. If $\Gamma$ is a real irrep, Schur's lemma gives $X = N\mathbbold{1}$. If $\Gamma = \Gamma_{\mathsf{R}}$ is decomplexified, the real matrices commuting with it are of the form $a\mathbbold{1} + bJ$ with $J^{2} = -\mathbbold{1}$, whose rows $(a,b)$ and $(-b,a)$ are orthogonal and of equal length $N = \sqrt{a^{2}+b^{2}}$. In either case,
\begin{equation}\label{eq:SchurOrtho}
\mathfrak{N}\mathfrak{N}^{\mathtt{T}} = XX^{\mathtt{T}} = N^{2}\,\mathbbold{1},
\end{equation} 
so the $\vec{N}_{a}$ are orthogonal to one another and of equal length $N$. Equation~(\ref{eq:choicePhi}) defines the elements corresponding to the spin point group symmetry of the resulting RMO.

\subsection{Reducible orthogonal order}\label{sec:perp_reducible}

Here, we consider the possibility that the order parameter transforms under a reducible representation $D = \Gamma_{S}^{(1-)}\otimes \,d$ with $d = \oplus_{\Gamma}\Gamma$, which is increasingly relevant at low temperatures. When $d$ is two- or three-dimensional, real, and reducible, additional RMOs arise.

Let us begin with the case of two-dimensional $d = \Gamma_{1}\oplus\Gamma_{2}$, formed from two one-dimensional irreps. There are 212 ways to pair one-dimensional $\Gamma$, unique up to conjugacy in $\mathrm{O(3)}$ (see Appendix~\ref{app:redundant_spgs}). In each case, $d(\mathsf{g})$ replaces $\Gamma(\mathsf{g})$ in Eq.~(\ref{eq:oz_coplanar}), giving a coplanar spin point group. Writing $\mathfrak{N}R^{\mathtt{T}} = [X\,|\,\vec{c}\,]$ as in Sec.~\ref{subsubsec:genericOrtho}, Eq.~(\ref{eq:commutingN}) becomes 
\begin{equation}\label{eq:reducible_commtingN}
X\,d(\mathsf{g}) = d(\mathsf{g})\,X, \qquad \Gamma_{1}(\mathsf{g})\Gamma_{2}(\mathsf{g})\,\vec{c} = d(\mathsf{g})\,\vec{c}.
\end{equation}
When $\Gamma_{1}\neq\Gamma_{2}$, Schur's lemma makes $X$ diagonal. The second condition sets $c_{1} = 0$ unless $\Gamma_{2}$ is the trivial irrep, and $c_{2} = 0$ unless $\Gamma_{1}$ is the trivial irrep, so at most one component of $\vec{c}$ is non-zero. Hence,
\begin{equation}\label{eq:reducible_schur2}
\mathfrak{N}\mathfrak{N}^{\mathtt{T}} = XX^{\mathtt{T}} + \vec{c}\,\vec{c}^{\,\mathtt{T}} = \mathrm{diag}(N_{1}^{2},N_{2}^{2}).
\end{equation}
Similar to Sec.~\ref{subsubsec:genericOrtho}, from Eq.~(\ref{eq:reducible_schur2}) it follows that $\vec{N}_{1}$ and $\vec{N}_{2}$ are orthogonal, but now they may have different lengths. When $d = \Gamma\oplus\Gamma$ contains the same irrep twice, $d(\mathsf{g}) = \Gamma(\mathsf{g})\,\mathbbold{1}_{2}$ and $X$ is unconstrained. In this case, the orthogonal configuration still realizes the coplanar spin point group constructed from $d$, but orthogonality is permitted, rather than enforced.

For three-dimensional reducible and real $d$, up to conjugacy in $\mathrm{O(3)}$, there are 451 unique ways to combine three one-dimensional irreps and 136 unique ways to combine a one-dimensional and a two-dimensional irrep (see Appendix~\ref{app:redundant_spgs}). In total, this yields 587 non-coplanar spin point groups. In each case, $\Phi_{0}(\mathsf{g}) = d(\mathsf{g})$, so that $X \equiv \mathfrak{N}R^{\mathtt{T}}$ commutes with $d(\mathsf{g})$ for all $\mathsf{g}$, as in Eq.~(\ref{eq:reducible_commtingN}). For $d$ arising from three distinct one-dimensional irreps, Schur's lemma makes $X$ diagonal, and
\begin{equation}\label{eq:reducible_schur3a}
\mathfrak{N}\mathfrak{N}^{\mathtt{T}} = \mathrm{diag}(N_{1}^{2}, N_{2}^{2}, N_{3}^{2}),
\end{equation}
while for the 136 $d$ arising from the combination of one- and two-dimensional irreps, $X$ is block diagonal, with the two-dimensional block of the form found in Sec.~\ref{subsubsec:genericOrtho}, so that
\begin{equation}\label{eq:reducible_schur3b}
\mathfrak{N}\mathfrak{N}^{\mathtt{T}} = \mathrm{diag}(N_{1}^{2}, N_{2}^{2}, N_{2}^{2}).
\end{equation}
In both Eqs.~(\ref{eq:reducible_schur3a}) and (\ref{eq:reducible_schur3b}) all
$\vec{N}_{a}$ are orthogonal. In the former, the $\vec{N}_{a}$ may all have
different lengths, while in the latter two of the three $\vec{N}_{a}$ are the
same length. When a one-dimensional irrep is repeated in $d$, the corresponding
block of $X$ is unconstrained, and orthogonality between those components is
permitted but not enforced.

On the kagom\'e lattice, a reducible RMO can be formed by allowing the
ferromagnetic $A_{1}$ component to align along the $\vec{\hat{z}}$ axis and the
coplanar $E_{2}$ component to order in the plane. This yields an umbrella state,
with moments given in Eq.~(\ref{eq:kagome_umbrella}),
\begin{align}\label{eq:kagome_umbrella}
\vec{m}_{A} &=\frac{N^{A_{1}}}{\sqrt{3}}\vec{\hat{z}}  + \sqrt{\frac{2}{3}}N^{E_{2}} \vec{\hat{x}} \nonumber \\
\vec{m}_{B} &=\frac{N^{A_{1}}}{\sqrt{3}} \vec{\hat{z}} -N^{E_{2}}\left(\frac{1}{\sqrt{6}}\vec{\hat{x}}+\frac{1}{\sqrt{2}}\vec{\hat{y}}\right)\\
\vec{m}_{C} &=\frac{N^{A_{1}}}{\sqrt{3}} \vec{\hat{z}} -N^{E_{2}}\left(\frac{1}{\sqrt{6}}\vec{\hat{x}}-\frac{1}{\sqrt{2}}\vec{\hat{y}}\right).\nonumber
\end{align}
Many non-coplanar RMOs can be formed from reducible representations. When both
reducible and irreducible RMOs are considered, every spin point group is
realized.


\section{Redundant SPGs from point-group irreps}\label{app:redundant_spgs} In the
construction of the spin point groups outlined in Sec.~\ref{sec:perp_reducible},
we want to count only groups unique up to conjugacy in
$\othree\times\mathrm{O(3)}$. Because the point groups $G$ are already unique up
to conjugacy in $\mathrm{O(3)}$, we need only ensure the spin parts of each spin
point group are unique in $\othree$. Recall that the elements
$[\Phi(\mathsf{g})\|\mathsf{g}]$ are formed by the choice $\Phi(\mathsf{g}) =
d(\mathsf{g})R$ where $d$ is decomposable. Uniqueness of the spin point group is
then determined by the unique ways to join irreps of $G$. 

The unique ways to join irreps of $G$ are determined by the orbits of the irreps
of $G$, under conjugacy by elements of the normalizer of $G$ in $\mathrm{O(3)}$.
The normalizer $N_{\mathrm{O(3)}}(G)$ is the largest subgroup of $\mathrm{O(3)}$
such that $G \triangleleft N_{\mathrm{O(3)}}(G)$ is normal. Then, conjugating
$G$ by elements $n\in N_{\mathrm{O(3)}}(G)$ sends $\Gamma(\mathsf{g})$ to
$\Gamma'(\mathsf{g}) \equiv \Gamma(n\mathsf{g}n^{-1})$. We say that $\Gamma'$
and $\Gamma$ are within the same orbit under conjugacy by
$N_{\mathrm{O(3)}}(G)$.

For $\Gamma$ and $\Gamma'$ in the same orbit under conjugacy by
$N_{\mathrm{O(3)}}(G)$, the representations $d = \Gamma \oplus \Gamma$ and $d' =
\Gamma' \oplus \Gamma'$ yield spin point groups that are equivalent under
conjugacy by $\othree\times \mathrm{O(3)}$. The unique RMOs will then be given
by $d$ that are unique under conjugacy by $N_{\mathrm{O(3)}}(G)$, and this is
determined by the orbits of the irreps of $G$ under $N_{\mathrm{O(3)}}(G)$. The
normalizers in $\mathrm{O(3)}$ for each point group $G$ are given in
\cite{ITA2002}, and are included in Table~\ref{tab:orbits}. We provide the orbit
of each decomplexified irreducible representation $\Gamma$ for each point group
$G$ under $N_{\mathrm{O(3)}}(G)$. In Table~\ref{tab:rep_counts}, we list the
number of unique reducible representations of each type for each normalizer.

\section{Spin conductivity}\label{app:spin_conductivity}

An electric field $\vec{E}$ may generate spin-polarized currents $\vec{J}^{a}$ at zero spin-orbit coupling according to 
\begin{equation}\label{eq:sc_def}
J^{a}_{i}(\vec{k}) = \sigma^{sc,a}_{ij}(\vec{k},\vec{S}) E_{j}(\vec{k}).\end{equation}
Here, $\vec{k}$ is the electric field wave-vector, $\vec{S}$ denotes the configuration of time-reversal (TR) breaking spin moments, and the static limit is $\vec{k}\rightarrow\vec{0}$. We use the Einstein summation convention. We call this spin-charge conductivity simply the spin conductivity. If an antiunitary operation $[UT\|R]$, with $U$ a spin rotation and $R$ a proper or improper real-space operation, is a symmetry, Onsager reciprocity relates the spin conductivity to the transpose of the inverse (charge-spin) conductivity $\sigma^{cs}$,
\begin{equation}\label{eq:sc_au}
    \sigma_{ij}^{sc,a}(\vec{S}) = -U^{ab}R_{jm}R_{in}\sigma^{cs,b}_{mn}(\vec{S}),
\end{equation}
while a unitary symmetry $[U\|R]$ gives
\begin{equation}\label{eq:sc_u}
    \sigma^{sc,a}_{ij}(\vec{S}) = U^{ab}R_{im}R_{jn}\sigma^{sc,b}_{mn}(\vec{S}).
\end{equation}
Quantities with definite TR parity thus combine $\sigma^{sc}$ and $\sigma^{cs}$,
\begin{equation}\label{eq:sc_tr_def}
    \sigma^{sc,a\pm}_{ij}(\vec{S}) = \frac{1}{2}\left( \sigma^{sc,a}_{ij}(\vec{S}) \mp \sigma^{cs,a}_{ji}(\vec{S}) \right),
\end{equation}
which indeed have definite parity 
\begin{equation}\label{eq:sc_tr_parity}
    \sigma_{ij}^{sc,a\pm}(-\vec{S}) = \pm \sigma^{sc,a\pm}_{ij}(\vec{S}).
\end{equation}
We may further symmetrize $[ij]$ or antisymmetrize $\{ij\}$ the spatial indices, as each $\sigma^{sc,a}$ is a rank-2 tensor. We introduce
\begin{align}
    \sigma^{a\pm}_{[ij]}(\vec{S}) &= \frac{1}{2}\left(\sigma_{ij}^{sc,a\pm}(\vec{S}) + \sigma_{ji}^{sc,a\pm}(\vec{S}) \right)\label{eq:sc_tr_[ij]}\\
    \sigma^{a\pm}_{\{ij\}}(\vec{S}) &= \frac{1}{2}\left(\sigma^{sc,a\pm}_{ij}(\vec{S}) - \sigma_{ji}^{sc,a\pm}(\vec{S}) \right)\label{eq:sc_tr_{ij}}.
\end{align}
This decomposition into parity under TR and spatial index permutation gives rise to \emph{four} spin-conductivity components, each transforming under the following representations:
\begin{align}
\sigma^{a+}_{[ij]}(\vec{S}) &: \quad \Gamma_{S}^{(1+)}\otimes [V^{2}],\label{eq:sc_++}\\
\sigma^{a-}_{[ij]}(\vec{S}) &: \quad \Gamma_{S}^{(1-)}\otimes [V^{2}],\label{eq:sc_-+}\\
\sigma^{a+}_{\{ij\}}(\vec{S}) &: \quad \Gamma_{S}^{(1+)}\otimes \{V^{2}\},\label{eq:sc_+-}\\
\sigma^{a-}_{\{ij\}}(\vec{S}) &: \quad \Gamma_{S}^{(1-)}\otimes \{V^{2}\}.\label{eq:sc_--}
\end{align}

In the terminology of Ref.~\cite{GonzalezHernandez2021_SpinCond}, the TR-even components constitute the spin Hall effect, traditionally referring only to $\sigma^{a+}_{\{ij\}}$, but now extended to the symmetric $\sigma^{a+}_{[ij]}$ in lower-symmetry crystals. The TR-odd components constitute the magnetic spin Hall effect~\cite{Zelezny2017_spinPolarizedCurrent, Kimata2019_magneticSHE, Mook2020_magneticSHE}. 

The antisymmetric components are equivalent to axial vectors, $\sigma^{a\pm}_{\{ij\}} = \epsilon_{ijk}h^{a\pm}_{k}$, and generate spin currents $\vec{J}^{a}\propto\vec{E}\times\vec{h}^{a\pm}$ transverse to the field. The symmetric components generate longitudinal spin-polarized currents for fields along their principal axes and acquire a transverse part otherwise, as in the (time-odd) spin-splitter effect of altermagnets~\cite{GonzalezHernandez2021_SpinCond}. 

Although the spin Hall effect is commonly attributed to spin-orbit coupling, at zero SOC it can arise from non-collinear order~\cite{Zhang2018_SHEnoSOC}, consistent with the vector-chirality form $\vec{N}_{a}\times\vec{N}_{b}$ of $\sigma^{a+}$. Both the symmetric and antisymmetric channels of the time-odd, magnetic spin Hall effect are allowed for collinear, coplanar and non-coplanar orders, restricted only by $\Gamma$, as described in Sec.~\ref{subsec:obs_odd_vectors}. In particular, the spin-splitter effect requires neither SOC nor non-collinear order.

\onecolumngrid

\begin{table}[htbp]
\centering
\caption{Normalizers $N_{\mathrm{O(3)}}(G)$ of the 32 crystallographic point groups in $\mathrm{O(3)}$, and the orbits of their (decomplexified) irreducible representations under conjugation by $N_{\mathrm{O(3)}}(G)$. Irreps within the same set $\{\ldots\}$ are mapped into each other.}
\label{tab:orbits}
\begin{tabular}{llll}
\hline\hline
Crystal system & $G$ & $N_{\mathrm{O(3)}}(G)$ & Irrep orbits \\
\hline
Triclinic & ${1}$ & $\mathrm{O(3)}$ & $A$ \\
 & ${\bar{1}}$ & $\mathrm{O(3)}$ & $A_g$, $A_u$ \\
\hline
Monoclinic & ${2}$ & ${\infty/mm}$ & $A$, $B$ \\
 & ${m}$ & ${\infty/mm}$ & $A'$, $A''$ \\
 & ${2/m}$ & ${\infty/mm}$ & $A_g$, $B_g$, $A_u$, $B_u$ \\
\hline
Orthorhombic & ${222}$ & ${m\bar{3}m}$ & $A$, $\{$$B_1$, $B_2$, $B_3$$\}$ \\
 & ${mm2}$ & ${4/mmm}$ & $A_1$, $A_2$, $\{$$B_1$, $B_2$$\}$ \\
 & ${mmm}$ & ${m\bar{3}m}$ & $A_g$, $\{$$B_{1g}$, $B_{2g}$, $B_{3g}$$\}$, $A_u$, $\{$$B_{1u}$, $B_{2u}$, $B_{3u}$$\}$ \\
\hline
Tetragonal & ${4}$ & ${\infty/mm}$ & $A$, $B$, $E$ \\
 & ${\bar{4}}$ & ${\infty/mm}$ & $A$, $B$, $E$ \\
 & ${4/m}$ & ${\infty/mm}$ & $A_g$, $B_g$, $A_u$, $B_u$, $E_g$, $E_u$ \\
 & ${422}$ & ${8/mmm}$ & $A_1$, $A_2$, $\{$$B_1$, $B_2$$\}$, $E$ \\
 & ${4mm}$ & ${8/mmm}$ & $A_1$, $A_2$, $\{$$B_1$, $B_2$$\}$, $E$ \\
 & ${\bar{4}2m}$ & ${4/mmm}$ & $A_1$, $A_2$, $B_1$, $B_2$, $E$ \\
 & ${4/mmm}$ & ${8/mmm}$ & $A_{1g}$, $A_{2g}$, $\{$$B_{1g}$, $B_{2g}$$\}$, $A_{1u}$, $A_{2u}$, $\{$$B_{1u}$, $B_{2u}$$\}$, $E_g$, $E_u$ \\
\hline
Trigonal & ${3}$ & ${\infty/mm}$ & $A$, $E$ \\
 & ${\bar{3}}$ & ${\infty/mm}$ & $A_g$, $A_u$, $E_g$, $E_u$ \\
 & ${32}$ & ${6/mmm}$ & $A_1$, $A_2$, $E$ \\
 & ${3m}$ & ${6/mmm}$ & $A_1$, $A_2$, $E$ \\
 & ${\bar{3}m}$ & ${6/mmm}$ & $A_{1g}$, $A_{2g}$, $A_{1u}$, $A_{2u}$, $E_g$, $E_u$ \\
\hline
Hexagonal & ${6}$ & ${\infty/mm}$ & $A$, $B$, $E_1$, $E_2$ \\
 & ${\bar{6}}$ & ${\infty/mm}$ & $A'$, $A''$, $E'$, $E''$ \\
 & ${6/m}$ & ${\infty/mm}$ & $A_g$, $B_g$, $A_u$, $B_u$, $E_{1g}$, $E_{2g}$, $E_{1u}$, $E_{2u}$ \\
 & ${622}$ & ${12/mmm}$ & $A_1$, $A_2$, $\{$$B_1$, $B_2$$\}$, $E_1$, $E_2$ \\
 & ${6mm}$ & ${12/mmm}$ & $A_1$, $A_2$, $\{$$B_1$, $B_2$$\}$, $E_1$, $E_2$ \\
 & ${\bar{6}2m}$ & ${6/mmm}$ & $A_1'$, $A_2'$, $A_1''$, $A_2''$, $E'$, $E''$ \\
 & ${6/mmm}$ & ${12/mmm}$ & $A_{1g}$, $A_{2g}$, $\{$$B_{1g}$, $B_{2g}$$\}$, $A_{1u}$, $A_{2u}$, $\{$$B_{1u}$, $B_{2u}$$\}$, $E_{1g}$, $E_{2g}$, $E_{1u}$, $E_{2u}$ \\
\hline
Cubic & ${23}$ & ${m\bar{3}m}$ & $A$, $E$, $T$ \\
 & ${m\bar{3}}$ & ${m\bar{3}m}$ & $A_g$, $A_u$, $E_g$, $E_u$, $T_g$, $T_u$ \\
 & ${432}$ & ${m\bar{3}m}$ & $A_1$, $A_2$, $E$, $T_1$, $T_2$ \\
 & ${\bar{4}3m}$ & ${m\bar{3}m}$ & $A_1$, $A_2$, $E$, $T_1$, $T_2$ \\
 & ${m\bar{3}m}$ & ${m\bar{3}m}$ & $A_{1g}$, $A_{2g}$, $A_{1u}$, $A_{2u}$, $E_g$, $E_u$, $T_{1g}$, $T_{2g}$, $T_{1u}$, $T_{2u}$ \\
\hline\hline
\end{tabular}
\end{table}

\begin{table}[htbp]
\centering
\caption{Number of distinct reducible representations by normalizer class, up to conjugacy by the normalizer $N_{\mathrm{O(3)}}(G)$.}
\label{tab:rep_counts}
\begin{tabular}{lccc}
\hline
$N_{\mathrm{O(3)}}(G)$ & 2D & $1\text{D}^3$ & $1\text{D} \oplus 2\text{D}$ \\
\hline
$\mathrm{O(3)}$        &   4 &   5 &   0 \\
${\infty/mm}$          &  52 &  89 &  41 \\
${m\bar{3}m}$          &  37 &  76 &  17 \\
${12/mmm}$             &  38 & 100 &  36 \\
${8/mmm}$              &  38 & 100 &  18 \\
${6/mmm}$              &  26 &  48 &  20 \\
${4/mmm}$              &  17 &  33 &   4 \\
\hline
Total                & 212 & 451 & 136 \\
\hline
\end{tabular}
\end{table}


\section{Supporting tables for epikernel analysis}\label{app:pyro_tables}

Here, we provide tables supplementing the epikernel analyses in Sec.~\ref{sec:examples}. In Table~\ref{tab:kagome_subduction}, we give a second method of checking the epikernel subgroups: one finds the spin point subgroups of $\mathcal{G} = \othree \times 6mm$ such that the order parameter irrep $\Gamma_{S}^{(1-)}\otimes \Gamma$, restricted to these subgroup elements, has a trivial component. Tables~\ref{tab:Pyrochlore_groups_(i)} and \ref{tab:pyro_epi_perp_groups} give the epikernels stabilizing the generalized collinear and mutually orthogonal families on the pyrochlore lattice, from Sec.~\ref{sec:pyrochlore}. 

In Tables~\ref{tab:Pyrochlore_groups_(i)} and \ref{tab:pyro_epi_perp_groups} the order-parameter
components are listed in a basis cyclically permuted with respect to
Eq.~(\ref{eq:Pyro_N_moments}): the entries $(N_1,N_2,N_3)$ in the tables correspond to
$(N_3,N_1,N_2)$ of Eq.~(\ref{eq:Pyro_N_moments}), i.e., they transform as $(xy,yz,zx)$ under
$m\bar{3}m$. In the orthogonal rows we take $\vhat{N}_1\parallel\vhat{z}$,
$\vhat{N}_2\parallel\vhat{y}$, $\vhat{N}_3\parallel\vhat{x}$.
At zero SOC, this choice of spin axes is a
convention. The single-$N$ state of Sec.~\ref{sec:pyrochlore} ($N_3\neq0$ in Eq.~(\ref{eq:Pyro_N_moments})) corresponds to
the row $(N_1,0,0)$.

\begin{table}[b]
\centering
\caption{For collinear (left) or coplanar (right) SPGs with parent group $6mm$, whether $\Gamma_{S}^{(1-)}\otimes E_{2}$ contains the trivial irrep $\Gamma_{0}$.}
\label{tab:kagome_subduction}
\small
\setlength{\tabcolsep}{5pt}
\renewcommand{\arraystretch}{1.1}
\begin{tabular}{cc|cc}
\hline\hline
\multicolumn{2}{c|}{Collinear: ${b_{\infty}}\times S$} & \multicolumn{2}{c}{Coplanar: ${b_{\mathbb{Z}_{2}}}\times S$} \\
$S$ & $\Gamma_{0}$? & $S$ & $\Gamma_{0}$? \\
\hline
${^{1}6^{1}m^{1}m}$ & \textcolor{red}{\newcrossmark} & ${^{1}6^{2}m^{2}m}$ & \textcolor{red}{\newcrossmark} \\
${^{\overline{1}}6^{1}m^{\overline{1}}m}$ & \textcolor{red}{\newcrossmark} & ${^{2}6^{1}m^{2}m}$ & \textcolor{red}{\newcrossmark} \\
${^{1}6^{\overline{1}}m^{\overline{1}}m}$ & \textcolor{red}{\newcrossmark} & ${^{2_{x}}6^{2_{y}}m^{2_{z}}m}$ & \textcolor{red}{\newcrossmark} \\
 & & ${^{3_{z}}6^{2_{x}}m^{2_{xy}}m}$ & \newcheckmark \\
 & & ${^{6_{z}}6^{2_{x}}m^{2_{1}}m}$ & \textcolor{red}{\newcrossmark} \\
\hline
${^{1}6}$ & \textcolor{red}{\newcrossmark} & ${^{6}6}$ & \textcolor{red}{\newcrossmark} \\
${^{\overline{1}}6}$ & \textcolor{red}{\newcrossmark} & ${^{3}6}$ & \newcheckmark \\
 & & ${^{2}6}$ & \textcolor{red}{\newcrossmark} \\
\hline
${^{1}3^{1}m}$ & \textcolor{red}{\newcrossmark} & ${^{1}3^{2}m}$ & \textcolor{red}{\newcrossmark} \\
${^{1}3^{\overline{1}}m}$ & \textcolor{red}{\newcrossmark} & ${^{3_{z}}3^{2_{x}}m}$ & \newcheckmark \\
\hline
${^{1}3}$ & \textcolor{red}{\newcrossmark} & ${^{3}3}$ & \newcheckmark \\
\hline
${^{1}m^{1}m^{1}2}$ & \newcheckmark & ${^{2}m^{2}m^{1}2}$ & \newcheckmark \\
${^{1}m^{\overline{1}}m^{\overline{1}}2}$ & \textcolor{red}{\newcrossmark} & ${^{1}m^{2}m^{2}2}$ & \newcheckmark \\
${^{\overline{1}}m^{\overline{1}}m^{1}2}$ & \newcheckmark & ${^{2_{x}}m^{2_{y}}m^{2_{z}}2}$ & \newcheckmark \\
\hline
${^{1}2}$ & \newcheckmark & ${^{2}2}$ & \newcheckmark \\
${^{\overline{1}}2}$ & \textcolor{red}{\newcrossmark} & & \\
\hline
${^{1}m}$ & \newcheckmark & ${^{2}m}$ & \newcheckmark \\
${^{\overline{1}}m}$ & \newcheckmark & & \\
\hline\hline
\end{tabular}
\end{table}

\clearpage

\onecolumngrid
\footnotesize
\begin{center}
\begin{longtable}[t!]{|c|c|c|c|c|}
\caption{Epikernels for collinear order on the pyrochlore lattice. The first column indicates the subspaces stabilized. The second column indicates the unitary SPG elements stabilizing the subspace. The third column indicates the unitary halving subgroup of the epikernel, and the fourth column indicates the total epikernel. The fifth column depicts a representative order on the pyrochlore unit cell, with colors accentuating differing moment lengths. Thick horizontal lines separate families of solutions related by broken symmetry. Here, $2_{\phi}$ is a $\pi$ rotation about an axis perpendicular to the collinear axis, obtained by rotating a fixed reference axis by $\phi$ about the spin axis (the reference axis is $\vec{\hat{x}}$ for spin axes $\vec{\hat{y}}$ and $\vec{\hat{z}}$, and $\vec{\hat{z}}$ for spin axis $\vec{\hat{x}}$).} \label{tab:Pyrochlore_groups_(i)}\\
\hline
Subspace & Unitary group elements & Unitary subgroup & Total group  & Order\\[0.5em]
\hline
$\begin{bmatrix} N_1 \\ N_2 \\ N_{3}\end{bmatrix}$ & 
$\{[R_z(\varphi)\|E], [R_z(\varphi)\|I]\}$ & 
$\mathrm{SO(2)} \times {\overline{1}}$ &
${b}_{\infty} \times {^{1}{\overline{1}}}$ &  \includegraphics[width=2cm,valign=m]{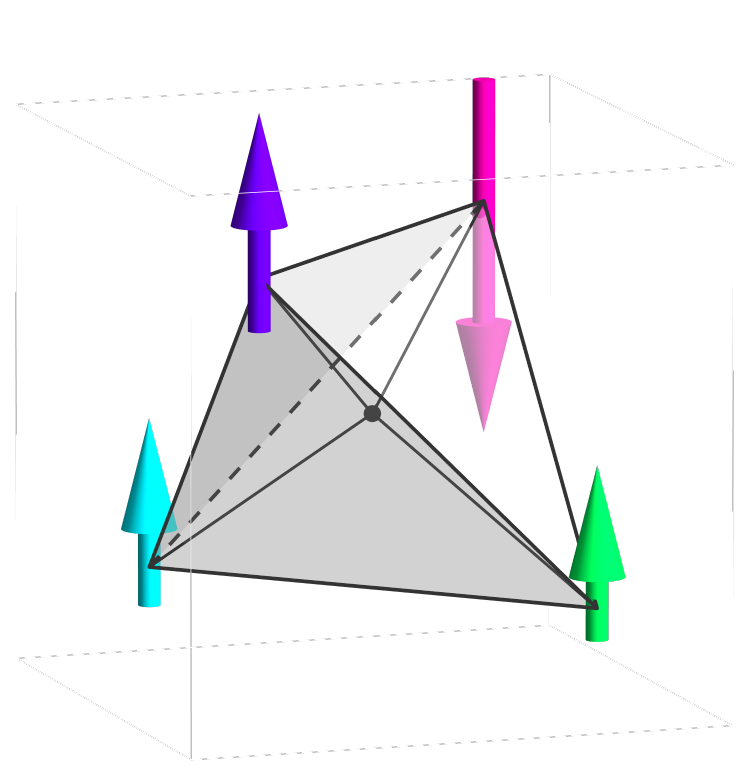} \\[0.5em]
\hline\hline\hline
$\begin{bmatrix} N_1 \\ N_{2} \\ N_{2} \end{bmatrix}$ & 
$\{[R_z(\varphi)\|E], [R_z(\varphi)\|2_{1\overline{1}0}], [R_z(\varphi)\|I], [R_z(\varphi)\|m_{1\overline{1}0}]\}$ & 
$\mathrm{SO(2)} \times {2/m}$ &
${b}_{\infty} \times {^{1}2/^{1}m}$ &  \includegraphics[width=2cm,valign=m]{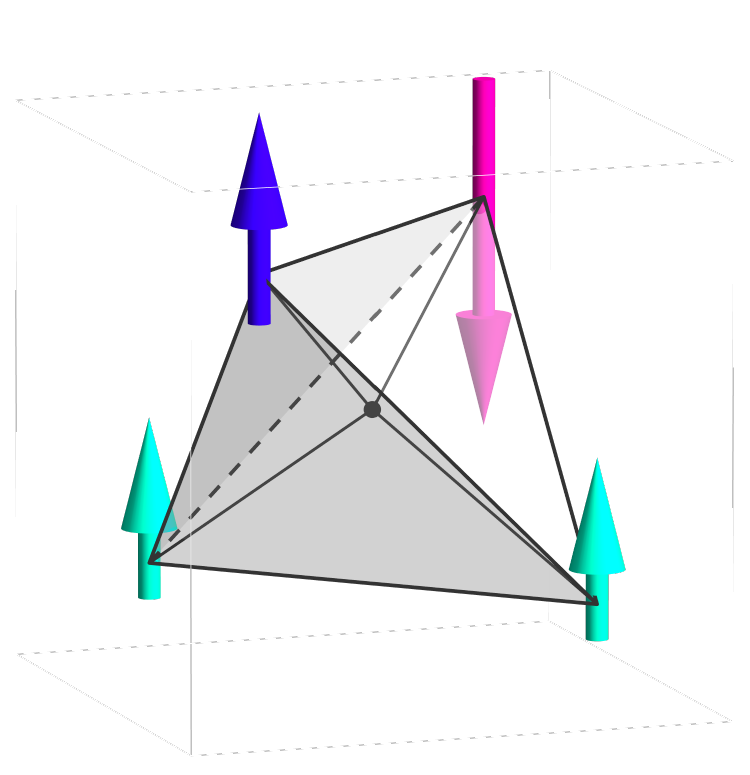} \\[0.5em]
\hline
$\begin{bmatrix} N_{1} \\ N_2\\ -N_{2} \end{bmatrix}$ & 
$\{[R_z(\varphi)\|E], [R_z(\varphi)\|2_{110}], [R_{z}(\varphi)\|I], [R_{z}(\varphi)\|m_{110}]\}$ & 
$\mathrm{SO(2)} \times {2/m}$ &
${b}_{\infty} \times {^{1}2/^{1}m}$  &  \includegraphics[width=2cm,valign=m]{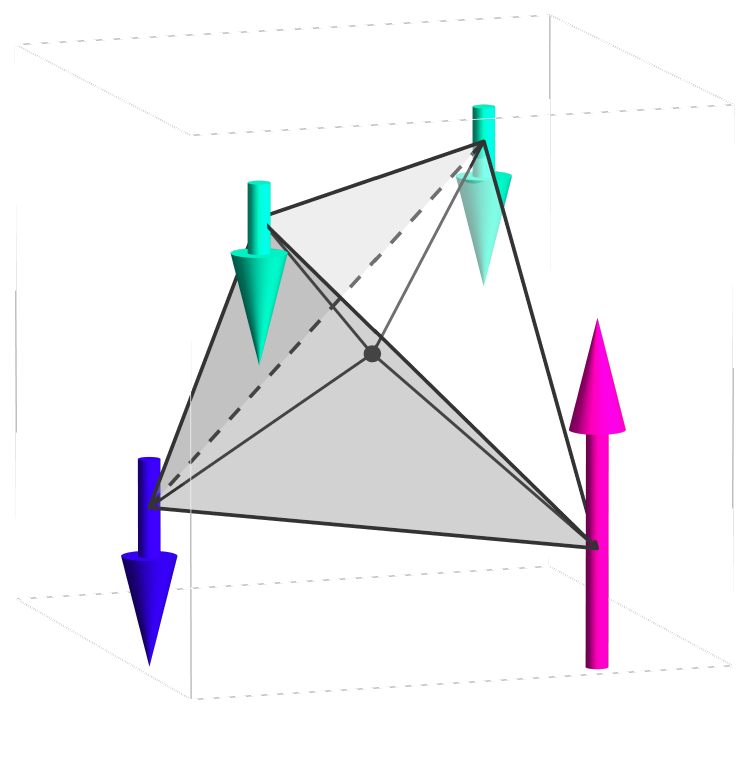} \\[0.5em]
\hline
$\begin{bmatrix} N_{1} \\ N_2\\ N_{1} \end{bmatrix}$ & 
$\{[R_z(\varphi)\|E], [R_z(\varphi)\|2_{01\overline{1}}], [R_{z}(\varphi)\|I], [R_{z}(\varphi)\|m_{01\overline{1}}]\}$ & 
$\mathrm{SO(2)} \times {2/m}$ &
${b}_{\infty} \times {^{1}2/^{1}m}$ &  \includegraphics[width=2cm,valign=m]{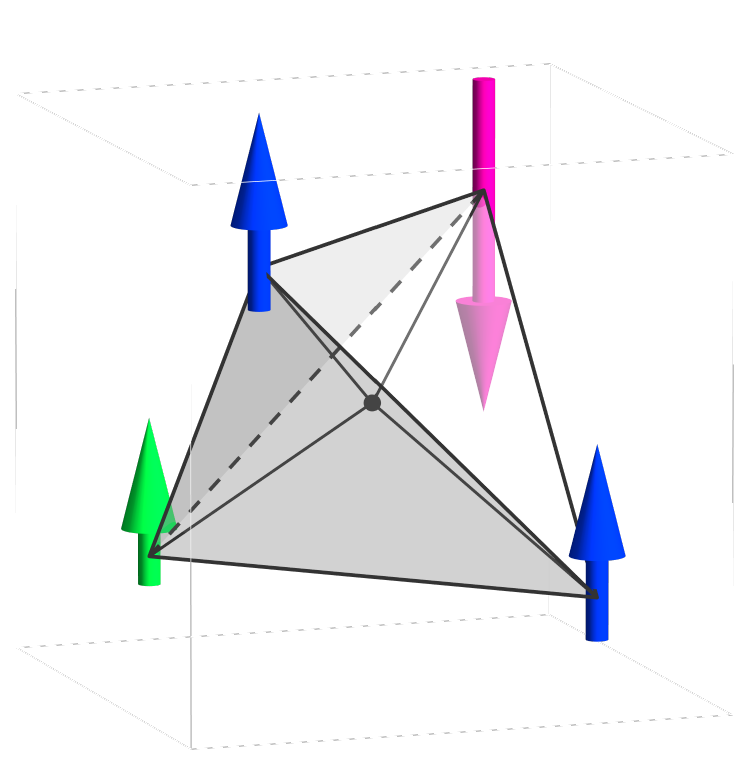} \\[0.5em]
\hline
$\begin{bmatrix} N_{1} \\ N_2\\ -N_{1} \end{bmatrix}$ & 
$\{[R_z(\varphi)\|E], [R_z(\varphi)\|2_{011}], [R_{z}(\varphi)\|I], [R_{z}(\varphi)\|m_{011}]\}$ & 
$\mathrm{SO(2)} \times {2/m}$ &
${b}_{\infty} \times {^{1}2/^{1}m}$ &  \includegraphics[width=2cm,valign=m]{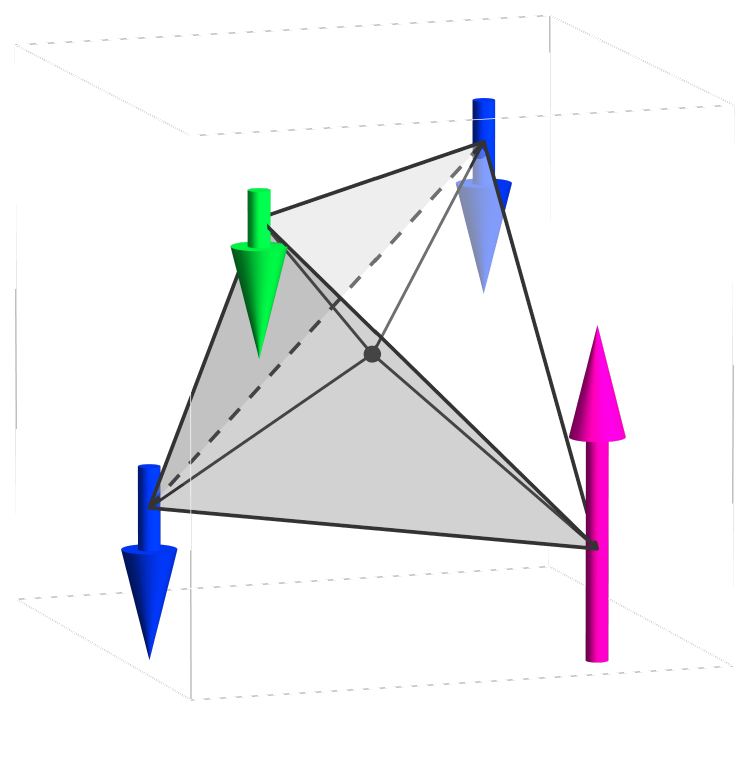} \\[0.5em]
\hline
$\begin{bmatrix} N_{1} \\ N_{1}\\ N_{3} \end{bmatrix}$ & 
$\{[R_z(\varphi)\|E], [R_z(\varphi)\|2_{10\overline{1}}], [R_{z}(\varphi)\|I], [R_{z}(\varphi)\|m_{10\overline{1}}]\}$ & 
$\mathrm{SO(2)} \times {2/m}$ &
${b}_{\infty} \times {^{1}2/^{1}m}$ &  \includegraphics[width=2cm,valign=m]{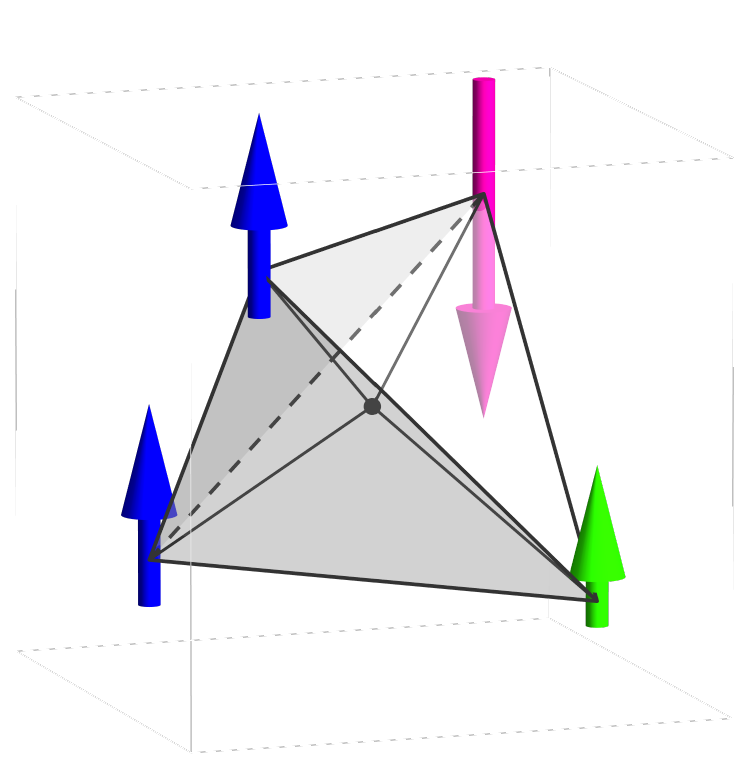} \\[0.5em]
\hline
$\begin{bmatrix} N_{1} \\ -N_{1}\\ N_{3} \end{bmatrix}$ & 
$\{[R_z(\varphi)\|E], [R_z(\varphi)\|2_{101}], [R_{z}(\varphi)\|I], [R_{z}(\varphi)\|m_{101}]\}$ & 
$\mathrm{SO(2)} \times {2/m}$ &
${b}_{\infty} \times {^{1}2/^{1}m}$  &  \includegraphics[width=2cm,valign=m]{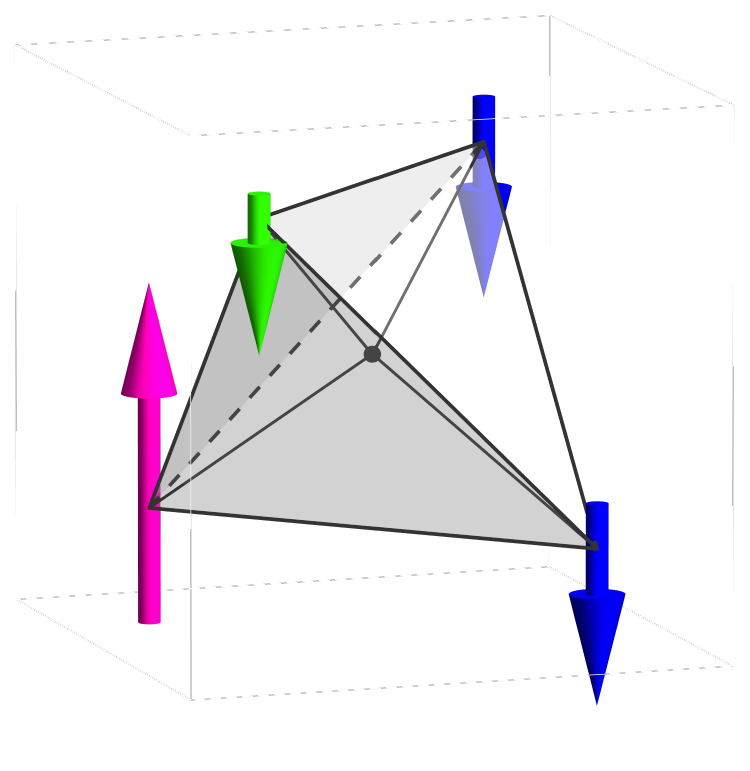}\\[0.5em]
\hline\hline\hline
$\begin{bmatrix} N_{1} \\ N_{1}\\ N_{1} \end{bmatrix}$ & 
\parbox[c]{0.55\textwidth}{\centering \{$[R_z(\varphi)\|E]$, $[R_z(\varphi)\|3_{111}]$,$[R_z(\varphi)\|3^{-1}_{111}]$, $[R_{z}(\varphi)\|I]$, $[R_z(\varphi)\|\overline{3}_{111}]$,$[R_z(\varphi)\|\overline{3}^{-1}_{111}]$,$[R_{z}(\varphi)\|2_{01\overline{1}}]$, $[R_{z}(\varphi)\|2_{10\overline{1}}]$, $[R_{z}(\varphi)\|2_{1\overline{1}0}]$,$[R_{z}(\varphi)\|m_{01\overline{1}}]$, $[R_{z}(\varphi)\|m_{10\overline{1}}]$, $[R_{z}(\varphi)\|m_{1\overline{1}0}]$\}} & 
$\mathrm{SO(2)} \times {\overline{3}m}$ &
${b}_{\infty} \times {^{1}\overline{3}m}$  &  \includegraphics[width=2cm,valign=m]{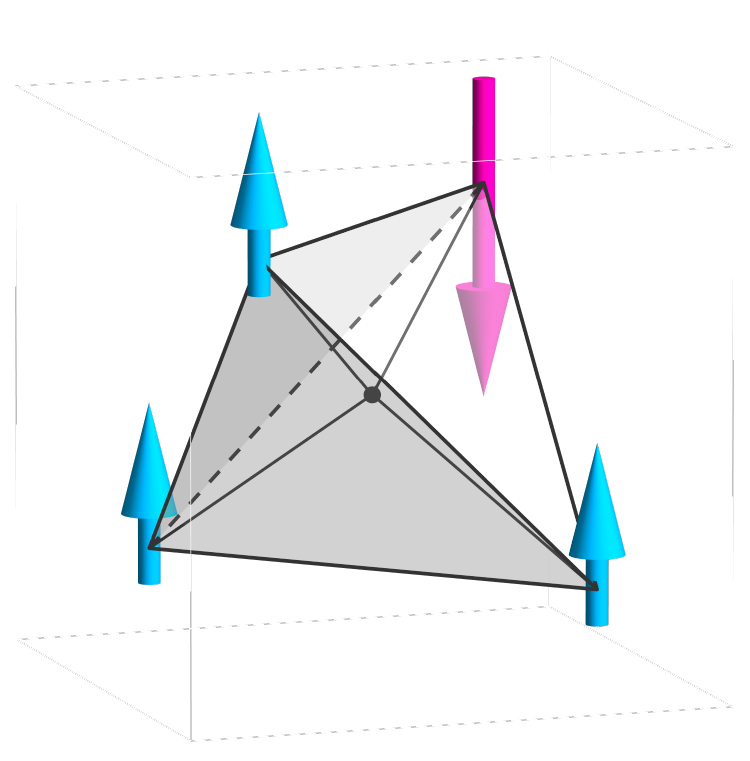}\\[.5em]
\hline
$\begin{bmatrix} N_{1} \\ N_{1}\\ -N_{1} \end{bmatrix}$ & 
\parbox[c]{0.55\textwidth}{\centering \{$[R_z(\varphi)\|E]$, $[R_z(\varphi)\|3_{1\overline{1}1}]$,$[R_z(\varphi)\|3^{-1}_{1\overline{1}1}]$, $[R_{z}(\varphi)\|I]$, $[R_z(\varphi)\|\overline{3}_{1\overline{1}1}]$,$[R_z(\varphi)\|\overline{3}^{-1}_{1\overline{1}1}]$,$[R_{z}(\varphi)\|2_{011}]$, $[R_{z}(\varphi)\|2_{10\overline{1}}]$, $[R_{z}(\varphi)\|2_{110}]$,$[R_{z}(\varphi)\|m_{011}]$, $[R_{z}(\varphi)\|m_{10\overline{1}}]$, $[R_{z}(\varphi)\|m_{110}]$\}} & 
$\mathrm{SO(2)} \times {\overline{3}m}$ &
${b}_{\infty} \times {^{1}\overline{3}m}$  &  \includegraphics[width=2cm,valign=m]{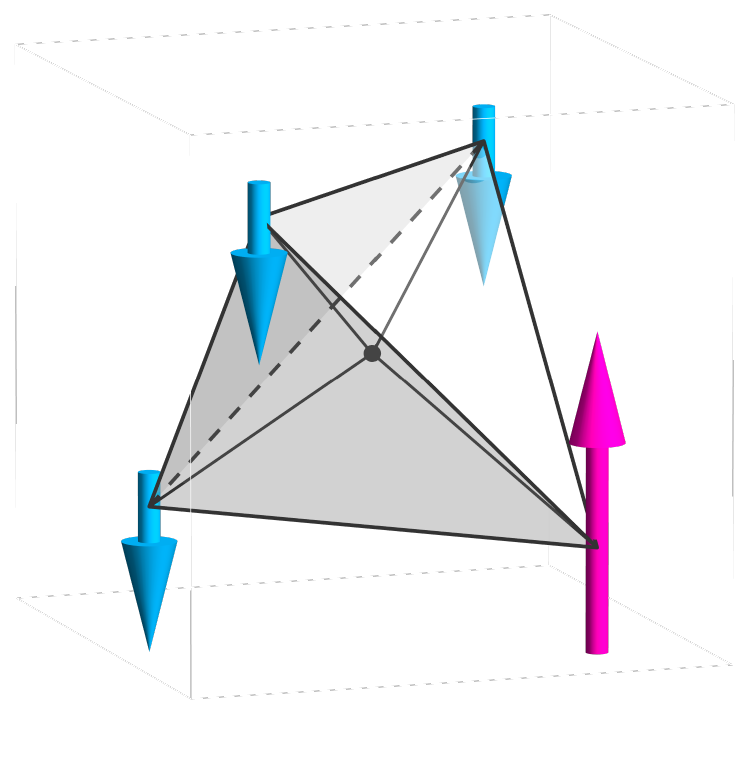}\\[0.5em]
\hline
$\begin{bmatrix} N_{1} \\ -N_{1}\\ N_{1} \end{bmatrix}$ & 
\parbox[c]{0.55\textwidth}{\centering \{$[R_z(\varphi)\|E]$,$ [R_z(\varphi)\|3_{1\overline{11}}]$,$[R_z(\varphi)\|3^{-1}_{1\overline{11}}]$, $[R_{z}(\varphi)\|I]$,$[R_z(\varphi)\|\overline{3}_{1\overline{11}}]$,$[R_z(\varphi)\|\overline{3}^{-1}_{1\overline{11}}]$,$[R_{z}(\varphi)\|2_{01\overline{1}}]$, $[R_{z}(\varphi)\|2_{101}]$, $[R_{z}(\varphi)\|2_{110}]$,$[R_{z}(\varphi)\|m_{01\overline{1}}]$, $[R_{z}(\varphi)\|m_{101}]$, $[R_{z}(\varphi)\|m_{110}]$\}} & 
$\mathrm{SO(2)} \times {\overline{3}m}$ &
${b}_{\infty} \times {^{1}\overline{3}m}$  &  \includegraphics[width=2cm,valign=m]{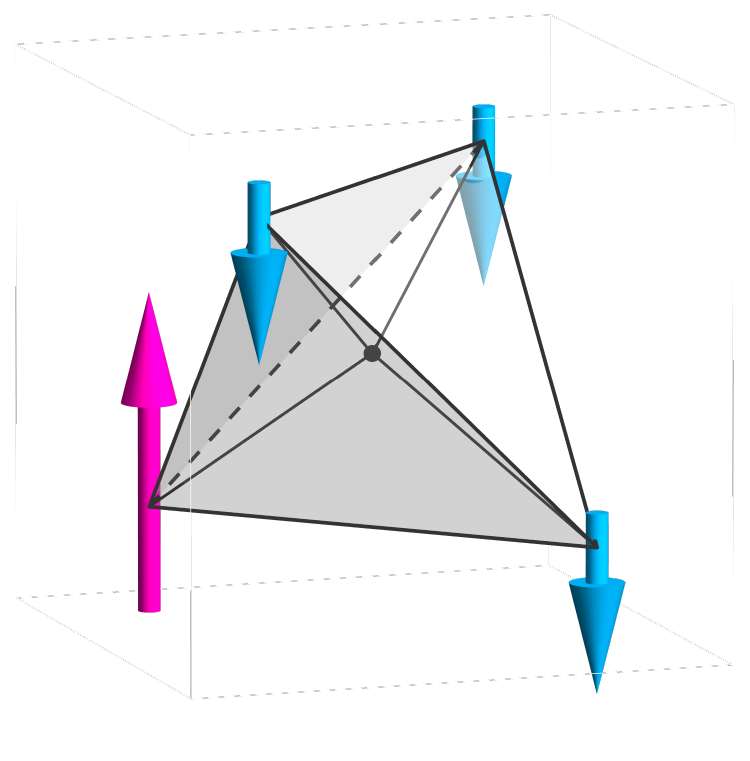}\\[0.5em]
\hline
$\begin{bmatrix} N_{1} \\ -N_{1}\\ -N_{1} \end{bmatrix}$ & 
\parbox[c]{0.55\textwidth}{\centering \{$[R_z(\varphi)\|E]$, $[R_z(\varphi)\|3_{11\overline{1}}]$,$[R_z(\varphi)\|3^{-1}_{11\overline{1}}]$, $[R_{z}(\varphi)\|I]$, $[R_z(\varphi)\|\overline{3}_{11\overline{1}}]$,\\$[R_z(\varphi)\|\overline{3}^{-1}_{11\overline{1}}]$,$[R_{z}(\varphi)\|2_{011}]$, $[R_{z}(\varphi)\|2_{101}]$, $[R_{z}(\varphi)\|2_{1\overline{1}0}]$,$[R_{z}(\varphi)\|m_{011}]$, \\$[R_{z}(\varphi)\|m_{101}]$, $[R_{z}(\varphi)\|m_{1\overline{1}0}]$\}} & 
$\mathrm{SO(2)} \times {\overline{3}m}$ &
${b}_{\infty} \times {^{1}\overline{3}m}$  &  \includegraphics[width=2cm,valign=m]{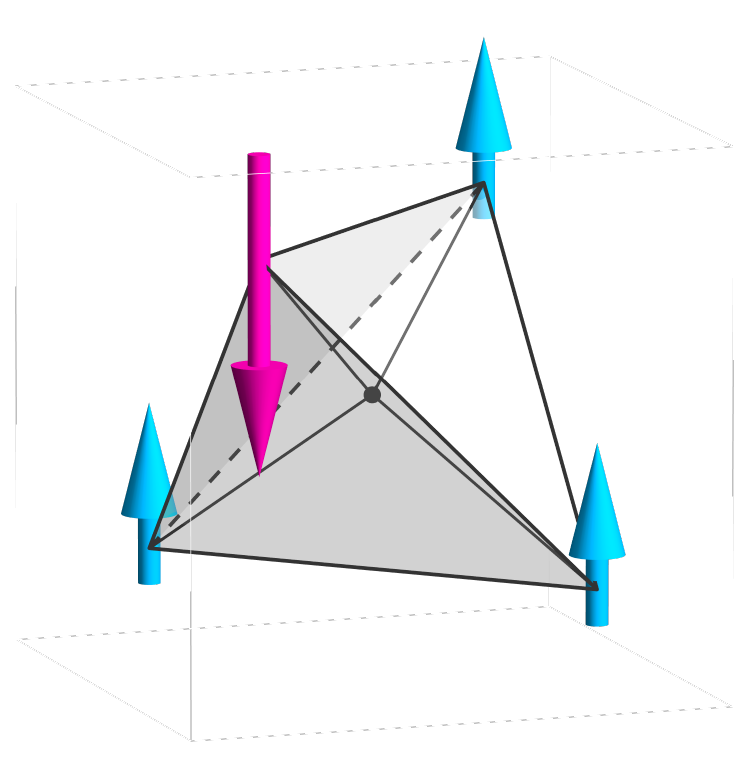} \\[0.5em]
\hline\hline\hline
$\begin{bmatrix} N_{1} \\ N_{2}\\ 0 \end{bmatrix}$ & 
$\{[R_z(\varphi)\|E], [2_{\phi}\|2_{y}],[R_z(\varphi)\|I], [2_{\phi}\|m_{y}]\}$ & 
$\mathrm{SO(2)} \times {^{2}\overline{2}/^{2}m}$ &
${b}_{\infty} \times {^{\overline{1}}2/^{\overline{1}}m}$ &\includegraphics[width=2cm,valign=m]{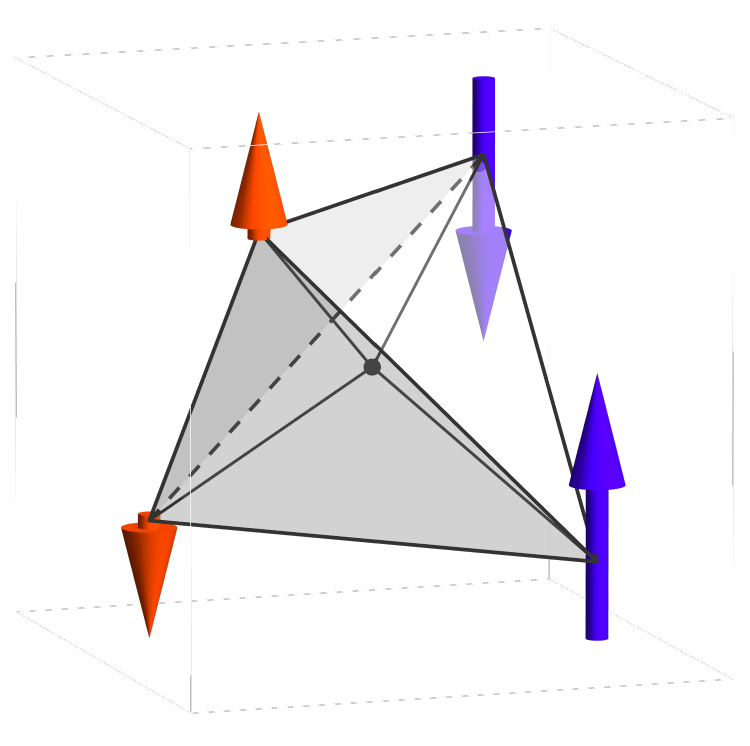}\\[0.5em]
\hline
$\begin{bmatrix} N_{1} \\ 0\\ N_{3} \end{bmatrix}$ & 
$\{[R_z(\varphi)\|E], [2_{\phi}\|2_{x}],[R_z(\varphi)\|I], [2_{\phi}\|m_{x}]\}$ & 
$\mathrm{SO(2)} \times {^{2}\overline{2}/^{2}m}$ &
${b}_{\infty} \times {^{\overline{1}}2/^{\overline{1}}m}$ &\includegraphics[width=2cm,valign=m]{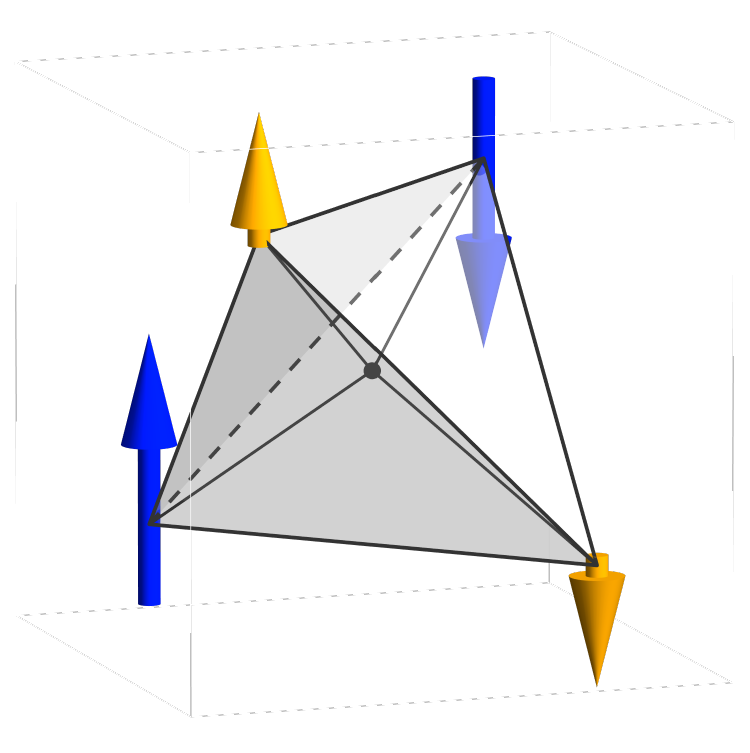}\\[0.5em]
\hline
$\begin{bmatrix} 0 \\ N_{2}\\ N_{3} \end{bmatrix}$ & 
$\{[R_z(\varphi)\|E], [2_{\phi}\|2_{z}],[R_z(\varphi)\|I], [2_{\phi}\|m_{z}]\}$ & 
$\mathrm{SO(2)} \times {^{2}\overline{2}/^{2}m}$ &
${b}_{\infty} \times {^{\overline{1}}2/^{\overline{1}}m}$ &\includegraphics[width=2cm,valign=m]{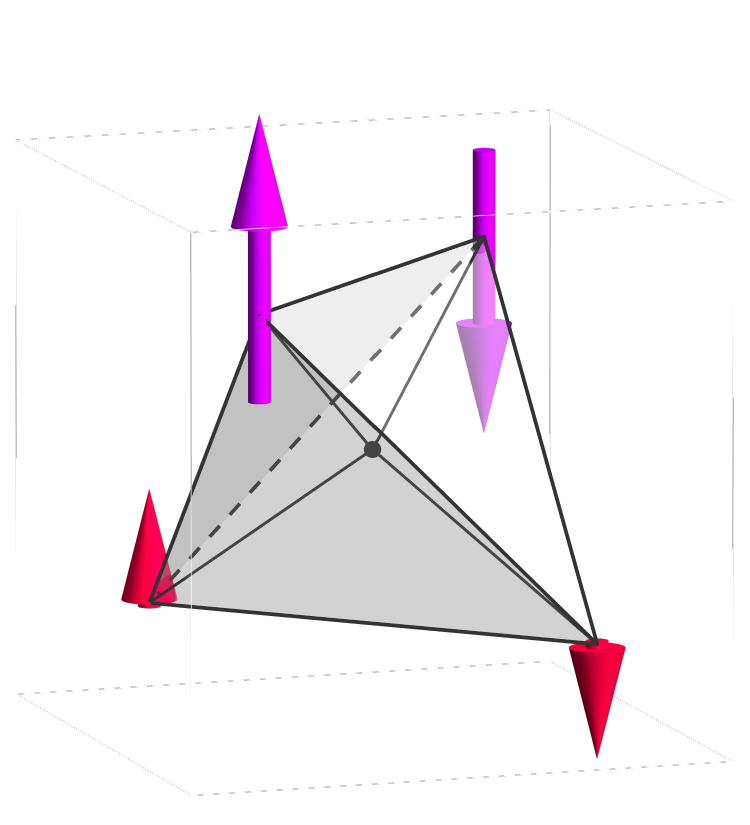}\\[0.5em]
\hline\hline\hline
$\begin{bmatrix} N_{1} \\ N_{1}\\ 0 \end{bmatrix}$ & 
\parbox[c]{0.55\textwidth}{\centering $\{[R_z(\varphi)\|E]$,$[2_{\phi}\|2_{101}]$,$ [R_{z}(\varphi)\|2_{10\overline{1}}]$, $[2_{\phi}\|2_{y}]$,$[R_z(\varphi)\|I]$, $[2_{\phi}\|m_{101}]$, $[R_{z}(\varphi)\|m_{10\overline{1}}]$, $[2_{\phi}\|m_{y}]\}$} & 
$\mathrm{SO(2)} \times {^{1}m^{2}m^{2}m}$ &
${b}_{\infty} \times {^{1}m^{\overline{1}}m^{\overline{1}}m}$&\includegraphics[width=2cm,valign=m]{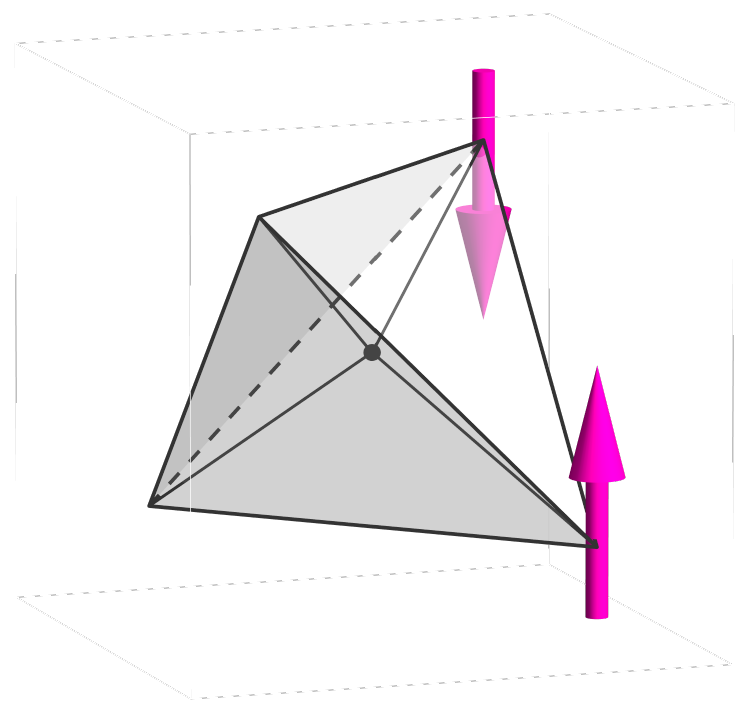} \\[0.5em]
\hline
$\begin{bmatrix} N_{1} \\ -N_{1}\\ 0 \end{bmatrix}$ & 
\parbox[c]{0.55\textwidth}{\centering $\{[R_z(\varphi)\|E]$,$[R_{z}(\varphi)\|2_{101}]$, $[2_{\phi}\|2_{10\overline{1}}]$, $[2_{\phi}\|2_{y}],[R_z(\varphi)\|I]$, $[R_{z}(\varphi)\|m_{101}]$, $[2_{\phi}\|m_{10\overline{1}}]$, $[2_{\phi}\|m_{y}]\}$} & 
$\mathrm{SO(2)} \times {^{1}m^{2}m^{2}m}$ &
${b}_{\infty} \times {^{1}m^{\overline{1}}m^{\overline{1}}m}$&\includegraphics[width=2cm,valign=m]{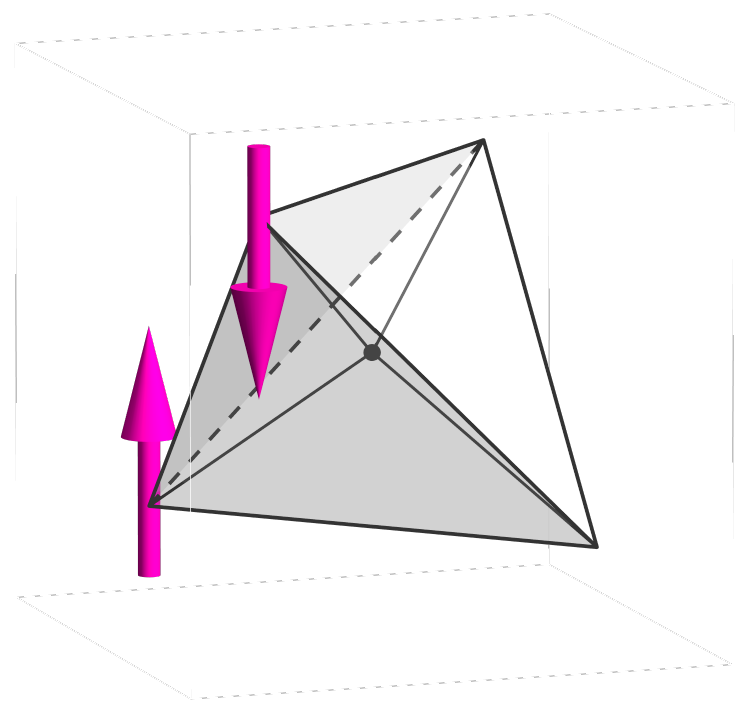} \\[0.5em]
\hline
$\begin{bmatrix} N_{1} \\ 0\\ N_{1} \end{bmatrix}$ & 
\parbox[c]{0.55\textwidth}{\centering $\{[R_z(\varphi)\|E],[R_{z}(\varphi)\|2_{011}]$, $[2_{\phi}\|2_{01\overline{1}}]$, $[2_{\phi}\|2_{x}],[R_z(\varphi)\|I]$, $[R_{z}(\varphi)\|m_{011}]$, $[2_{\phi}\|m_{01\overline{1}}]$,$ [2_{\phi}\|m_{x}]\}$} & 
$\mathrm{SO(2)} \times {^{1}m^{2}m^{2}m}$ &
${b}_{\infty} \times {^{1}m^{\overline{1}}m^{\overline{1}}m}$&\includegraphics[width=2cm,valign=m]{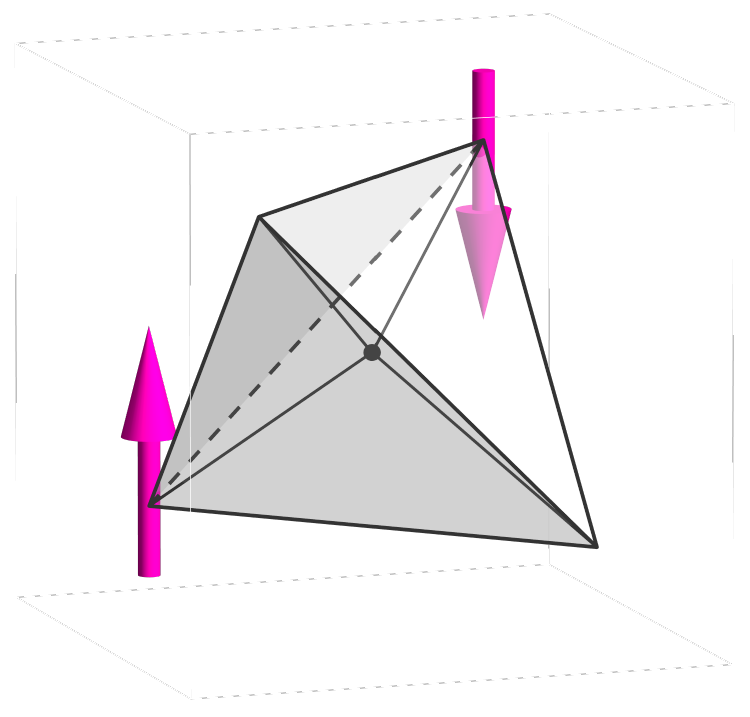} \\[0.5em]
\hline
$\begin{bmatrix} N_{1} \\ 0\\ -N_{1} \end{bmatrix}$ & 
\parbox[c]{0.55\textwidth}{\centering $\{[R_z(\varphi)\|E]$,$[2_{\phi}\|2_{011}]$, $[R_{z}(\varphi)\|2_{01\overline{1}}]$, $[2_{\phi}\|2_{x}]$,$[R_z(\varphi)\|I]$, $[2_{\phi}\|m_{011}]$, $[R_{z}(\varphi)\|m_{01\overline{1}}]$, $[2_{\phi}\|m_{x}]\}$} & 
$\mathrm{SO(2)} \times {^{1}m^{2}m^{2}m}$ &
${b}_{\infty} \times {^{1}m^{\overline{1}}m^{\overline{1}}m}$&\includegraphics[width=2cm,valign=m]{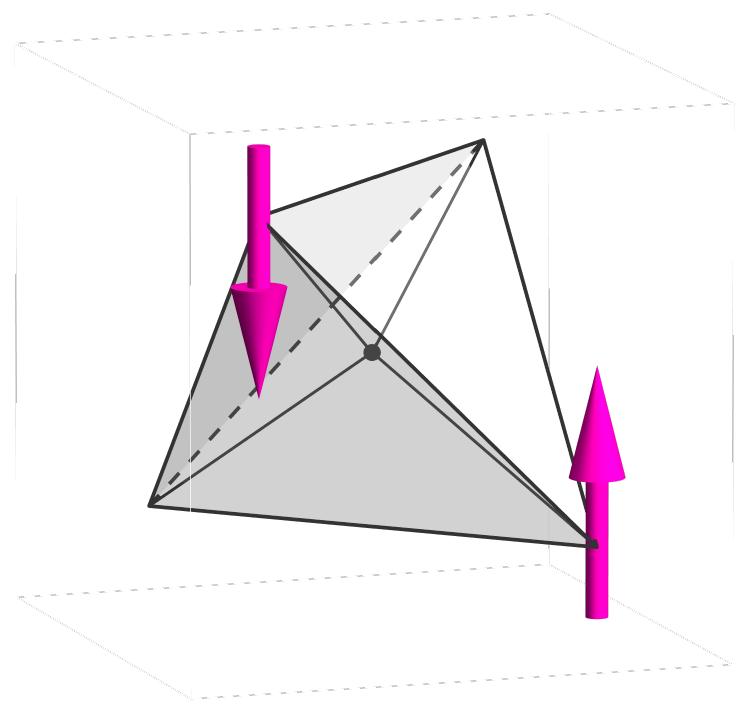} \\[0.5em]
\hline
$\begin{bmatrix} 0 \\ N_{2}\\ N_{2} \end{bmatrix}$ & 
\parbox[c]{0.55\textwidth}{\centering $\{[R_z(\varphi)\|E]$,$[2_{\phi}\|2_{110}]$,$ [R_{z}(\varphi)\|2_{1\overline{1}0}]$,$ [2_{\phi}\|2_{z}]$,$[R_z(\varphi)\|I]$,\\$ [2_{\phi}\|m_{110}]$,$ [R_{z}(\varphi)\|m_{1\overline{1}0}]$,$ [2_{\phi}\|m_{z}]\}$} & 
$\mathrm{SO(2)} \times {^{1}m^{2}m^{2}m}$ &
${b}_{\infty} \times {^{1}m^{\overline{1}}m^{\overline{1}}m}$&\includegraphics[width=2cm,valign=m]{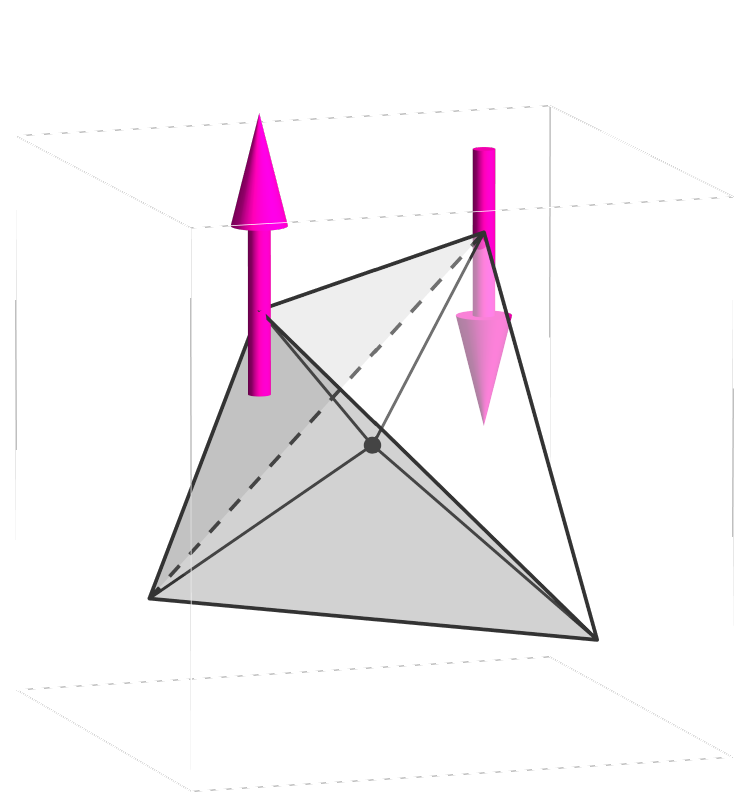} \\[0.5em]
\hline
$\begin{bmatrix} 0 \\ N_{2}\\ -N_{2} \end{bmatrix}$ & 
\parbox[c]{0.55\textwidth}{\centering $\{[R_z(\varphi)\|E]$,$[R_{z}(\varphi)\|2_{110}]$,$ [2_{\phi}\|2_{1\overline{1}0}]$, $ [2_{\phi}\|2_{z}]$,$[R_z(\varphi)\|I]$,\\$ [R_{z}(\varphi)\|m_{110}]$,$ [2_{\phi}\|m_{1\overline{1}0}]$,$ [2_{\phi}\|m_{z}]\}$} & 
$\mathrm{SO(2)} \times {^{1}m^{2}m^{2}m}$ &
${b}_{\infty} \times {^{1}m^{\overline{1}}m^{\overline{1}}m}$ &\includegraphics[width=2cm,valign=m]{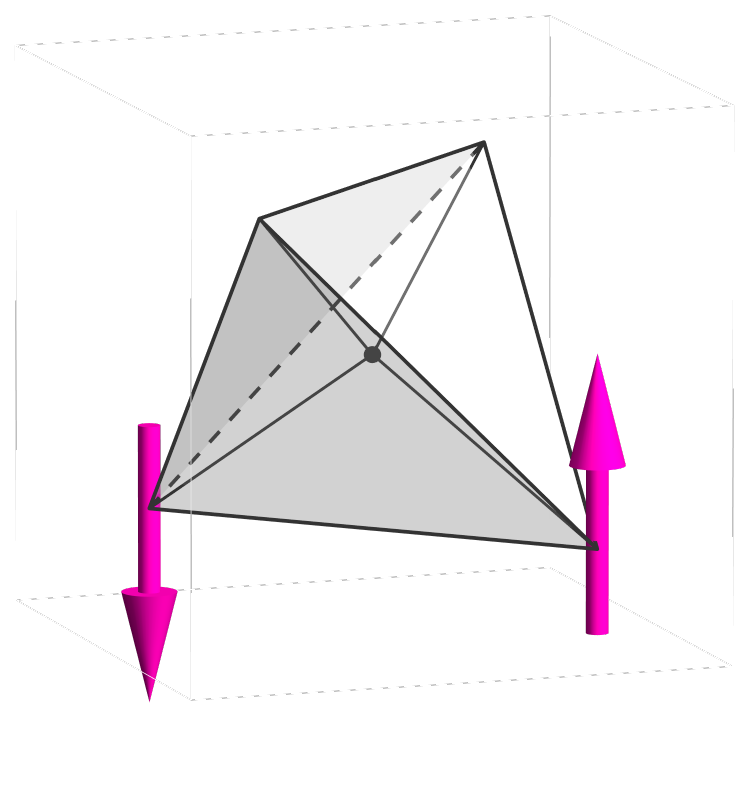}\\[0.5em]
\hline\hline\hline
$\begin{bmatrix} N_{1} \\ 0\\ 0 \end{bmatrix}$ & 
\parbox[c]{0.55\textwidth}{\centering \{$[R_{z}(\varphi) \|  E]$, $[R_{z}(\varphi) \|  I]$, $[ 2_{\phi}R_{z}(\varphi) \|  2_{x}]$, $[ 2_{\phi}R_{z}(\varphi) \|  2_{y}]$, $[R_{z}(\varphi) \|  2_{z}]$, $[R_{z}(\varphi) \|  2_{110}]$, $[R_{z}(\varphi) \|  2_{1\bar{1}0}]$, $[ 2_{\phi}R_{z}(\varphi)  \|  m_{x}]$, $[ 2_{\phi}R_{z}(\varphi) \|  m_{y}]$, $[R_{z}(\varphi) \|  m_{z}]$, $[R_{z}(\varphi) \|  m_{110}]$, $[R_{z}(\varphi) \|  m_{1\bar{1}0}]$, $[ 2_{\phi}R_{z}(\varphi) \|  4^{-}_{z}]$, $[ 2_{\phi}R_{z}(\varphi) \|  4^{+}_{z}]$, $[ 2_{\phi}R_{z}(\varphi) \|  \bar{4}^{-}_{z}]$, $[ 2_{\phi}R_{z}(\varphi) \|  \bar{4}^{+}_{z}]$\}}&
$\mathrm{SO(2)} \times {^{2}4/^{1}m^{2}m^{1}m}$ &
${b_{\infty}^{z}}\times{^{\overline{1}}4/^{1}m^{\overline{1}}m^{1}m}$  &\includegraphics[width=2cm,valign=m]{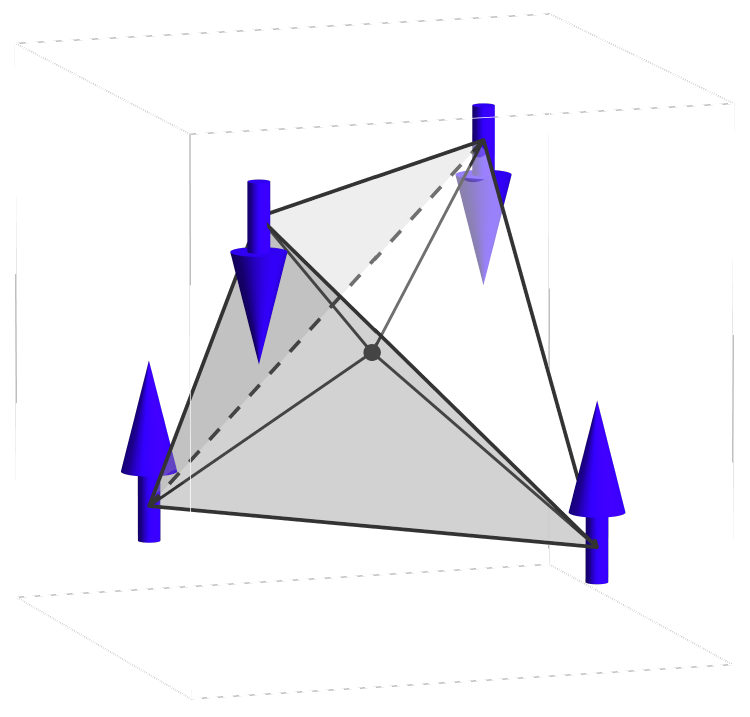}\\[0.5em]
\hline
$\begin{bmatrix} 0 \\ N_{2}\\ 0 \end{bmatrix}$ & 
\parbox[c]{0.55\textwidth}{\centering \{$[R_{y}(\varphi) \|  E]$, $[R_{y}(\varphi) \|  I]$, $[R_{y}(\varphi) \|  2_{x}]$, $[ 2_{\phi}R_{y}(\varphi) \|  2_{y}]$, $[ 2_{\phi}R_{y}(\varphi) \|  2_{z}]$, $[ R_{y}(\varphi) \|  2_{011}]$, $[ R_{y}(\varphi) \|  2_{01\bar{1}}]$, $[  R_{y}(\varphi) \|  m_{x}]$, $[ 2_{\phi} R_{y}(\varphi) \|  m_{y}]$, $[ 2_{\phi} R_{y}(\varphi) \|  m_{z}]$, $[ R_{y}(\varphi) \|  m_{011}]$, $[ R_{y}(\varphi) \|  m_{01\bar{1}}]$, $[ 2_{\phi} R_{y}(\varphi) \|  4^{-}_{x}]$, $[ 2_{\phi} R_{y}(\varphi) \|  4^{+}_{x}]$, $[ 2_{\phi} R_{y}(\varphi) \|  \bar{4}^{-}_{x}]$, $[ 2_{\phi} R_{y}(\varphi) \|  \bar{4}^{+}_{x}]$\}} & $\mathrm{SO(2)} \times {^{2}4/^{1}m^{2}m^{1}m}$ &
${b_{\infty}^{y}}\times{^{\overline{1}}4/^{1}m^{\overline{1}}m^{1}m}$ &\includegraphics[width=2cm,valign=m]{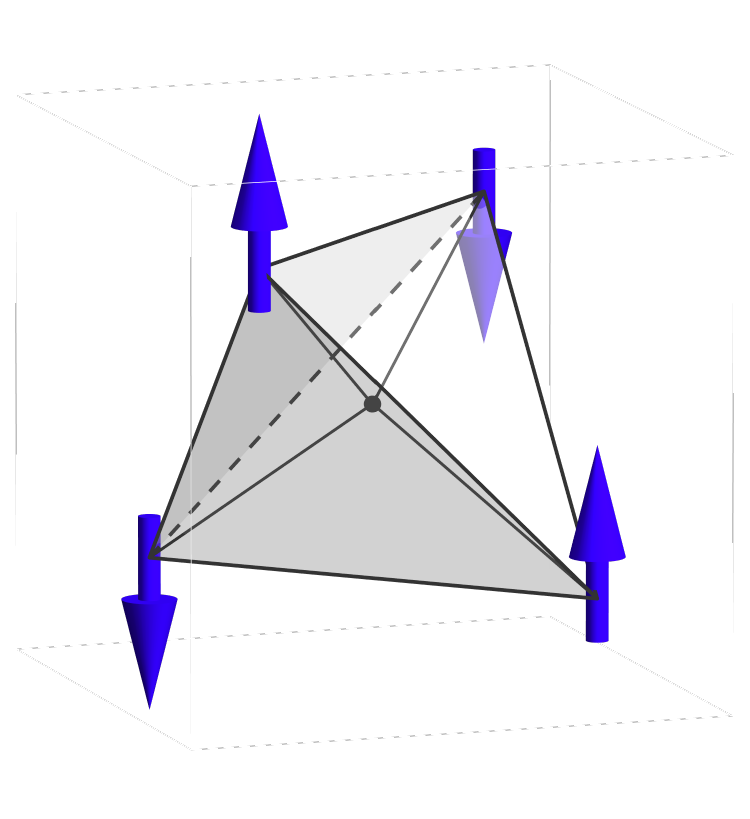} \\[0.5em]
\hline
$\begin{bmatrix} 0 \\ 0\\ N_{3} \end{bmatrix}$ & 
\parbox[c]{0.55\textwidth}{\centering \{$[R_{x}(\varphi) \|  E]$, $[R_{x}(\varphi) \|  I]$, $[ 2_{\phi}R_{x}(\varphi) \|  2_{x}]$, $[R_{x}(\varphi) \|  2_{y}]$, $[ 2_{\phi}R_{x}(\varphi) \|  2_{z}]$, $[R_{x}(\varphi) \|  2_{101}]$, $[R_{x}(\varphi) \|  2_{10\bar{1}}]$, $[ 2_{\phi}R_{x}(\varphi) \|  m_{x}]$, $[R_{x}(\varphi) \|  m_{y}]$, $[ 2_{\phi}R_{x}(\varphi) \|  m_{z}]$, $[R_{x}(\varphi) \|  m_{101}]$, $[R_{x}(\varphi) \|  m_{10\bar{1}}]$, $[ 2_{\phi}R_{x}(\varphi) \|  4^{-}_{y}]$, $[ 2_{\phi}R_{x}(\varphi) \|  4^{+}_{y}]$, $[ 2_{\phi}R_{x}(\varphi) \|  \bar{4}^{-}_{y}]$, $[ 2_{\phi}R_{x}(\varphi) \|  \bar{4}^{+}_{y}]$\}} & 
$\mathrm{SO(2)} \times {^{2}4/^{1}m^{2}m^{1}m}$ &
${b_{\infty}^{x}}\times{^{\overline{1}}4/^{1}m^{\overline{1}}m^{1}m}$ &\includegraphics[width=2cm,valign=m]{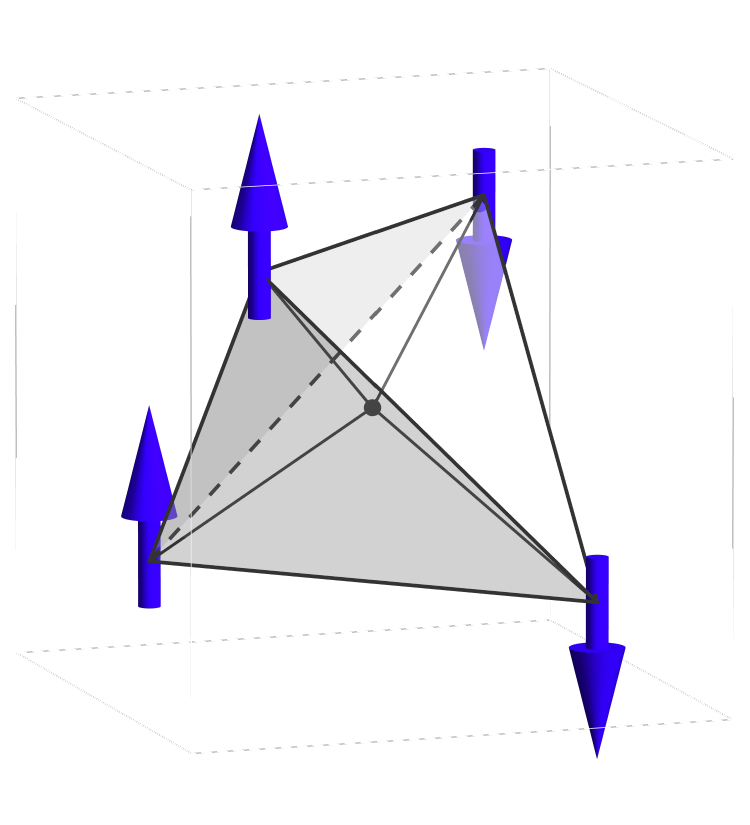}\\[0.5em]
\hline
\end{longtable}
\end{center}

\begin{longtable}{ |c|c| c|c|}
\caption{Epikernels for orthogonal order on the pyrochlore lattice. The first column indicates the stabilized subspace, the second column indicates the SPG elements stabilizing the subspace, the third column indicates the total SPG of the symmetry-broken phase, and the fourth column depicts a representative order on the pyrochlore unit cell. Thick lines separate families of solutions related by broken symmetry. Here, ${b_{\mathbb{Z}_{2}}^{\alpha}}$ indicates the coplanar spin-only group with plane perpendicular to the axis $\alpha$.}\label{tab:pyro_epi_perp_groups}\\
\hline
Subspace & SPG elements & SPG & Moments \\
\hline
\endfirsthead
\hline
Subspace & SPG elements & SPG & Moments \\
\endhead
$\begin{bmatrix} N_{1} \\ N_{2} \\ N_{3} \end{bmatrix}$ & \parbox[t]{0.55\textwidth}{\{$[E \|  E]$, $[E \|  I]$, $[2_{y} \|  2_{x}]$, $[2_{x} \|  2_{y}]$, $[2_{z} \|  2_{z}]$, $[2_{y} \|  m_{x}]$, $[2_{x} \|  m_{y}]$, $[2_{z} \|  m_{z}]$\}} & ${{^{2_{z}}m^{2_{y}}m^{2_{x}}m}}$  & \includegraphics[width=2cm,valign=m]{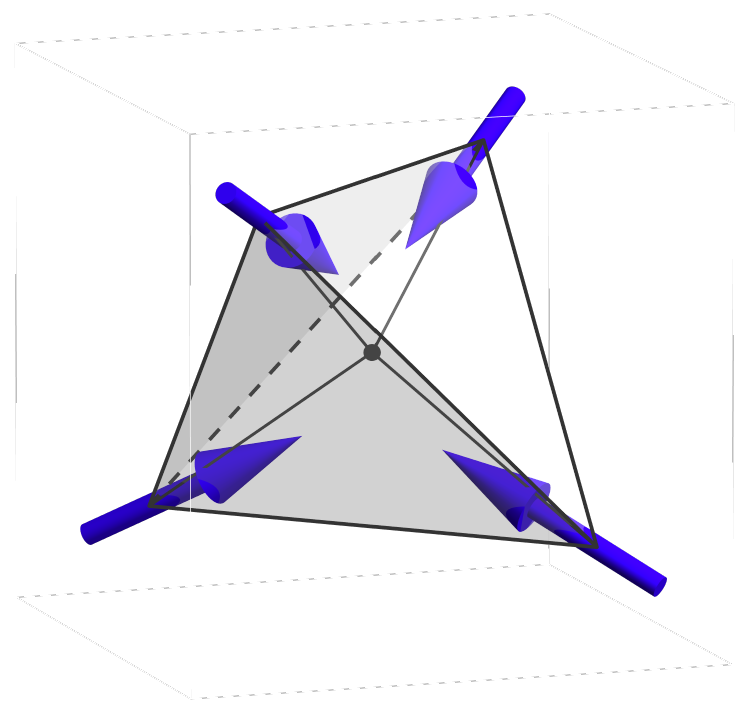}\\
\hline\hline\hline
$\begin{bmatrix} 0 \\ N_{2} \\ N_{3} \end{bmatrix}$ & \parbox[c]{0.55\textwidth}{\{$[m_{z} \|  E]$, $[E \|  E]$, $[m_{z} \|  I]$, $[E \|  I]$, $[2_{y} \|  2_{x}]$, $[m_{x} \|  2_{x}]$, $[2_{x} \|  2_{y}]$, $[m_{y} \|  2_{y}]$, $[T \|  2_{z}]$, $[2_{z} \|  2_{z}]$, $[2_{y} \|  m_{x}]$, $[m_{x} \|  m_{x}]$, $[2_{x} \|  m_{y}]$, $[m_{y} \|  m_{y}]$, $[T \|  m_{z}]$, $[2_{z} \|  m_{z}]$\}} & ${b_{\mathbb{Z}_{2}}^{z} \times \,^{2_{z}}m^{2_{y}}m^{2_{x}}m}$& \includegraphics[width=2cm,valign=m]{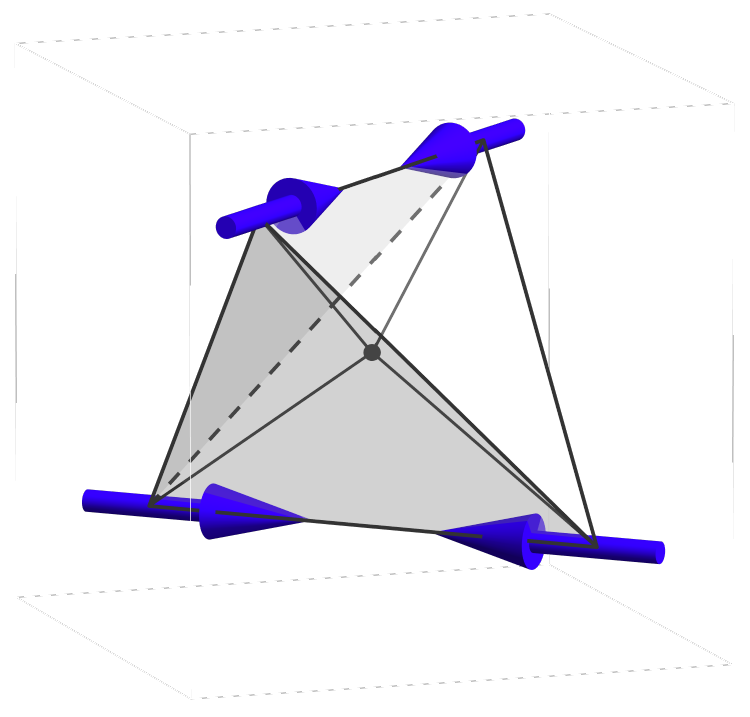} \\
\hline
$\begin{bmatrix} N_{1} \\ 0 \\ N_{3} \end{bmatrix}$ & \parbox[c]{0.55\textwidth}{\{$[m_{y} \|  E]$, $[E \|  E]$, $[m_{y} \|  I]$, $[E \|  I]$, $[T \|  2_{x}]$, $[2_{y} \|  2_{x}]$, $[2_{x} \|  2_{y}]$, $[m_{z} \|  2_{y}]$, $[2_{z} \|  2_{z}]$, $[m_{x} \|  2_{z}]$, $[T \|  m_{x}]$, $[2_{y} \|  m_{x}]$, $[2_{x} \|  m_{y}]$, $[m_{z} \|  m_{y}]$, $[2_{z} \|  m_{z}]$, $[m_{x} \|  m_{z}]$\}} &${b_{\mathbb{Z}_{2}}^{y} \times \,^{2_{z}}m^{2_{y}}m^{2_{x}}m}$ & \includegraphics[width=2cm,valign=m]{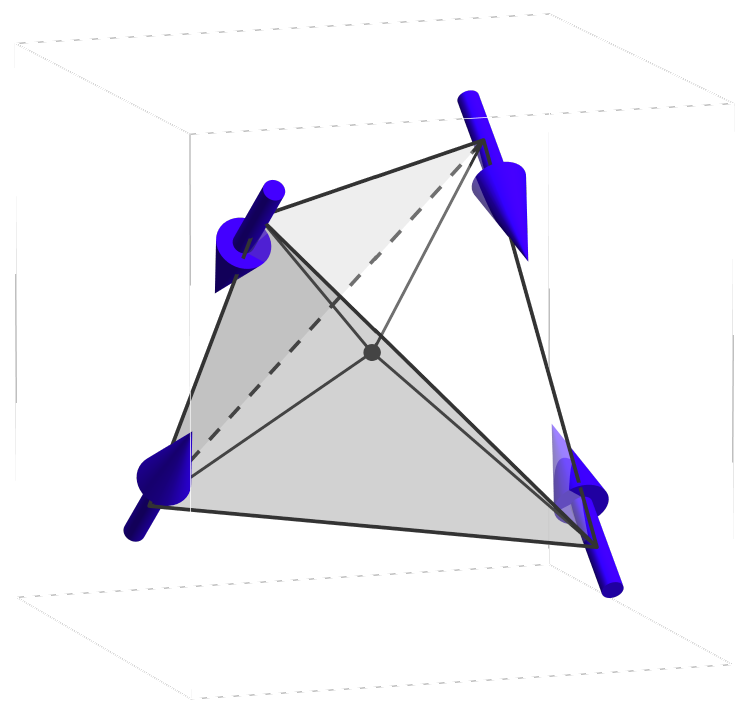} \\
\hline
$\begin{bmatrix} N_{1} \\ N_{2} \\ 0 \end{bmatrix}$ & \parbox[c]{0.55\textwidth}{\{$[m_{x} \|  E]$, $[E \|  E]$, $[m_{x} \|  I]$, $[E \|  I]$, $[2_{y} \|  2_{x}]$, $[m_{z} \|  2_{x}]$, $[T \|  2_{y}]$, $[2_{x} \|  2_{y}]$, $[2_{z} \|  2_{z}]$, $[m_{y} \|  2_{z}]$, $[2_{y} \|  m_{x}]$, $[m_{z} \|  m_{x}]$, $[T \|  m_{y}]$, $[2_{x} \|  m_{y}]$, $[2_{z} \|  m_{z}]$, $[m_{y} \|  m_{z}]$\}} &${b_{\mathbb{Z}_{2}}^{x} \times \,^{2_{z}}m^{2_{y}}m^{2_{x}}m}$& \includegraphics[width=2cm,valign=m]{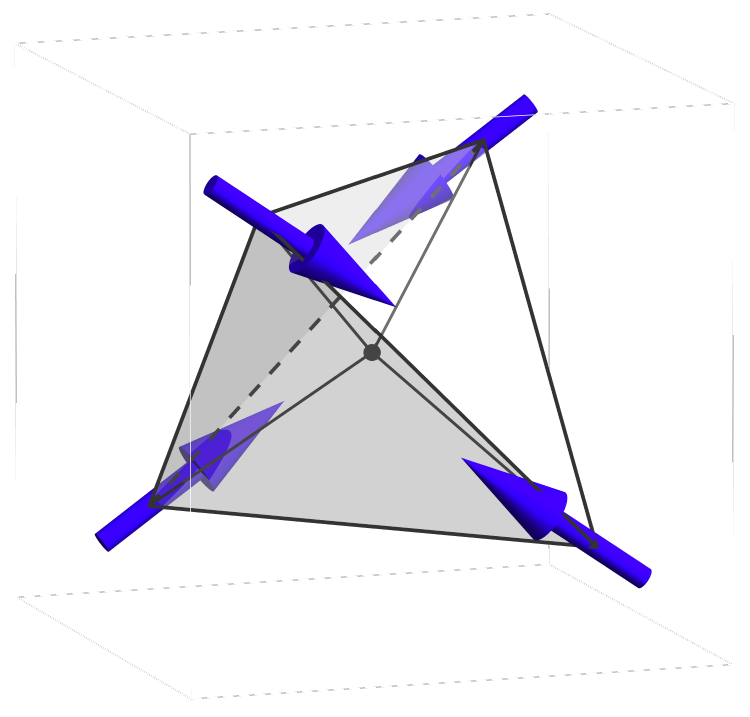} \\
\hline\hline\hline
$\begin{bmatrix} N_{1} \\ -N_{1} \\ N_{3} \end{bmatrix}$ & \parbox[c]{0.55\textwidth}{\{$[E \|  E]$, $[E \|  I]$, $[2_{y} \|  2_{x}]$, $[2_{x} \|  2_{y}]$, $[2_{z} \|  2_{z}]$, $[m_{01\bar{1}} \|  2_{101}]$, $[m_{011} \|  2_{10\bar{1}}]$, $[2_{y} \|  m_{x}]$, $[2_{x} \|  m_{y}]$, $[2_{z} \|  m_{z}]$, $[m_{01\bar{1}} \|  m_{101}]$, $[m_{011} \|  m_{10\bar{1}}]$, $[\bar{4}^{-}_{x} \|  4^{-}_{y}]$, $[\bar{4}^{+}_{x} \|  4^{+}_{y}]$, $[\bar{4}^{-}_{x} \|  \bar{4}^{-}_{y}]$, $[\bar{4}^{+}_{x} \|  \bar{4}^{+}_{y}]$\}} &${^{\overline{4}_{x}}4/^{2_{x}}m^{2_{y}}m^{m_{xz}}m}$ & \includegraphics[width=2cm,valign=m]{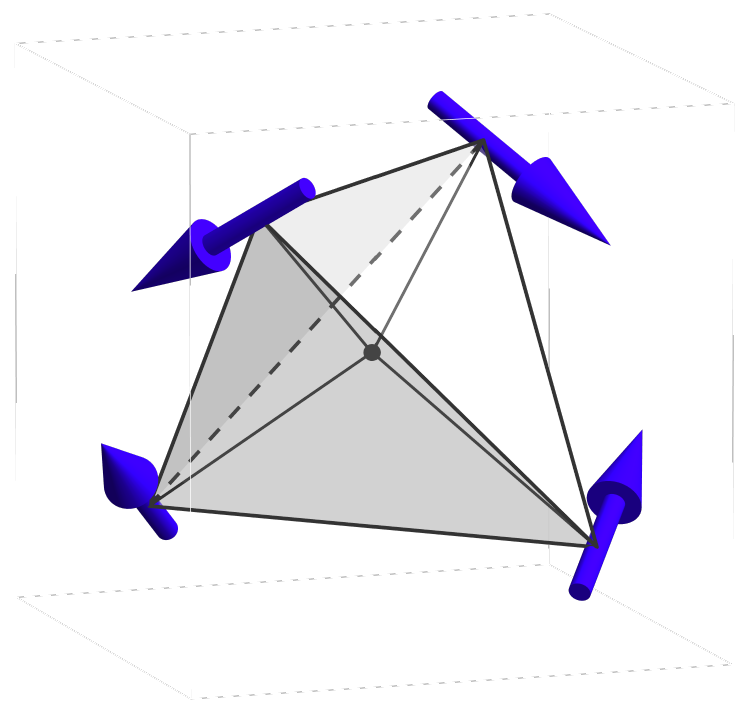} \\
\hline
$\begin{bmatrix} N_{1} \\ N_{1} \\ N_{3} \end{bmatrix}$ & \parbox[c]{0.55\textwidth}{\{$[E \|  E]$, $[E \|  I]$, $[2_{y} \|  2_{x}]$, $[2_{x} \|  2_{y}]$, $[2_{z} \|  2_{z}]$, $[m_{011} \|  2_{101}]$, $[m_{01\bar{1}} \|  2_{10\bar{1}}]$, $[2_{y} \|  m_{x}]$, $[2_{x} \|  m_{y}]$, $[2_{z} \|  m_{z}]$, $[m_{011} \|  m_{101}]$, $[m_{01\bar{1}} \|  m_{10\bar{1}}]$, $[\bar{4}^{+}_{x} \|  4^{-}_{y}]$, $[\bar{4}^{-}_{x} \|  4^{+}_{y}]$, $[\bar{4}^{+}_{x} \|  \bar{4}^{-}_{y}]$, $[\bar{4}^{-}_{x} \|  \bar{4}^{+}_{y}]$\}}&${^{\overline{4}_{x}}4/^{2_{x}}m^{2_{y}}m^{m_{xz}}m}$& \includegraphics[width=2cm,valign=m]{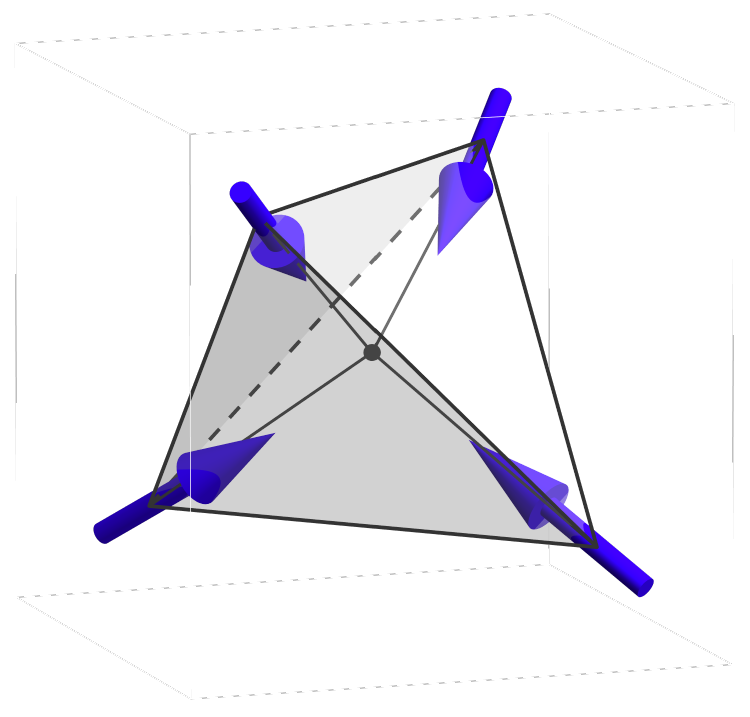} \\
\hline
$\begin{bmatrix} N_{1} \\ N_{2} \\ -N_{1} \end{bmatrix}$ & \parbox[c]{0.55\textwidth}{\{$[E \|  E]$, $[E \|  I]$, $[2_{y} \|  2_{x}]$, $[2_{x} \|  2_{y}]$, $[2_{z} \|  2_{z}]$, $[m_{10\bar{1}} \|  2_{011}]$, $[m_{101} \|  2_{01\bar{1}}]$, $[2_{y} \|  m_{x}]$, $[2_{x} \|  m_{y}]$, $[2_{z} \|  m_{z}]$, $[m_{10\bar{1}} \|  m_{011}]$, $[m_{101} \|  m_{01\bar{1}}]$, $[\bar{4}^{-}_{y} \|  4^{-}_{x}]$, $[\bar{4}^{+}_{y} \|  4^{+}_{x}]$, $[\bar{4}^{-}_{y} \|  \bar{4}^{-}_{x}]$, $[\bar{4}^{+}_{y} \|  \bar{4}^{+}_{x}]$\}}&${ ^{\overline{4}_{y}}4/^{2_{y}}m^{2_{x}}m^{m_{yz}}m }$ &\includegraphics[width=2cm,valign=m]{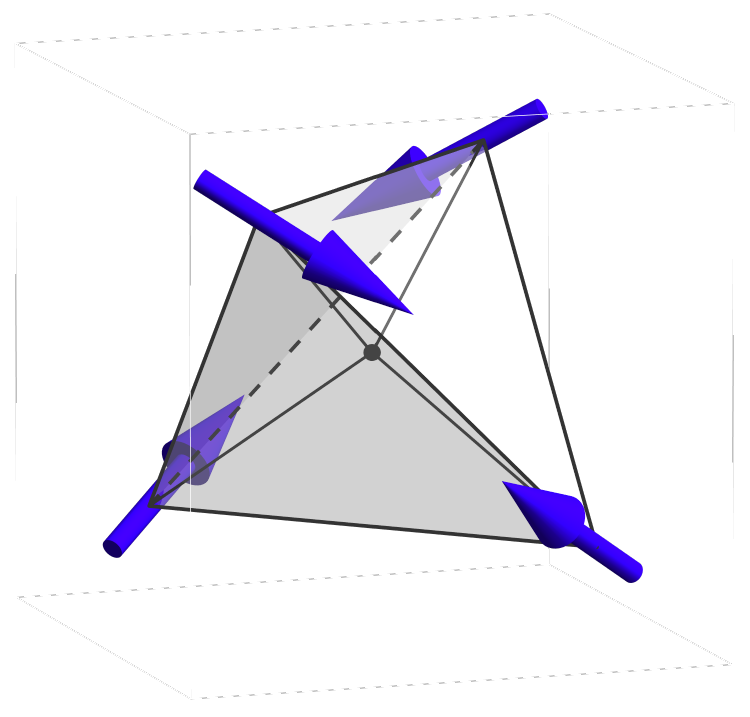} \\
\hline
$\begin{bmatrix} N_{1} \\ N_{2} \\ -N_{2} \end{bmatrix}$ & \parbox[c]{0.55\textwidth}{\{$[E \|  E]$, $[E \|  I]$, $[2_{y} \|  2_{x}]$, $[2_{x} \|  2_{y}]$, $[2_{z} \|  2_{z}]$, $[m_{1\bar{1}0} \|  2_{110}]$, $[m_{110} \|  2_{1\bar{1}0}]$, $[2_{y} \|  m_{x}]$, $[2_{x} \|  m_{y}]$, $[2_{z} \|  m_{z}]$, $[m_{1\bar{1}0} \|  m_{110}]$, $[m_{110} \|  m_{1\bar{1}0}]$, $[\bar{4}^{-}_{z} \|  4^{-}_{z}]$, $[\bar{4}^{+}_{z} \|  4^{+}_{z}]$, $[\bar{4}^{-}_{z} \|  \bar{4}^{-}_{z}]$, $[\bar{4}^{+}_{z} \|  \bar{4}^{+}_{z}]$\}}&${^{\overline{4}_{z}}4/^{2_{z}}m^{2_{x}}m^{m_{xy}}m}$&\includegraphics[width=2cm,valign=m]{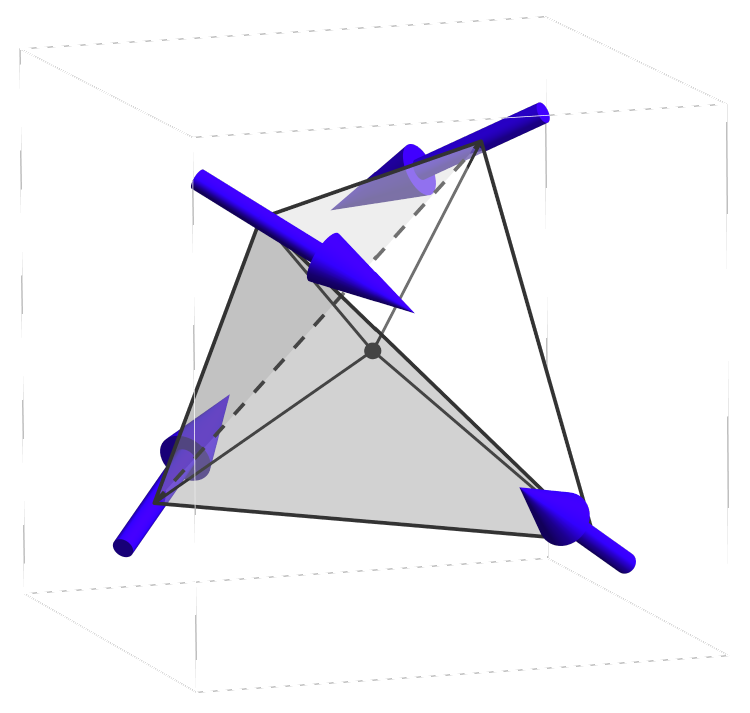} \\
\hline
$\begin{bmatrix} N_{1} \\ N_{2} \\ N_{1} \end{bmatrix}$ & \parbox[c]{0.55\textwidth}{\{$[E \|  E]$, $[E \|  I]$, $[2_{y} \|  2_{x}]$, $[2_{x} \|  2_{y}]$, $[2_{z} \|  2_{z}]$, $[m_{101} \|  2_{011}]$, $[m_{10\bar{1}} \|  2_{01\bar{1}}]$, $[2_{y} \|  m_{x}]$, $[2_{x} \|  m_{y}]$, $[2_{z} \|  m_{z}]$, $[m_{101} \|  m_{011}]$, $[m_{10\bar{1}} \|  m_{01\bar{1}}]$, $[\bar{4}^{+}_{y} \|  4^{-}_{x}]$, $[\bar{4}^{-}_{y} \|  4^{+}_{x}]$, $[\bar{4}^{+}_{y} \|  \bar{4}^{-}_{x}]$, $[\bar{4}^{-}_{y} \|  \bar{4}^{+}_{x}]$\}} &${ ^{\overline{4}_{y}}4/^{2_{y}}m^{2_{x}}m^{m_{yz}}m }$ & \includegraphics[width=2cm,valign=m]{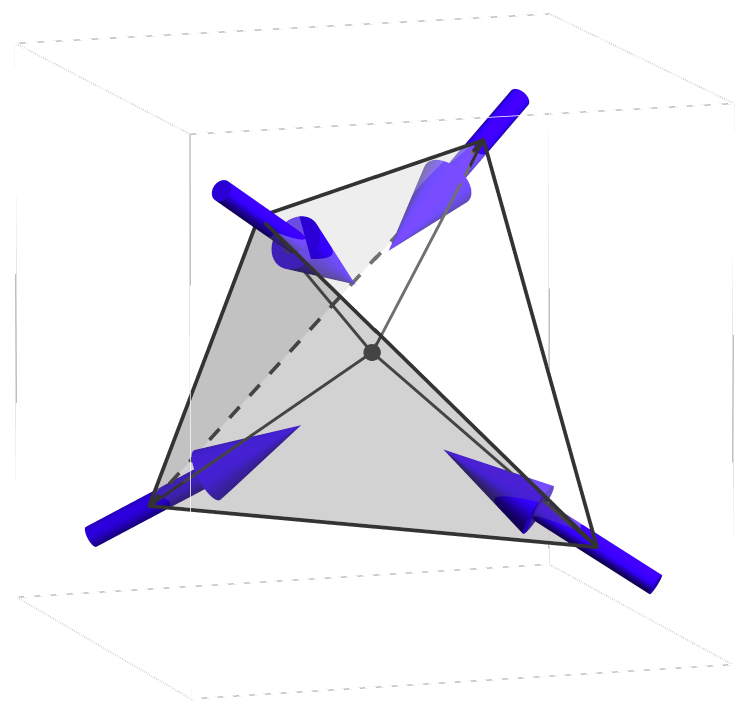}\\
\hline
$\begin{bmatrix} N_{1} \\ N_{2} \\ N_{2} \end{bmatrix}$ & \parbox[c]{0.55\textwidth}{\{$[E \|  E]$, $[E \|  I]$, $[2_{y} \|  2_{x}]$, $[2_{x} \|  2_{y}]$, $[2_{z} \|  2_{z}]$, $[m_{110} \|  2_{110}]$, $[m_{1\bar{1}0} \|  2_{1\bar{1}0}]$, $[2_{y} \|  m_{x}]$, $[2_{x} \|  m_{y}]$, $[2_{z} \|  m_{z}]$, $[m_{110} \|  m_{110}]$, $[m_{1\bar{1}0} \|  m_{1\bar{1}0}]$, $[\bar{4}^{+}_{z} \|  4^{-}_{z}]$, $[\bar{4}^{-}_{z} \|  4^{+}_{z}]$, $[\bar{4}^{+}_{z} \|  \bar{4}^{-}_{z}]$, $[\bar{4}^{-}_{z} \|  \bar{4}^{+}_{z}]$\}} & ${^{\overline{4}_{z}}4/^{2_{z}}m^{2_{x}}m^{m_{xy}}m}$ & \includegraphics[width=2cm,valign=m]{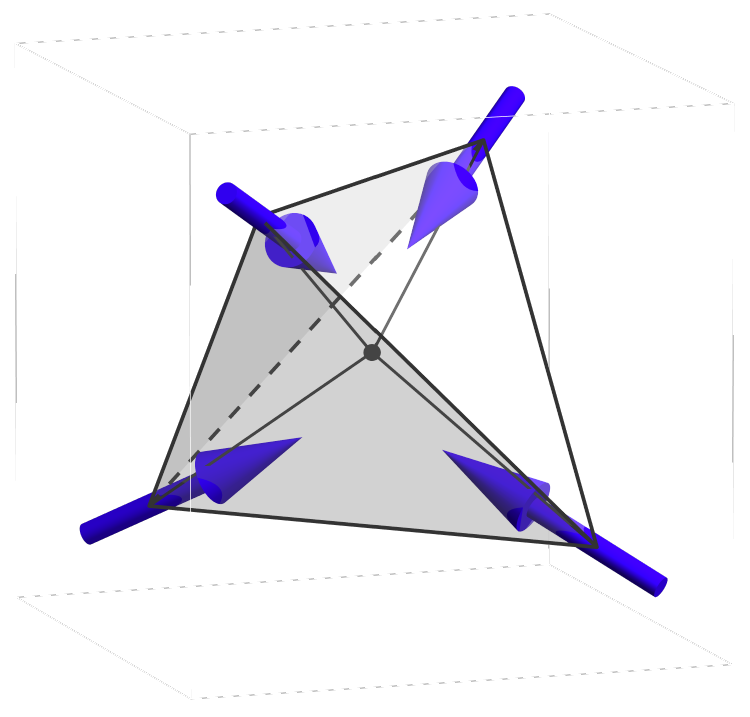}\\
\hline\hline\hline
$\begin{bmatrix} 0 \\ N_{2} \\ -N_{2} \end{bmatrix}$ & \parbox[c]{0.55\textwidth}{\{$[m_{z} \|  E]$, $[E \|  E]$, $[m_{z} \|  I]$, $[E \|  I]$, $[2_{y} \|  2_{x}]$, $[m_{x} \|  2_{x}]$, $[2_{x} \|  2_{y}]$, $[m_{y} \|  2_{y}]$, $[T \|  2_{z}]$, $[2_{z} \|  2_{z}]$, $[2_{110} \|  2_{110}]$, $[m_{1\bar{1}0} \|  2_{110}]$, $[2_{1\bar{1}0} \|  2_{1\bar{1}0}]$, $[m_{110} \|  2_{1\bar{1}0}]$, $[2_{y} \|  m_{x}]$, $[m_{x} \|  m_{x}]$, $[2_{x} \|  m_{y}]$, $[m_{y} \|  m_{y}]$, $[T \|  m_{z}]$, $[2_{z} \|  m_{z}]$, $[2_{110} \|  m_{110}]$, $[m_{1\bar{1}0} \|  m_{110}]$, $[2_{1\bar{1}0} \|  m_{1\bar{1}0}]$, $[m_{110} \|  m_{1\bar{1}0}]$, $[\bar{4}^{-}_{z} \|  4^{-}_{z}]$, $[4^{+}_{z} \|  4^{-}_{z}]$, $[\bar{4}^{+}_{z} \|  4^{+}_{z}]$, $[4^{-}_{z} \|  4^{+}_{z}]$, $[\bar{4}^{-}_{z} \|  \bar{4}^{-}_{z}]$, $[4^{+}_{z} \|  \bar{4}^{-}_{z}]$, $[\bar{4}^{+}_{z} \|  \bar{4}^{+}_{z}]$, $[4^{-}_{z} \|  \bar{4}^{+}_{z}]$\}}&${b_{\mathbb{Z}_{2}}^{z}}\times{^{\overline{4}_{z}}4/^{2_{z}}m^{2_{x}}m^{m_{xy}}m}$ & \includegraphics[width=2cm,valign=m]{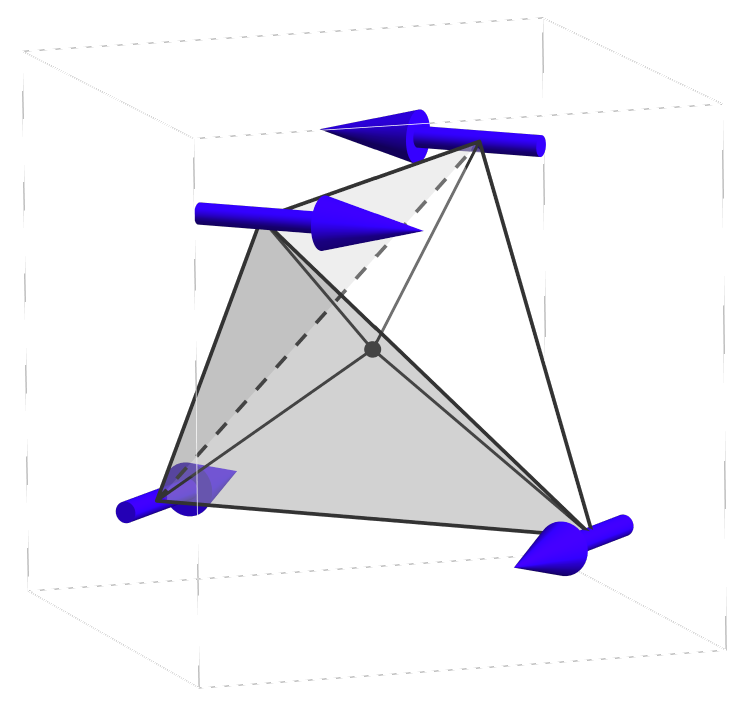}\\
\hline
$\begin{bmatrix} 0 \\ N_{2} \\ N_{2} \end{bmatrix}$ & \parbox[c]{0.55\textwidth}{\{$[m_{z} \|  E]$, $[E \|  E]$, $[m_{z} \|  I]$, $[E \|  I]$, $[2_{y} \|  2_{x}]$, $[m_{x} \|  2_{x}]$, $[2_{x} \|  2_{y}]$, $[m_{y} \|  2_{y}]$, $[T \|  2_{z}]$, $[2_{z} \|  2_{z}]$, $[2_{1\bar{1}0} \|  2_{110}]$, $[m_{110} \|  2_{110}]$, $[2_{110} \|  2_{1\bar{1}0}]$, $[m_{1\bar{1}0} \|  2_{1\bar{1}0}]$, $[2_{y} \|  m_{x}]$, $[m_{x} \|  m_{x}]$, $[2_{x} \|  m_{y}]$, $[m_{y} \|  m_{y}]$, $[T \|  m_{z}]$, $[2_{z} \|  m_{z}]$, $[2_{1\bar{1}0} \|  m_{110}]$, $[m_{110} \|  m_{110}]$, $[2_{110} \|  m_{1\bar{1}0}]$, $[m_{1\bar{1}0} \|  m_{1\bar{1}0}]$, $[\bar{4}^{+}_{z} \|  4^{-}_{z}]$, $[4^{-}_{z} \|  4^{-}_{z}]$, $[\bar{4}^{-}_{z} \|  4^{+}_{z}]$, $[4^{+}_{z} \|  4^{+}_{z}]$, $[\bar{4}^{+}_{z} \|  \bar{4}^{-}_{z}]$, $[4^{-}_{z} \|  \bar{4}^{-}_{z}]$, $[\bar{4}^{-}_{z} \|  \bar{4}^{+}_{z}]$, $[4^{+}_{z} \|  \bar{4}^{+}_{z}]$\}} &${b_{\mathbb{Z}_{2}}^{z}}\times{^{\overline{4}_{z}}4/^{2_{z}}m^{2_{x}}m^{m_{xy}}m}$ & \includegraphics[width=2cm,valign=m]{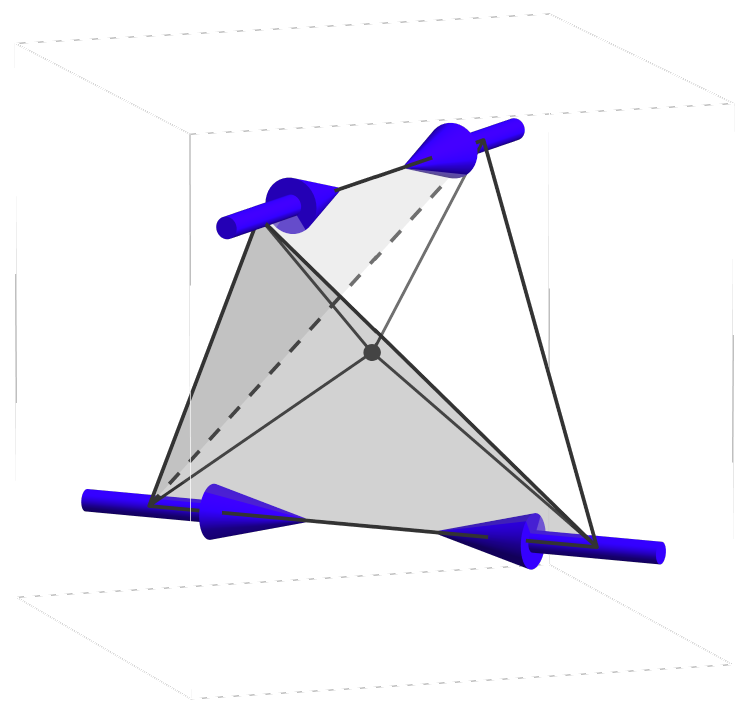}\\
\hline
$\begin{bmatrix} N_{1} \\ 0 \\ -N_{1} \end{bmatrix}$ & \parbox[c]{0.55\textwidth}{\{$[m_{y} \|  E]$, $[E \|  E]$, $[m_{y} \|  I]$, $[E \|  I]$, $[T \|  2_{x}]$, $[2_{y} \|  2_{x}]$, $[2_{x} \|  2_{y}]$, $[m_{z} \|  2_{y}]$, $[2_{z} \|  2_{z}]$, $[m_{x} \|  2_{z}]$, $[2_{101} \|  2_{011}]$, $[m_{10\bar{1}} \|  2_{011}]$, $[2_{10\bar{1}} \|  2_{01\bar{1}}]$, $[m_{101} \|  2_{01\bar{1}}]$, $[T \|  m_{x}]$, $[2_{y} \|  m_{x}]$, $[2_{x} \|  m_{y}]$, $[m_{z} \|  m_{y}]$, $[2_{z} \|  m_{z}]$, $[m_{x} \|  m_{z}]$, $[2_{101} \|  m_{011}]$, $[m_{10\bar{1}} \|  m_{011}]$, $[2_{10\bar{1}} \|  m_{01\bar{1}}]$, $[m_{101} \|  m_{01\bar{1}}]$, $[\bar{4}^{-}_{y} \|  4^{-}_{x}]$, $[4^{+}_{y} \|  4^{-}_{x}]$, $[\bar{4}^{+}_{y} \|  4^{+}_{x}]$, $[4^{-}_{y} \|  4^{+}_{x}]$, $[\bar{4}^{-}_{y} \|  \bar{4}^{-}_{x}]$, $[4^{+}_{y} \|  \bar{4}^{-}_{x}]$, $[\bar{4}^{+}_{y} \|  \bar{4}^{+}_{x}]$, $[4^{-}_{y} \|  \bar{4}^{+}_{x}]$\}}& ${b_{\mathbb{Z}_{2}}^{y}}\times{ ^{\overline{4}_{y}}4/^{2_{y}}m^{2_{x}}m^{m_{yz}}m }$ & \includegraphics[width=2cm,valign=m]{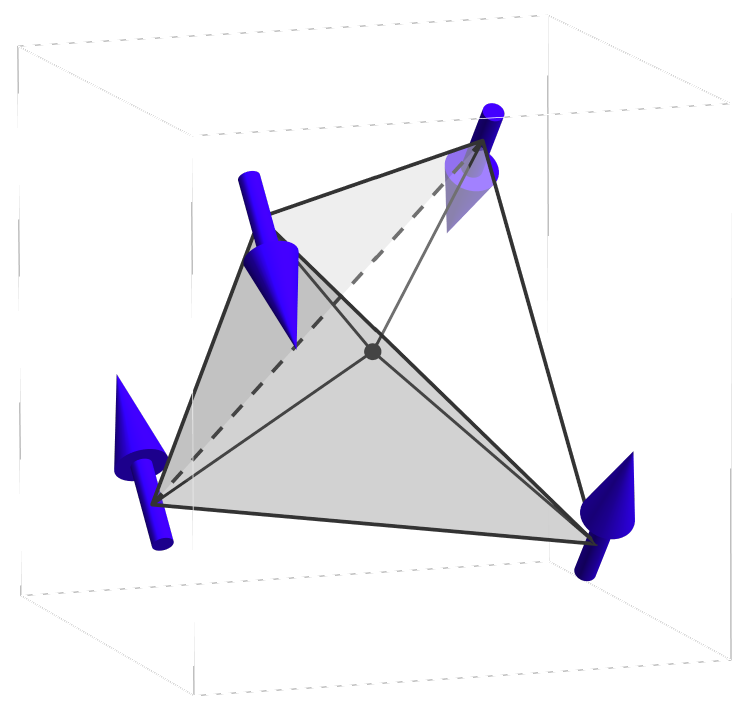}\\
\hline
$\begin{bmatrix} N_{1} \\ 0 \\ N_{1} \end{bmatrix}$ & \parbox[c]{0.55\textwidth}{\{$[m_{y} \|  E]$, $[E \|  E]$, $[m_{y} \|  I]$, $[E \|  I]$, $[T \|  2_{x}]$, $[2_{y} \|  2_{x}]$, $[2_{x} \|  2_{y}]$, $[m_{z} \|  2_{y}]$, $[2_{z} \|  2_{z}]$, $[m_{x} \|  2_{z}]$, $[2_{10\bar{1}} \|  2_{011}]$, $[m_{101} \|  2_{011}]$, $[2_{101} \|  2_{01\bar{1}}]$, $[m_{10\bar{1}} \|  2_{01\bar{1}}]$, $[T \|  m_{x}]$, $[2_{y} \|  m_{x}]$, $[2_{x} \|  m_{y}]$, $[m_{z} \|  m_{y}]$, $[2_{z} \|  m_{z}]$, $[m_{x} \|  m_{z}]$, $[2_{10\bar{1}} \|  m_{011}]$, $[m_{101} \|  m_{011}]$, $[2_{101} \|  m_{01\bar{1}}]$, $[m_{10\bar{1}} \|  m_{01\bar{1}}]$, $[\bar{4}^{+}_{y} \|  4^{-}_{x}]$, $[4^{-}_{y} \|  4^{-}_{x}]$, $[\bar{4}^{-}_{y} \|  4^{+}_{x}]$, $[4^{+}_{y} \|  4^{+}_{x}]$, $[\bar{4}^{+}_{y} \|  \bar{4}^{-}_{x}]$, $[4^{-}_{y} \|  \bar{4}^{-}_{x}]$, $[\bar{4}^{-}_{y} \|  \bar{4}^{+}_{x}]$, $[4^{+}_{y} \|  \bar{4}^{+}_{x}]$\}} & ${b_{\mathbb{Z}_{2}}^{y}}\times{ ^{\overline{4}_{y}}4/^{2_{y}}m^{2_{x}}m^{m_{yz}}m }$  & \includegraphics[width=2cm,valign=m]{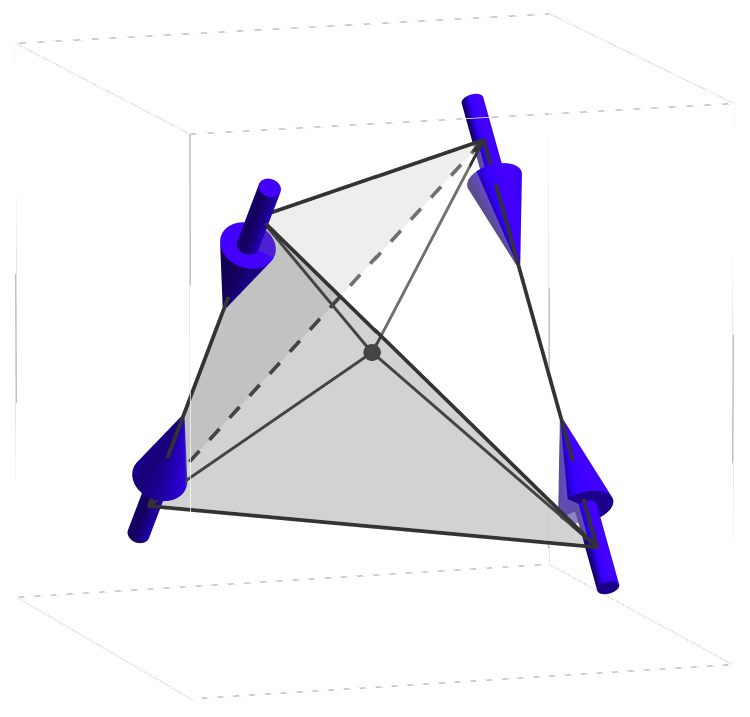}\\
\hline
$\begin{bmatrix} N_{1} \\ -N_{1} \\ 0 \end{bmatrix}$ & \parbox[c]{0.55\textwidth}{\{$[m_{x} \|  E]$, $[E \|  E]$, $[m_{x} \|  I]$, $[E \|  I]$, $[2_{y} \|  2_{x}]$, $[m_{z} \|  2_{x}]$, $[T \|  2_{y}]$, $[2_{x} \|  2_{y}]$, $[2_{z} \|  2_{z}]$, $[m_{y} \|  2_{z}]$, $[2_{011} \|  2_{101}]$, $[m_{01\bar{1}} \|  2_{101}]$, $[2_{01\bar{1}} \|  2_{10\bar{1}}]$, $[m_{011} \|  2_{10\bar{1}}]$, $[2_{y} \|  m_{x}]$, $[m_{z} \|  m_{x}]$, $[T \|  m_{y}]$, $[2_{x} \|  m_{y}]$, $[2_{z} \|  m_{z}]$, $[m_{y} \|  m_{z}]$, $[2_{011} \|  m_{101}]$, $[m_{01\bar{1}} \|  m_{101}]$, $[2_{01\bar{1}} \|  m_{10\bar{1}}]$, $[m_{011} \|  m_{10\bar{1}}]$, $[\bar{4}^{-}_{x} \|  4^{-}_{y}]$, $[4^{+}_{x} \|  4^{-}_{y}]$, $[\bar{4}^{+}_{x} \|  4^{+}_{y}]$, $[4^{-}_{x} \|  4^{+}_{y}]$, $[\bar{4}^{-}_{x} \|  \bar{4}^{-}_{y}]$, $[4^{+}_{x} \|  \bar{4}^{-}_{y}]$, $[\bar{4}^{+}_{x} \|  \bar{4}^{+}_{y}]$, $[4^{-}_{x} \|  \bar{4}^{+}_{y}]$\}}& ${b_{\mathbb{Z}_{2}}^{x}}\times{^{\overline{4}_{x}}4/^{2_{x}}m^{2_{y}}m^{m_{xz}}m}$  & \includegraphics[width=2cm,valign=m]{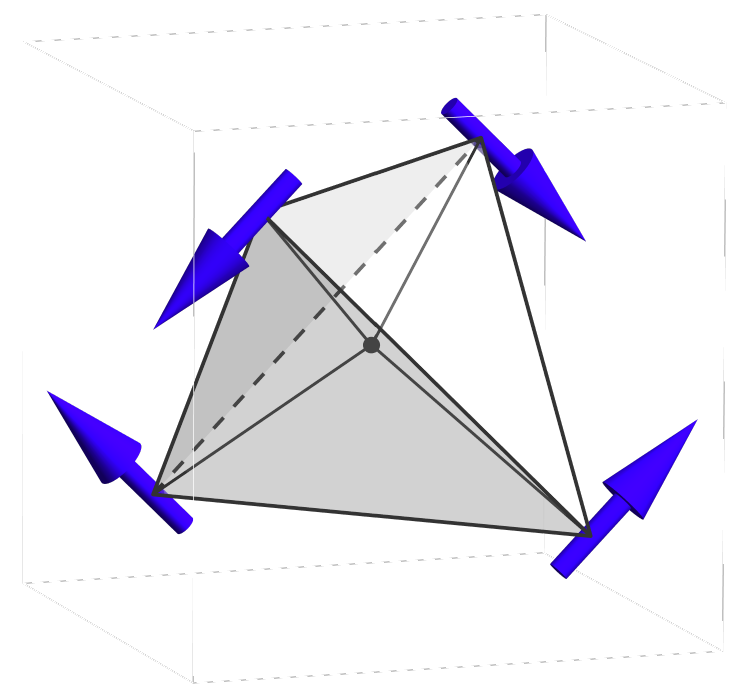}\\
\hline
$\begin{bmatrix} N_{1} \\ N_{1} \\ 0 \end{bmatrix}$ & \parbox[c]{0.55\textwidth}{\{$[m_{x} \|  E]$, $[E \|  E]$, $[m_{x} \|  I]$, $[E \|  I]$, $[2_{y} \|  2_{x}]$, $[m_{z} \|  2_{x}]$, $[T \|  2_{y}]$, $[2_{x} \|  2_{y}]$, $[2_{z} \|  2_{z}]$, $[m_{y} \|  2_{z}]$, $[2_{01\bar{1}} \|  2_{101}]$, $[m_{011} \|  2_{101}]$, $[2_{011} \|  2_{10\bar{1}}]$, $[m_{01\bar{1}} \|  2_{10\bar{1}}]$, $[2_{y} \|  m_{x}]$, $[m_{z} \|  m_{x}]$, $[T \|  m_{y}]$, $[2_{x} \|  m_{y}]$, $[2_{z} \|  m_{z}]$, $[m_{y} \|  m_{z}]$, $[2_{01\bar{1}} \|  m_{101}]$, $[m_{011} \|  m_{101}]$, $[2_{011} \|  m_{10\bar{1}}]$, $[m_{01\bar{1}} \|  m_{10\bar{1}}]$, $[\bar{4}^{+}_{x} \|  4^{-}_{y}]$, $[4^{-}_{x} \|  4^{-}_{y}]$, $[\bar{4}^{-}_{x} \|  4^{+}_{y}]$, $[4^{+}_{x} \|  4^{+}_{y}]$, $[\bar{4}^{+}_{x} \|  \bar{4}^{-}_{y}]$, $[4^{-}_{x} \|  \bar{4}^{-}_{y}]$, $[\bar{4}^{-}_{x} \|  \bar{4}^{+}_{y}]$, $[4^{+}_{x} \|  \bar{4}^{+}_{y}]$\}} & ${b_{\mathbb{Z}_{2}}^{x}}\times{^{\overline{4}_{x}}4/^{2_{x}}m^{2_{y}}m^{m_{xz}}m}$& \includegraphics[width=2cm,valign=m]{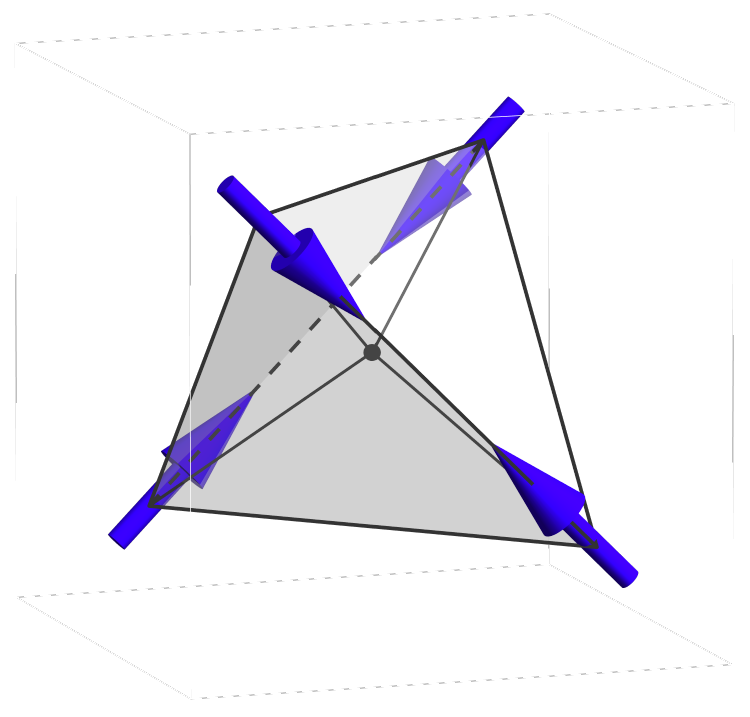}\\
\hline\hline\hline
$\begin{bmatrix} N_{1} \\ -N_{1} \\ -N_{1} \end{bmatrix}$ & \parbox[c]{0.55\textwidth}{\{$[E \|  E]$, $[E \|  I]$, $[2_{y} \|  2_{x}]$, $[2_{x} \|  2_{y}]$, $[2_{z} \|  2_{z}]$, $[m_{10\bar{1}} \|  2_{011}]$, $[m_{101} \|  2_{01\bar{1}}]$, $[m_{01\bar{1}} \|  2_{101}]$, $[m_{011} \|  2_{10\bar{1}}]$, $[m_{110} \|  2_{110}]$, $[m_{1\bar{1}0} \|  2_{1\bar{1}0}]$, $[2_{y} \|  m_{x}]$, $[2_{x} \|  m_{y}]$, $[2_{z} \|  m_{z}]$, $[m_{10\bar{1}} \|  m_{011}]$, $[m_{101} \|  m_{01\bar{1}}]$, $[m_{01\bar{1}} \|  m_{101}]$, $[m_{011} \|  m_{10\bar{1}}]$, $[m_{110} \|  m_{110}]$, $[m_{1\bar{1}0} \|  m_{1\bar{1}0}]$, $[3^{-}_{11\bar{1}} \|  3^{-}_{111}]$, $[3^{+}_{11\bar{1}} \|  3^{+}_{111}]$, $[3^{-}_{111} \|  3^{-}_{11\bar{1}}]$, $[3^{+}_{111} \|  3^{+}_{11\bar{1}}]$, $[3^{+}_{1\bar{1}1} \|  3^{-}_{1\bar{1}1}]$, $[3^{-}_{1\bar{1}1} \|  3^{+}_{1\bar{1}1}]$, $[3^{+}_{1\bar{1}\bar{1}} \|  3^{-}_{1\bar{1}\bar{1}}]$, $[3^{-}_{1\bar{1}\bar{1}} \|  3^{+}_{1\bar{1}\bar{1}}]$, $[\bar{4}^{-}_{y} \|  4^{-}_{x}]$, $[\bar{4}^{+}_{y} \|  4^{+}_{x}]$, $[\bar{4}^{-}_{x} \|  4^{-}_{y}]$, $[\bar{4}^{+}_{x} \|  4^{+}_{y}]$, $[\bar{4}^{+}_{z} \|  4^{-}_{z}]$, $[\bar{4}^{-}_{z} \|  4^{+}_{z}]$, $[3^{-}_{11\bar{1}} \|  \bar{3}^{-}_{111}]$, $[3^{+}_{11\bar{1}} \|  \bar{3}^{+}_{111}]$, $[3^{-}_{111} \|  \bar{3}^{-}_{11\bar{1}}]$, $[3^{+}_{111} \|  \bar{3}^{+}_{11\bar{1}}]$, $[3^{+}_{1\bar{1}1} \|  \bar{3}^{-}_{1\bar{1}1}]$, $[3^{-}_{1\bar{1}1} \|  \bar{3}^{+}_{1\bar{1}1}]$, $[3^{+}_{1\bar{1}\bar{1}} \|  \bar{3}^{-}_{1\bar{1}\bar{1}}]$, $[3^{-}_{1\bar{1}\bar{1}} \|  \bar{3}^{+}_{1\bar{1}\bar{1}}]$, $[\bar{4}^{-}_{y} \|  \bar{4}^{-}_{x}]$, $[\bar{4}^{+}_{y} \|  \bar{4}^{+}_{x}]$, $[\bar{4}^{-}_{x} \|  \bar{4}^{-}_{y}]$, $[\bar{4}^{+}_{x} \|  \bar{4}^{+}_{y}]$, $[\bar{4}^{+}_{z} \|  \bar{4}^{-}_{z}]$, $[\bar{4}^{-}_{z} \|  \bar{4}^{+}_{z}]$\}}& ${^{\overline{4}_{z}}4/^{2_{z}}m^{3_{xy\overline{z}}}3 ^{m_{xy}}2/^{m_{xy}}m}$ & \includegraphics[width=2cm,valign=m]{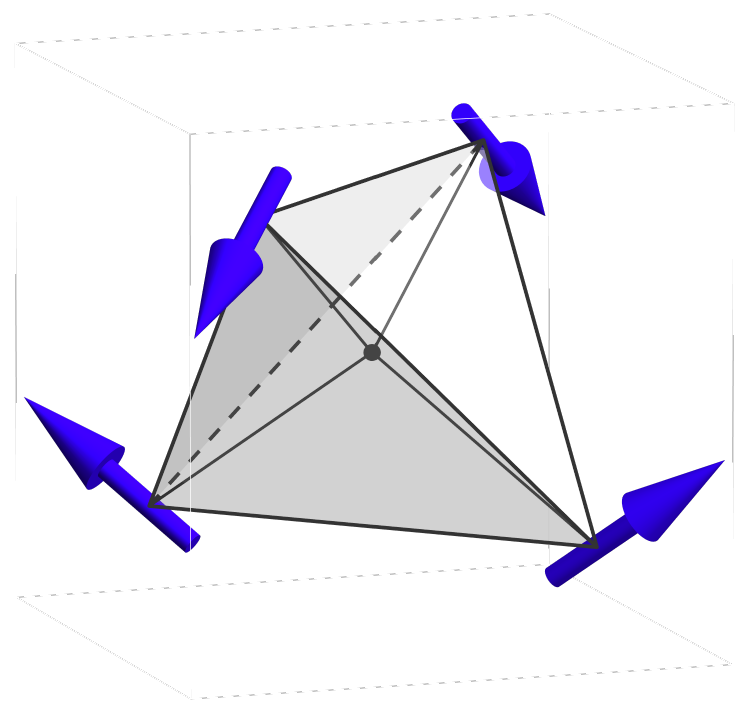} \\
\hline
$\begin{bmatrix} N_{1} \\ -N_{1} \\ N_{1} \end{bmatrix}$ & \parbox[c]{0.55\textwidth}{\{$[E \|  E]$, $[E \|  I]$, $[2_{y} \|  2_{x}]$, $[2_{x} \|  2_{y}]$, $[2_{z} \|  2_{z}]$, $[m_{101} \|  2_{011}]$, $[m_{10\bar{1}} \|  2_{01\bar{1}}]$, $[m_{01\bar{1}} \|  2_{101}]$, $[m_{011} \|  2_{10\bar{1}}]$, $[m_{1\bar{1}0} \|  2_{110}]$, $[m_{110} \|  2_{1\bar{1}0}]$, $[2_{y} \|  m_{x}]$, $[2_{x} \|  m_{y}]$, $[2_{z} \|  m_{z}]$, $[m_{101} \|  m_{011}]$, $[m_{10\bar{1}} \|  m_{01\bar{1}}]$, $[m_{01\bar{1}} \|  m_{101}]$, $[m_{011} \|  m_{10\bar{1}}]$, $[m_{1\bar{1}0} \|  m_{110}]$, $[m_{110} \|  m_{1\bar{1}0}]$, $[3^{-}_{1\bar{1}1} \|  3^{-}_{111}]$, $[3^{+}_{1\bar{1}1} \|  3^{+}_{111}]$, $[3^{-}_{1\bar{1}\bar{1}} \|  3^{-}_{11\bar{1}}]$, $[3^{+}_{1\bar{1}\bar{1}} \|  3^{+}_{11\bar{1}}]$, $[3^{+}_{11\bar{1}} \|  3^{-}_{1\bar{1}1}]$, $[3^{-}_{11\bar{1}} \|  3^{+}_{1\bar{1}1}]$, $[3^{+}_{111} \|  3^{-}_{1\bar{1}\bar{1}}]$, $[3^{-}_{111} \|  3^{+}_{1\bar{1}\bar{1}}]$, $[\bar{4}^{+}_{y} \|  4^{-}_{x}]$, $[\bar{4}^{-}_{y} \|  4^{+}_{x}]$, $[\bar{4}^{-}_{x} \|  4^{-}_{y}]$, $[\bar{4}^{+}_{x} \|  4^{+}_{y}]$, $[\bar{4}^{-}_{z} \|  4^{-}_{z}]$, $[\bar{4}^{+}_{z} \|  4^{+}_{z}]$, $[3^{-}_{1\bar{1}1} \|  \bar{3}^{-}_{111}]$, $[3^{+}_{1\bar{1}1} \|  \bar{3}^{+}_{111}]$, $[3^{-}_{1\bar{1}\bar{1}} \|  \bar{3}^{-}_{11\bar{1}}]$, $[3^{+}_{1\bar{1}\bar{1}} \|  \bar{3}^{+}_{11\bar{1}}]$, $[3^{+}_{11\bar{1}} \|  \bar{3}^{-}_{1\bar{1}1}]$, $[3^{-}_{11\bar{1}} \|  \bar{3}^{+}_{1\bar{1}1}]$, $[3^{+}_{111} \|  \bar{3}^{-}_{1\bar{1}\bar{1}}]$, $[3^{-}_{111} \|  \bar{3}^{+}_{1\bar{1}\bar{1}}]$, $[\bar{4}^{+}_{y} \|  \bar{4}^{-}_{x}]$, $[\bar{4}^{-}_{y} \|  \bar{4}^{+}_{x}]$, $[\bar{4}^{-}_{x} \|  \bar{4}^{-}_{y}]$, $[\bar{4}^{+}_{x} \|  \bar{4}^{+}_{y}]$, $[\bar{4}^{-}_{z} \|  \bar{4}^{-}_{z}]$, $[\bar{4}^{+}_{z} \|  \bar{4}^{+}_{z}]$\}}& ${^{\overline{4}_{z}}4/^{2_{z}}m^{3_{x\overline{y}z}}3 ^{m_{x\overline{y}}}2/^{m_{x\overline{y}}}m}$ & \includegraphics[width=2cm,valign=m]{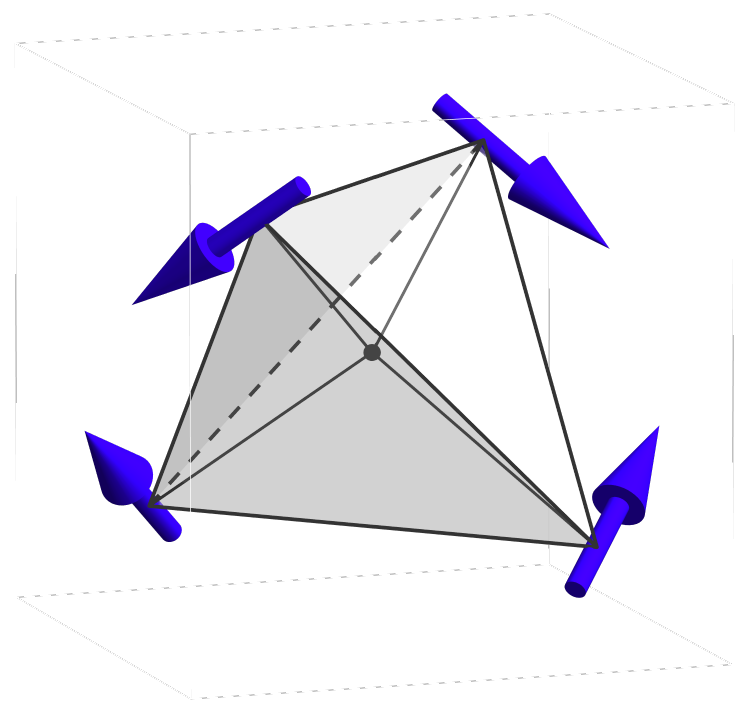} \\
\hline
$\begin{bmatrix} N_{1} \\ N_{1} \\ -N_{1} \end{bmatrix}$ & \parbox[c]{0.55\textwidth}{\{$[E \|  E]$, $[E \|  I]$, $[2_{y} \|  2_{x}]$, $[2_{x} \|  2_{y}]$, $[2_{z} \|  2_{z}]$, $[m_{10\bar{1}} \|  2_{011}]$, $[m_{101} \|  2_{01\bar{1}}]$, $[m_{011} \|  2_{101}]$, $[m_{01\bar{1}} \|  2_{10\bar{1}}]$, $[m_{1\bar{1}0} \|  2_{110}]$, $[m_{110} \|  2_{1\bar{1}0}]$, $[2_{y} \|  m_{x}]$, $[2_{x} \|  m_{y}]$, $[2_{z} \|  m_{z}]$, $[m_{10\bar{1}} \|  m_{011}]$, $[m_{101} \|  m_{01\bar{1}}]$, $[m_{011} \|  m_{101}]$, $[m_{01\bar{1}} \|  m_{10\bar{1}}]$, $[m_{1\bar{1}0} \|  m_{110}]$, $[m_{110} \|  m_{1\bar{1}0}]$, $[3^{+}_{1\bar{1}\bar{1}} \|  3^{-}_{111}]$, $[3^{-}_{1\bar{1}\bar{1}} \|  3^{+}_{111}]$, $[3^{+}_{1\bar{1}1} \|  3^{-}_{11\bar{1}}]$, $[3^{-}_{1\bar{1}1} \|  3^{+}_{11\bar{1}}]$, $[3^{-}_{111} \|  3^{-}_{1\bar{1}1}]$, $[3^{+}_{111} \|  3^{+}_{1\bar{1}1}]$, $[3^{-}_{11\bar{1}} \|  3^{-}_{1\bar{1}\bar{1}}]$, $[3^{+}_{11\bar{1}} \|  3^{+}_{1\bar{1}\bar{1}}]$, $[\bar{4}^{-}_{y} \|  4^{-}_{x}]$, $[\bar{4}^{+}_{y} \|  4^{+}_{x}]$, $[\bar{4}^{+}_{x} \|  4^{-}_{y}]$, $[\bar{4}^{-}_{x} \|  4^{+}_{y}]$, $[\bar{4}^{-}_{z} \|  4^{-}_{z}]$, $[\bar{4}^{+}_{z} \|  4^{+}_{z}]$, $[3^{+}_{1\bar{1}\bar{1}} \|  \bar{3}^{-}_{111}]$, $[3^{-}_{1\bar{1}\bar{1}} \|  \bar{3}^{+}_{111}]$, $[3^{+}_{1\bar{1}1} \|  \bar{3}^{-}_{11\bar{1}}]$, $[3^{-}_{1\bar{1}1} \|  \bar{3}^{+}_{11\bar{1}}]$, $[3^{-}_{111} \|  \bar{3}^{-}_{1\bar{1}1}]$, $[3^{+}_{111} \|  \bar{3}^{+}_{1\bar{1}1}]$, $[3^{-}_{11\bar{1}} \|  \bar{3}^{-}_{1\bar{1}\bar{1}}]$, $[3^{+}_{11\bar{1}} \|  \bar{3}^{+}_{1\bar{1}\bar{1}}]$, $[\bar{4}^{-}_{y} \|  \bar{4}^{-}_{x}]$, $[\bar{4}^{+}_{y} \|  \bar{4}^{+}_{x}]$, $[\bar{4}^{+}_{x} \|  \bar{4}^{-}_{y}]$, $[\bar{4}^{-}_{x} \|  \bar{4}^{+}_{y}]$, $[\bar{4}^{-}_{z} \|  \bar{4}^{-}_{z}]$, $[\bar{4}^{+}_{z} \|  \bar{4}^{+}_{z}]$\}}& ${^{\overline{4}_{z}}4/^{2_{z}}m^{3_{x\overline{yz}}}3 ^{m_{x\overline{y}}}2/^{m_{x\overline{y}}}m}$  & \includegraphics[width=2cm,valign=m]{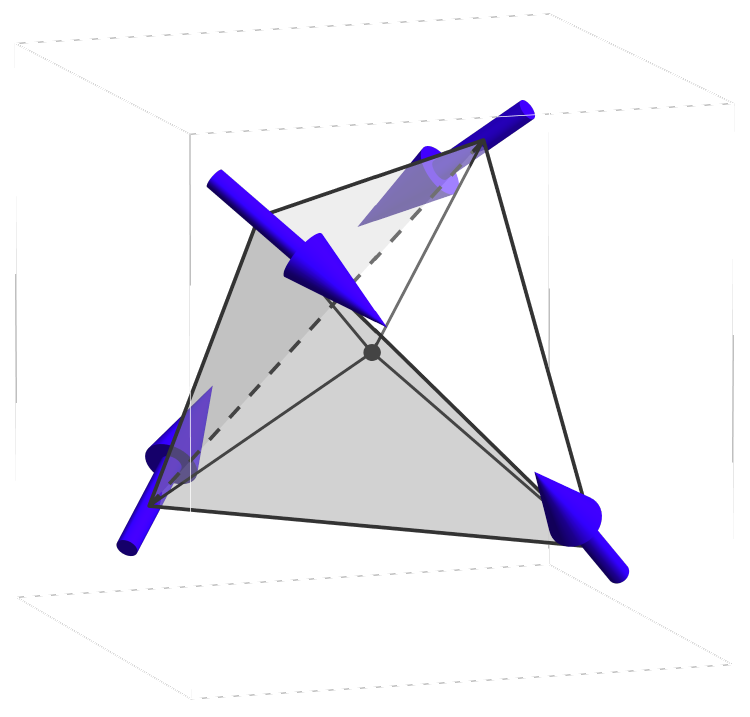} \\
\hline
$\begin{bmatrix} N_{1} \\ N_{1} \\ N_{1} \end{bmatrix}$ & \parbox[c]{0.55\textwidth}{\{$[E \|  E]$, $[E \|  I]$, $[2_{y} \|  2_{x}]$, $[2_{x} \|  2_{y}]$, $[2_{z} \|  2_{z}]$, $[m_{101} \|  2_{011}]$, $[m_{10\bar{1}} \|  2_{01\bar{1}}]$, $[m_{011} \|  2_{101}]$, $[m_{01\bar{1}} \|  2_{10\bar{1}}]$, $[m_{110} \|  2_{110}]$, $[m_{1\bar{1}0} \|  2_{1\bar{1}0}]$, $[2_{y} \|  m_{x}]$, $[2_{x} \|  m_{y}]$, $[2_{z} \|  m_{z}]$, $[m_{101} \|  m_{011}]$, $[m_{10\bar{1}} \|  m_{01\bar{1}}]$, $[m_{011} \|  m_{101}]$, $[m_{01\bar{1}} \|  m_{10\bar{1}}]$, $[m_{110} \|  m_{110}]$, $[m_{1\bar{1}0} \|  m_{1\bar{1}0}]$, $[3^{+}_{111} \|  3^{-}_{111}]$, $[3^{-}_{111} \|  3^{+}_{111}]$, $[3^{+}_{11\bar{1}} \|  3^{-}_{11\bar{1}}]$, $[3^{-}_{11\bar{1}} \|  3^{+}_{11\bar{1}}]$, $[3^{-}_{1\bar{1}\bar{1}} \|  3^{-}_{1\bar{1}1}]$, $[3^{+}_{1\bar{1}\bar{1}} \|  3^{+}_{1\bar{1}1}]$, $[3^{-}_{1\bar{1}1} \|  3^{-}_{1\bar{1}\bar{1}}]$, $[3^{+}_{1\bar{1}1} \|  3^{+}_{1\bar{1}\bar{1}}]$, $[\bar{4}^{+}_{y} \|  4^{-}_{x}]$, $[\bar{4}^{-}_{y} \|  4^{+}_{x}]$, $[\bar{4}^{+}_{x} \|  4^{-}_{y}]$, $[\bar{4}^{-}_{x} \|  4^{+}_{y}]$, $[\bar{4}^{+}_{z} \|  4^{-}_{z}]$, $[\bar{4}^{-}_{z} \|  4^{+}_{z}]$, $[3^{+}_{111} \|  \bar{3}^{-}_{111}]$, $[3^{-}_{111} \|  \bar{3}^{+}_{111}]$, $[3^{+}_{11\bar{1}} \|  \bar{3}^{-}_{11\bar{1}}]$, $[3^{-}_{11\bar{1}} \|  \bar{3}^{+}_{11\bar{1}}]$, $[3^{-}_{1\bar{1}\bar{1}} \|  \bar{3}^{-}_{1\bar{1}1}]$, $[3^{+}_{1\bar{1}\bar{1}} \|  \bar{3}^{+}_{1\bar{1}1}]$, $[3^{-}_{1\bar{1}1} \|  \bar{3}^{-}_{1\bar{1}\bar{1}}]$, $[3^{+}_{1\bar{1}1} \|  \bar{3}^{+}_{1\bar{1}\bar{1}}]$, $[\bar{4}^{+}_{y} \|  \bar{4}^{-}_{x}]$, $[\bar{4}^{-}_{y} \|  \bar{4}^{+}_{x}]$, $[\bar{4}^{+}_{x} \|  \bar{4}^{-}_{y}]$, $[\bar{4}^{-}_{x} \|  \bar{4}^{+}_{y}]$, $[\bar{4}^{+}_{z} \|  \bar{4}^{-}_{z}]$, $[\bar{4}^{-}_{z} \|  \bar{4}^{+}_{z}]$\}}&${^{\overline{4}_{z}}4/^{2_{z}}m^{3_{xyz}}3 ^{m_{xy}}2/^{m_{xy}}m}$ & \includegraphics[width=2cm,valign=m]{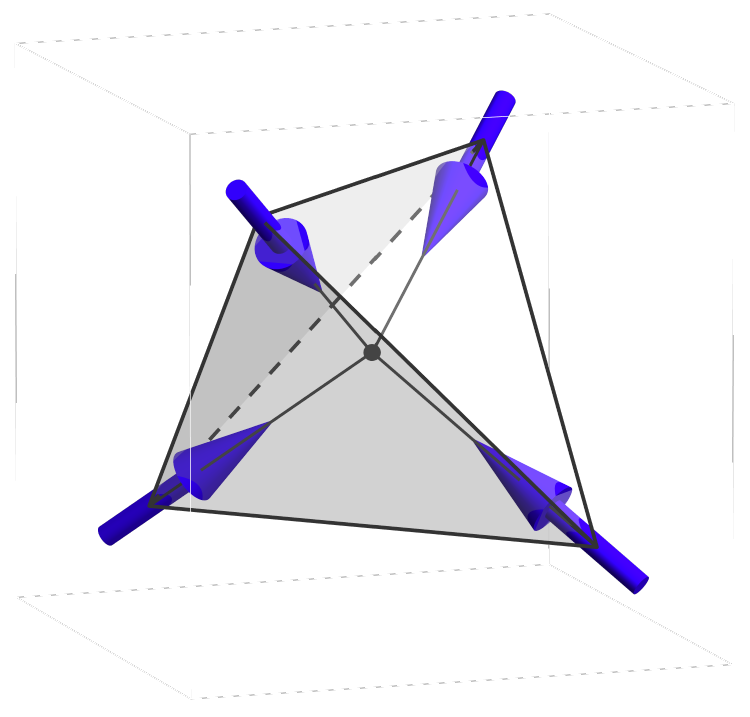}  \\
\hline\hline\hline
$\begin{bmatrix} N_{1} \\ 0 \\ 0 \end{bmatrix}$ & \parbox[c]{0.55\textwidth}{\{$[R_{z}(\varphi) \|  E]$, $[R_{z}(\varphi) \|  I]$, $[T R_{z}(\varphi) \|  2_{x}]$, $[T R_{z}(\varphi) \|  2_{y}]$, $[R_{z}(\varphi) \|  2_{z}]$, $[R_{z}(\varphi) \|  2_{110}]$, $[R_{z}(\varphi) \|  2_{1\bar{1}0}]$, $[T R_{z}(\varphi)  \|  m_{x}]$, $[T R_{z}(\varphi) \|  m_{y}]$, $[R_{z}(\varphi) \|  m_{z}]$, $[R_{z}(\varphi) \|  m_{110}]$, $[R_{z}(\varphi) \|  m_{1\bar{1}0}]$, $[T R_{z}(\varphi) \|  4^{-}_{z}]$, $[T R_{z}(\varphi) \|  4^{+}_{z}]$, $[T R_{z}(\varphi) \|  \bar{4}^{-}_{z}]$, $[T R_{z}(\varphi) \|  \bar{4}^{+}_{z}]$\}} & ${b_{\infty}^{z}}\times{^{\overline{1}}4/^{1}m^{\overline{1}}m^{1}m}$ & \includegraphics[width=2cm,valign=m]{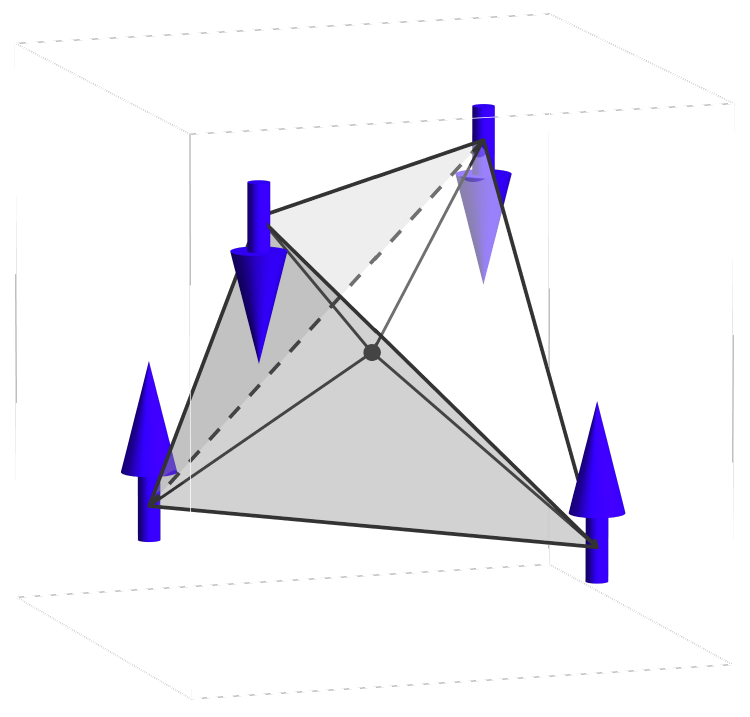} \\
\hline
$\begin{bmatrix} 0 \\ N_{2} \\ 0 \end{bmatrix}$ & \parbox[c]{0.55\textwidth}{\{$[R_{y}(\varphi) \|  E]$, $[R_{y}(\varphi) \|  I]$, $[R_{y}(\varphi) \|  2_{x}]$, $[T R_{y}(\varphi) \|  2_{y}]$, $[T R_{y}(\varphi) \|  2_{z}]$, $[ R_{y}(\varphi) \|  2_{011}]$, $[ R_{y}(\varphi) \|  2_{01\bar{1}}]$, $[ R_{y}(\varphi) \|  m_{x}]$, $[ T R_{y}(\varphi) \|  m_{y}]$, $[ T R_{y}(\varphi) \|  m_{z}]$, $[ R_{y}(\varphi) \|  m_{011}]$, $[ R_{y}(\varphi) \|  m_{01\bar{1}}]$, $[ T R_{y}(\varphi) \|  4^{-}_{x}]$, $[ T R_{y}(\varphi) \|  4^{+}_{x}]$, $[ T R_{y}(\varphi) \|  \bar{4}^{-}_{x}]$, $[ T R_{y}(\varphi) \|  \bar{4}^{+}_{x}]$\}} &${b_{\infty}^{y}}\times{ ^{\overline{1}}4/^{1}m^{\overline{1}}m^{1}m }$ & \includegraphics[width=2cm,valign=m]{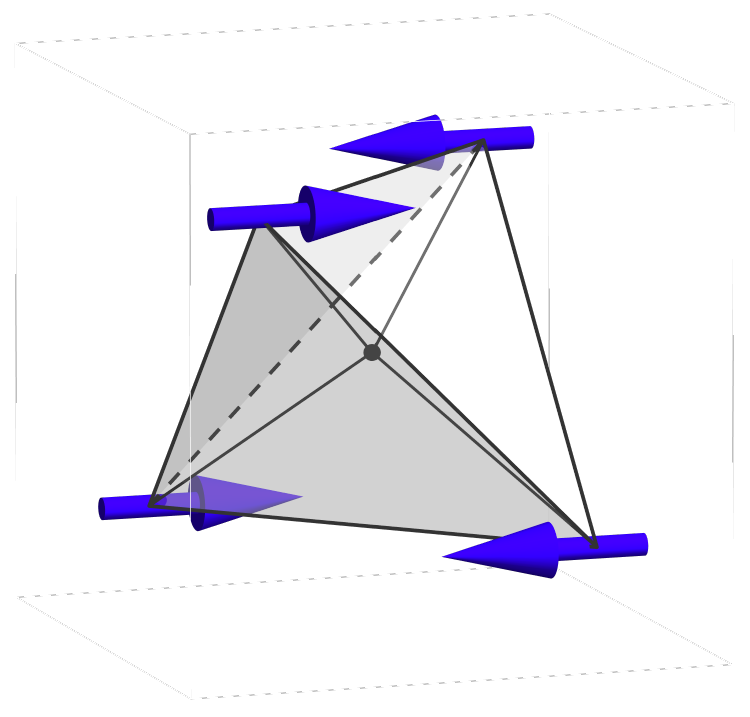}\\
\hline
$\begin{bmatrix} 0 \\ 0 \\ N_{3} \end{bmatrix}$ & \parbox[c]{0.55\textwidth}{\{$[R_{x}(\varphi) \|  E]$, $[R_{x}(\varphi) \|  I]$, $[T R_{x}(\varphi) \|  2_{x}]$, $[R_{x}(\varphi) \|  2_{y}]$, $[T R_{x}(\varphi) \|  2_{z}]$, $[R_{x}(\varphi) \|  2_{101}]$, $[R_{x}(\varphi) \|  2_{10\bar{1}}]$, $[T R_{x}(\varphi) \|  m_{x}]$, $[R_{x}(\varphi) \|  m_{y}]$, $[T R_{x}(\varphi) \|  m_{z}]$, $[R_{x}(\varphi) \|  m_{101}]$, $[R_{x}(\varphi) \|  m_{10\bar{1}}]$, $[T R_{x}(\varphi) \|  4^{-}_{y}]$, $[T R_{x}(\varphi) \|  4^{+}_{y}]$, $[T R_{x}(\varphi) \|  \bar{4}^{-}_{y}]$, $[T R_{x}(\varphi) \|  \bar{4}^{+}_{y}]$\}}&${b_{\infty}^{x}}\times{^{\overline{1}}4/^{1}m^{\overline{1}}m^{1}m}$ & \includegraphics[width=2cm,valign=m]{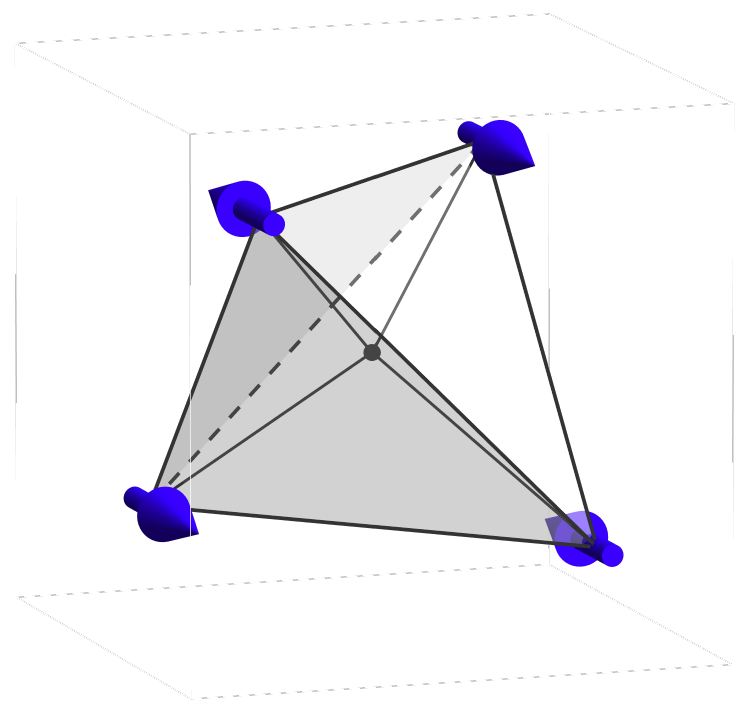}\\
\hline
\end{longtable}

\normalsize
\twocolumngrid


\onecolumngrid 
\section{Decompositions of common representations into point group irreps}\label{app:rep_decomps}

For completeness and to accompany Sec.~\ref{sec:observables}, we include here decompositions of common (point group) representations into their irreps. In Table~\ref{tab:decomps}, parentheses indicate two complex-conjugate irreps that join into a single real irrep.

\begin{table}[t]
\centering
\scriptsize
\setlength{\tabcolsep}{3pt}
\renewcommand{\arraystretch}{1.3}
\resizebox{\textwidth}{!}{%
\begin{tabular}{@{}c l l l l l@{}}
\hline
PG & $V$ & $eV=\{V^2\}$ & $[V^2]$ & $[V^3]$ & $[V^4]$ \\
\hline
${1}$         & $3A$ & $3A$ & $6A$ & $10A$ & $15A$ \\
${\bar{1}}$   & $3A_u$ & $3A_g$ & $6A_g$ & $10A_u$ & $15A_g$ \\
${2}$         & $A+2B$ & $A+2B$ & $4A+2B$ & $4A+6B$ & $9A+6B$ \\
${m}$         & $2A'+A''$ & $A'+2A''$ & $4A'+2A''$ & $6A'+4A''$ & $9A'+6A''$ \\
${2/m}$       & $A_u+2B_u$ & $A_g+2B_g$ & $4A_g+2B_g$ & $4A_u+6B_u$ & $9A_g+6B_g$ \\
${222}$       & $B_1+B_2+B_3$ & $B_1+B_2+B_3$ & $3A+B_1+B_2+B_3$ & $A+3B_1+3B_2+3B_3$ & $6A+3B_1+3B_2+3B_3$ \\
${mm2}$       & $A_1+B_1+B_2$ & $A_2+B_1+B_2$ & $3A_1+A_2+B_1+B_2$ & $3A_1+A_2+3B_1+3B_2$ & $6A_1+3A_2+3B_1+3B_2$ \\
${mmm}$       & $B_{1u}+B_{2u}+B_{3u}$ & $B_{1g}+B_{2g}+B_{3g}$ & $3A_g+B_{1g}+B_{2g}+B_{3g}$ & $A_u+3B_{1u}+3B_{2u}+3B_{3u}$ & $6A_g+3B_{1g}+3B_{2g}+3B_{3g}$ \\
${4}$         & $({}^{1}E+{}^{2}E)+A$ & $({}^{1}E+{}^{2}E)+A$ & $({}^{1}E+{}^{2}E)+2A+2B$ & $3({}^{1}E+{}^{2}E)+2A+2B$ & $3({}^{1}E+{}^{2}E)+5A+4B$ \\
${\bar{4}}$   & $({}^{1}E+{}^{2}E)+B$ & $({}^{1}E+{}^{2}E)+A$ & $({}^{1}E+{}^{2}E)+2A+2B$ & $3({}^{1}E+{}^{2}E)+2A+2B$ & $3({}^{1}E+{}^{2}E)+5A+4B$ \\
${4/m}$       & $({}^{1}E_u+{}^{2}E_u)+A_u$ & $({}^{1}E_g+{}^{2}E_g)+A_g$ & $({}^{1}E_g+{}^{2}E_g)+2A_g+2B_g$ & $3({}^{1}E_u+{}^{2}E_u)+2A_u+2B_u$ & $3({}^{1}E_g+{}^{2}E_g)+5A_g+4B_g$ \\
${422}$       & $A_2+E$ & $A_2+E$ & $2A_1+B_1+B_2+E$ & $2A_2+B_1+B_2+3E$ & $4A_1+A_2+2B_1+2B_2+3E$ \\
${4mm}$       & $A_1+E$ & $A_2+E$ & $2A_1+B_1+B_2+E$ & $2A_1+B_1+B_2+3E$ & $4A_1+A_2+2B_1+2B_2+3E$ \\
${\bar{4}2m}$ & $B_2+E$ & $A_2+E$ & $2A_1+B_1+B_2+E$ & $A_1+A_2+2B_2+3E$ & $4A_1+A_2+2B_1+2B_2+3E$ \\
${\bar{4}m2}$ & $B_1+E$ & $B_2+E$ & $2A_1+A_2+B_1+E$ & $A_1+2B_1+B_2+3E$ & $4A_1+2A_2+2B_1+B_2+3E$ \\
${4/mmm}$     & $A_{2u}+E_u$ & $A_{2g}+E_g$ & $2A_{1g}+B_{1g}+B_{2g}+E_g$ & $2A_{2u}+B_{1u}+B_{2u}+3E_u$ & $4A_{1g}+A_{2g}+2B_{1g}+2B_{2g}+3E_g$ \\
${3}$         & $({}^{1}E+{}^{2}E)+A_1$ & $({}^{1}E+{}^{2}E)+A_1$ & $2({}^{1}E+{}^{2}E)+2A_1$ & $3({}^{1}E+{}^{2}E)+4A_1$ & $5({}^{1}E+{}^{2}E)+5A_1$ \\
${\bar{3}}$   & $({}^{1}E_u+{}^{2}E_u)+A_u$ & $({}^{1}E_g+{}^{2}E_g)+A_g$ & $2({}^{1}E_g+{}^{2}E_g)+2A_g$ & $3({}^{1}E_u+{}^{2}E_u)+4A_u$ & $5({}^{1}E_g+{}^{2}E_g)+5A_g$ \\
${32}$        & $A_2+E$ & $A_2+E$ & $2A_1+2E$ & $A_1+3A_2+3E$ & $4A_1+A_2+5E$ \\
${3m}$        & $A_1+E$ & $A_2+E$ & $2A_1+2E$ & $3A_1+A_2+3E$ & $4A_1+A_2+5E$ \\
${\bar{3}m}$  & $A_{2u}+E_u$ & $A_{2g}+E_g$ & $2A_{1g}+2E_g$ & $A_{1u}+3A_{2u}+3E_u$ & $4A_{1g}+A_{2g}+5E_g$ \\
${6}$         & $({}^{1}E_2+{}^{2}E_2)+A$ & $({}^{1}E_2+{}^{2}E_2)+A$ & $({}^{1}E_1+{}^{2}E_1)$ & $({}^{1}E_1+{}^{2}E_1)+2({}^{1}E_2+{}^{2}E_2)$ & $3({}^{1}E_1+{}^{2}E_1)+2({}^{1}E_2+{}^{2}E_2)$ \\
              &  &  & $\,\,{}+({}^{1}E_2+{}^{2}E_2)+2A$ & $\,\,{}+2A+2B$ & $\,\,{}+3A+2B$ \\
${\bar{6}}$   & $({}^{1}E'+{}^{2}E')+A''$ & $({}^{1}E''+{}^{2}E'')+A'$ & $({}^{1}E'+{}^{2}E')$ & $2({}^{1}E'+{}^{2}E')+({}^{1}E''+{}^{2}E'')$ & $3({}^{1}E'+{}^{2}E')+2({}^{1}E''+{}^{2}E'')$ \\
              &  &  & $\,\,{}+({}^{1}E''+{}^{2}E'')+2A'$ & $\,\,{}+2A'+2A''$ & $\,\,{}+3A'+2A''$ \\
${6/m}$       & $({}^{1}E_{2u}+{}^{2}E_{2u})+A_u$ & $({}^{1}E_{2g}+{}^{2}E_{2g})+A_g$ & $({}^{1}E_{1g}+{}^{2}E_{1g})$ & $({}^{1}E_{1u}+{}^{2}E_{1u})+2({}^{1}E_{2u}+{}^{2}E_{2u})$ & $3({}^{1}E_{1g}+{}^{2}E_{1g})+2({}^{1}E_{2g}+{}^{2}E_{2g})$ \\
              &  &  & $\,\,{}+({}^{1}E_{2g}+{}^{2}E_{2g})+2A_g$ & $\,\,{}+2A_u+2B_u$ & $\,\,{}+3A_g+2B_g$ \\
${622}$       & $A_2+E_1$ & $A_2+E_1$ & $2A_1+E_1+E_2$ & $2A_2+B_1+B_2+2E_1+E_2$ & $3A_1+B_1+B_2+2E_1+3E_2$ \\
${6mm}$       & $A_1+E_1$ & $A_2+E_1$ & $2A_1+E_1+E_2$ & $2A_1+B_1+B_2+2E_1+E_2$ & $3A_1+B_1+B_2+2E_1+3E_2$ \\
${\bar{6}m2}$ & $A_2''+E'$ & $A_1''+E''$ & $2A_1'+E'+E''$ & $A_1'+A_1''+2A_2''+2E'+E''$ & $3A_1'+A_2'+A_2''+3E'+2E''$ \\
${\bar{6}2m}$ & $A_2''+E'$ & $A_2'+E''$ & $2A_1'+E'+E''$ & $A_1'+A_2'+2A_2''+2E'+E''$ & $3A_1'+A_1''+A_2''+3E'+2E''$ \\
${6/mmm}$     & $A_{2u}+E_{1u}$ & $A_{2g}+E_{1g}$ & $2A_{1g}+E_{1g}+E_{2g}$ & $2A_{2u}+B_{1u}+B_{2u}+2E_{1u}+E_{2u}$ & $3A_{1g}+B_{1g}+B_{2g}+2E_{1g}+3E_{2g}$ \\
${23}$        & $T$ & $T$ & $({}^{1}E+{}^{2}E)+A+T$ & $A+3T$ & $2({}^{1}E+{}^{2}E)+2A+3T$ \\
${m\bar{3}}$  & $T_u$ & $T_g$ & $({}^{1}E_g+{}^{2}E_g)+A_g+T_g$ & $A_u+3T_u$ & $2({}^{1}E_g+{}^{2}E_g)+2A_g+3T_g$ \\
${432}$       & $T_1$ & $T_1$ & $A_1+E+T_2$ & $A_2+2T_1+T_2$ & $2A_1+2E+T_1+2T_2$ \\
${\bar{4}3m}$ & $T_2$ & $T_1$ & $A_1+E+T_2$ & $A_1+T_1+2T_2$ & $2A_1+2E+T_1+2T_2$ \\
${m\bar{3}m}$ & $T_{1u}$ & $T_{1g}$ & $A_{1g}+E_g+T_{2g}$ & $A_{2u}+2T_{1u}+T_{2u}$ & $2A_{1g}+2E_g+T_{1g}+2T_{2g}$ \\
\hline
\end{tabular}%
}
\caption{Decompositions of $V$, $eV$, $[V^2]$, $\{V^2\}$, $[V^3]$ and $[V^4]$ into irreducible representations for each point group (PG). $eV$ and $\{V^2\}$ coincide for every point group and share a column.}
\label{tab:decomps}
\end{table}

\normalsize

\newpage
\twocolumngrid

\bibliography{main}

@misc{jaeschke2025atomic,
  title={{Atomic altermagnetism}},
  author={Jaeschke-Ubiergo, Rodrigo and Bharadwaj, Venkata-Krishna and Campos, Warlley and Zarzuela, Ricardo and Biniskos, Nikolaos and Fernandes, Rafael M and Jungwirth, Tomas and Sinova, Jairo and {\v{S}}mejkal, Libor},
eprint={2503.10797},
      archivePrefix={arXiv},      
year          = {2025},
      primaryClass={cond-mat.str-el},
 url={https://arxiv.org/abs/2503.10797} 
}

@misc{d2025altermagnetism,
  title={{Altermagnetism Without Crystal Symmetry}},
  author={d'Ornellas, Peru and Leeb, Valentin and Grushin, Adolfo G and Knolle, Johannes},
eprint={2504.08597},
      archivePrefix={arXiv},      
year          = {2025},
      primaryClass={cond-mat.str-el},
 url={https://arxiv.org/abs/2504.08597} 
}

@article{hu2025spin,
  title={{Spin Hall and Edelstein effects in chiral non-collinear altermagnets}},
  author={Hu, Mengli and Janson, Oleg and Felser, Claudia and McClarty, Paul and van den Brink, Jeroen and G. Vergniory, Maia},
  journal={Nature Communications},
  volume={16},
  number={1},
  pages={8529},
  year={2025},
  doi={10.1038/s41467-025-64271-8},
  publisher={Nature Publishing Group UK London}
}

@article{smejkal2022emerging,
  title = {{Emerging Research Landscape of Altermagnetism}},
  author = {\ifmmode \check{S}\else \v{S}\fi{}mejkal, Libor and Sinova, Jairo and Jungwirth, Tomas},
  journal = {Phys. Rev. X},
  volume = {12},
  issue = {4},
  pages = {040501},
  numpages = {27},
  year = {2022},
  month = {Dec},
  publisher = {American Physical Society},
  doi = {10.1103/PhysRevX.12.040501},
  url = {https://link.aps.org/doi/10.1103/PhysRevX.12.040501}
}

@article{smejkal2022beyond,
  title = {{Beyond Conventional Ferromagnetism and Antiferromagnetism: A Phase with Nonrelativistic Spin and Crystal Rotation Symmetry}},
  author = {\ifmmode \check{S}\else \v{S}\fi{}mejkal, Libor and Sinova, Jairo and Jungwirth, Tomas},
  journal = {Phys. Rev. X},
  volume = {12},
  issue = {3},
  pages = {031042},
  numpages = {16},
  year = {2022},
  month = {Sep},
  publisher = {American Physical Society},
  doi = {10.1103/PhysRevX.12.031042},
  url = {https://link.aps.org/doi/10.1103/PhysRevX.12.031042}
}

@article{mcclarty2024landau,
  title = {{Landau Theory of Altermagnetism}},
  author = {McClarty, Paul A. and Rau, Jeffrey G.},
  journal = {Phys. Rev. Lett.},
  volume = {132},
  issue = {17},
  pages = {176702},
  numpages = {8},
  year = {2024},
  month = {Apr},
  publisher = {American Physical Society},
  doi = {10.1103/PhysRevLett.132.176702},
  url = {https://link.aps.org/doi/10.1103/PhysRevLett.132.176702}
}

@article{schiff2025collinear,
  title = {{Collinear altermagnets and their Landau theories}},
  author = {Schiff, Hana and McClarty, Paul and Rau, Jeffrey G. and Romh\'anyi, Judit},
  journal = {Phys. Rev. Res.},
  volume = {7},
  issue = {3},
  pages = {033301},
  numpages = {31},
  year = {2025},
  month = {Sep},
  publisher = {American Physical Society},
  doi = {10.1103/q44z-ynbr},
  url = {https://link.aps.org/doi/10.1103/q44z-ynbr}
}

@article{song2025altermagnets,
  title={{Altermagnets as a new class of functional materials}},
  author={Song, Cheng and Bai, Hua and Zhou, Zhiyuan and Han, Lei and Reichlova, Helena and Dil, J Hugo and Liu, Junwei and Chen, Xianzhe and Pan, Feng},
  journal={Nature Reviews Materials},
  pages={473-485},
  volume={10},
  doi={10.1038/s41578-025-00779-1},
  year={2025},
  publisher={Nature Publishing Group UK London}
}

@misc{jungwirth2025altermagnetic,
  title={{Altermagnetic spintronics}},
  author={Jungwirth, Tomas and Sinova, J and Wadley, P and Kriegner, D and Reichlova, H and Krizek, F and Ohno, H and Smejkal, L},
eprint={2508.09748},
      archivePrefix={arXiv},      
year          = {2025},
      primaryClass={cond-mat.str-el},
 url={https://arxiv.org/abs/2508.09748} 
}

@article{dal2024antiferromagnetic,
  title={{Antiferromagnetic spintronics and beyond}},
  author={Dal Din, A and Amin, OJ and Wadley, P and Edmonds, KW},
  journal={npj Spintronics},
  volume={2},
  number={1},
  pages={25},
  year={2024},
  publisher={Nature Publishing Group UK London}
}

@article{yu2025odd,
  title = {{Odd-Parity Magnetism Driven by Antiferromagnetic Exchange}},
  author = {Yu, Yue and Lyngby, Magnus B. and Shishidou, Tatsuya and Roig, Merc\`e and Kreisel, Andreas and Weinert, Michael and Andersen, Brian M. and Agterberg, Daniel F.},
  journal = {Phys. Rev. Lett.},
  volume = {135},
  issue = {4},
  pages = {046701},
  numpages = {7},
  year = {2025},
  month = {Jul},
  publisher = {American Physical Society},
  doi = {10.1103/zk69-k6b2},
  url = {https://link.aps.org/doi/10.1103/zk69-k6b2}
}

@misc{luo2025spin,
  title={{Spin Group Symmetry Criteria for Odd-parity Magnets}},
  author={Luo, Xun-Jiang and Hu, Jin-Xin and  Hu, Meng-Li and Law, KT}, 
      eprint={2510.05512},
      archivePrefix={arXiv},      
      primaryClass={cond-mat.str-el},
year          = {2025},
 url={https://arxiv.org/abs/2510.05512}  
}

@misc{hellenes2023p,
  title={{P-wave magnets}},
  author={Hellenes, Anna Birk and Jungwirth, Tom{\'a}{\v{s}} and Jaeschke-Ubiergo, Rodrigo and Chakraborty, Atasi and Sinova, Jairo and {\v{S}}mejkal, Libor},
  eprint={2309.01607},
 archivePrefix={arXiv},      
year          = {2023},
 primaryClass={cond-mat.str-el},
 url={https://arxiv.org/abs/2309.01607}  
}

@article{brekke2024minimal,
  title = {{Minimal Models and Transport Properties of Unconventional $p$-Wave Magnets}},
  author = {Brekke, Bj\o{}rnulf and Sukhachov, Pavlo and Giil, Hans Gl\o{}ckner and Brataas, Arne and Linder, Jacob},
  journal = {Phys. Rev. Lett.},
  volume = {133},
  issue = {23},
  pages = {236703},
  numpages = {9},
  year = {2024},
  month = {Dec},
  publisher = {American Physical Society},
  doi = {10.1103/PhysRevLett.133.236703},
  url = {https://link.aps.org/doi/10.1103/PhysRevLett.133.236703}
}

@article{hannappel2016,
  title = {Stability of a spin-triplet nematic state near to a quantum critical point},
  author = {Hannappel, G. and Pedder, C. J. and Kr\"uger, F. and Green, A. G.},
  journal = {Phys. Rev. B},
  volume = {93},
  issue = {23},
  pages = {235105},
  numpages = {15},
  year = {2016},
  month = {Jun},
  publisher = {American Physical Society},
  doi = {10.1103/PhysRevB.93.235105},
  url = {https://link.aps.org/doi/10.1103/PhysRevB.93.235105}
}

@article{fradkin2010nematic,
  title={{Nematic Fermi fluids in condensed matter physics}},
  author={Fradkin, Eduardo and Kivelson, Steven A and Lawler, Michael J and Eisenstein, James P and Mackenzie, Andrew P},
  journal={Annu. Rev. Condens. Matter Phys.},
  volume={1},
  number={1},
  pages={153--178},
  year={2010},
  doi={10.1146/annurev-conmatphys-070909-103925},
  publisher={Annual Reviews}
}

@article{wu2004dynamic,
  title = {{Dynamic Generation of Spin-Orbit Coupling}},
  author = {Wu, Congjun and Zhang, Shou-Cheng},
  journal = {Phys. Rev. Lett.},
  volume = {93},
  issue = {3},
  pages = {036403},
  numpages = {4},
  year = {2004},
  month = {Jul},
  publisher = {American Physical Society},
  doi = {10.1103/PhysRevLett.93.036403},
  url = {https://link.aps.org/doi/10.1103/PhysRevLett.93.036403}
}

@article{wu2007fermi,
  title = {{Fermi liquid instabilities in the spin channel}},
  author = {Wu, Congjun and Sun, Kai and Fradkin, Eduardo and Zhang, Shou-Cheng},
  journal = {Phys. Rev. B},
  volume = {75},
  issue = {11},
  pages = {115103},
  numpages = {25},
  year = {2007},
  month = {Mar},
  publisher = {American Physical Society},
  doi = {10.1103/PhysRevB.75.115103},
  url = {https://link.aps.org/doi/10.1103/PhysRevB.75.115103}
}

@book{vollhardt2013superfluid,
  title={{The Superfluid Phases of Helium 3}},
  author={Vollhardt, Dieter and Wolfle, Peter},
  year={2013},
  publisher={Courier Corporation}
}

@article{balian1963superconductivity,
  title = {Superconductivity with Pairs in a Relative $p$ Wave},
  author = {Balian, R. and Werthamer, N. R.},
  journal = {Phys. Rev.},
  volume = {131},
  issue = {4},
  pages = {1553--1564},
  numpages = {0},
  year = {1963},
  month = {Aug},
  publisher = {American Physical Society},
  doi = {10.1103/PhysRev.131.1553},
  url = {https://link.aps.org/doi/10.1103/PhysRev.131.1553}
}

@book{ITA2002,
  author    = {{International Union of Crystallography}},
  editor    = {Hahn, Theo},
  title     = {International Tables for Crystallography, Volume A: Space-Group Symmetry},
  edition   = {5th revised},
  publisher = {Kluwer Academic Publishers},
  address   = {Dordrecht, Boston, London},
  year      = {2002},
  isbn      = {0-7923-6590-9}
}

@article{litvin2008tables,
  title={Tables of crystallographic properties of magnetic space groups},
  author={Litvin, DB},
  journal={Foundations of Crystallography},
  volume={64},
  number={3},
  pages={419--424},
  doi={10.1107/S010876730800768X},
  year={2008},
  publisher={International Union of Crystallography}
}

@book{bradley2009mathematical,
  title={{The mathematical theory of symmetry in solids: representation theory for point groups and space groups}},
  author={Bradley, Christopher and Cracknell, Arthur},
  year={2009},
  doi={10.1093/oso/9780199582587.001.0001},
  publisher={Oxford University Press}
}

@book{wigner2012group,
  title={Group Theory: And its Application to the Quantum Mechanics of Atomic Spectra},
  author={Wigner, E.},
  isbn={9780323152785},
  series={Pure and applied physics},
  url={https://books.google.com/books?id=ENZzI49uZMcC},
  year={2012},
  publisher={Elsevier Science}
}

@book{toledano1987landau,
author = {Tolédano, J C and Tolédano, P},
title = {The Landau Theory of Phase Transitions},
publisher = {World Scientific},
year = {1987},
doi = {10.1142/0215}
}

@article{landau1936theory,
  title={{The theory of phase transitions}},
  author={Landau, Lev},
  journal={Nature},
  volume={138},
  number={3498},
  pages={840--841},
  year={1936},
  doi={10.1038/138840a0},
  publisher={Nature Publishing Group UK London}
}

@book{landau2013statistical,
author = {L.D. Landau and E.M. Lifshitz},
title = {{Statistical Physics}},
publisher = {Pergamon},
year = {1980},
volume = {5},
series = {Course of Theoretical Physics},
isbn = {978-0-08-023039-9},
doi = {https://doi.org/10.1016/B978-0-08-023039-9.50001-9},
url = {https://www.sciencedirect.com/science/article/pii/B9780080230399500019}
}

@article{Birman1966_LT,
  title = {Simplified Theory of Symmetry Change in Second-Order Phase Transitions: Application to ${\mathrm{V}}_{3}$Si},
  author = {Birman, Joseph L.},
  journal = {Phys. Rev. Lett.},
  volume = {17},
  issue = {24},
  pages = {1216--1219},
  numpages = {0},
  year = {1966},
  month = {Dec},
  publisher = {American Physical Society},
  doi = {10.1103/PhysRevLett.17.1216},
  url = {https://link.aps.org/doi/10.1103/PhysRevLett.17.1216}
}

@article{Goldrich1968_LT,
  title = {Theory of Symmetry Change in Second-Order Phase Transitions in Perovskite Structure},
  author = {Goldrich, Fredric E. and Birman, Joseph L.},
  journal = {Phys. Rev.},
  volume = {167},
  issue = {2},
  pages = {528--532},
  numpages = {0},
  year = {1968},
  month = {Mar},
  publisher = {American Physical Society},
  doi = {10.1103/PhysRev.167.528},
  url = {https://link.aps.org/doi/10.1103/PhysRev.167.528}
}

@article{Jaric1977_LT,
  title = {Group theory of phase transitions in $A\ensuremath{-}15$ ${O}_{h}^{3}\ensuremath{-}\mathrm{Pm}3n$) structure},
  author = {Jari\ifmmode \acute{c}\else \'{c}\fi{}, Marko V. and Birman, Joseph L.},
  journal = {Phys. Rev. B},
  volume = {16},
  issue = {6},
  pages = {2564--2568},
  numpages = {0},
  year = {1977},
  month = {Sep},
  publisher = {American Physical Society},
  doi = {10.1103/PhysRevB.16.2564},
  url = {https://link.aps.org/doi/10.1103/PhysRevB.16.2564}
}

@article{Hatch1981_LT,
  title = {Structural phase transitions in crystals of $R\overline{3}c$ symmetry},
  author = {Hatch, Dorian M.},
  journal = {Phys. Rev. B},
  volume = {23},
  issue = {5},
  pages = {2346--2349},
  numpages = {0},
  year = {1981},
  month = {Mar},
  publisher = {American Physical Society},
  doi = {10.1103/PhysRevB.23.2346},
  url = {https://link.aps.org/doi/10.1103/PhysRevB.23.2346}
}

@InProceedings{Hatch1984_LTIso,
author="Hatch, Dorian M.",
editor="Denardo, G.
and Ghirardi, G.
and Weber, T.",
title="Isotropy groups of space groups --- A simple method for their determination",
booktitle="Group Theoretical Methods in Physics",
year="1984",
publisher="Springer Berlin Heidelberg",
address="Berlin, Heidelberg",
pages="390--393",
isbn="978-3-540-38859-3"
}

@article{Ascher1977_LProb,
doi = {10.1088/0022-3719/10/9/009},
url = {https://doi.org/10.1088/0022-3719/10/9/009},
year = {1977},
month = {may},
publisher = {},
volume = {10},
number = {9},
pages = {1349},
author = {E Ascher and J Kobayashi},
title = {Symmetry and phase transitions: The inverse Landau problem},
journal = {Journal of Physics C: Solid State Physics}
}

@article{Ascher1977_Epikernel,
doi = {10.1088/0022-3719/10/9/010},
url = {https://doi.org/10.1088/0022-3719/10/9/010},
year = {1977},
month = {may},
publisher = {},
volume = {10},
number = {9},
pages = {1365},
author = {E Ascher},
title = {Permutation representations, epikernels and phase transitions},
journal = {Journal of Physics C: Solid State Physics}
}

@Book{DamnjanovicSymm,
  author = 	 {Damnjanovi\'c, M.},
  title = 	 {{Symmetry in Quantum Non-Relativistic Physics}},
  year = 	 2014,
  journal-URL =  "http://www.ff.bg.ac.rs/Katedre/QMF/SiteQMF/pdf/sqnp2e.pdf"
}

@article{Toledano1976_OParamSymm1,
  title = {Order-parameter symmetries for improper ferroelectric nonferroelastic transitions},
  author = {Tol\'edano, Pierre and Tol\'edano, Jean-Claude},
  journal = {Phys. Rev. B},
  volume = {14},
  issue = {7},
  pages = {3097--3109},
  numpages = {0},
  year = {1976},
  month = {Oct},
  publisher = {American Physical Society},
  doi = {10.1103/PhysRevB.14.3097},
  url = {https://link.aps.org/doi/10.1103/PhysRevB.14.3097}
}

@article{Toledano1977_OParSymm2,
  title = {Order-parameter symmetries for the phase transitions of nonmagnetic secondary and higher-order ferroics},
  author = {Tol\'edano, Pierre and Tol\'edano, Jean-Claude},
  journal = {Phys. Rev. B},
  volume = {16},
  issue = {1},
  pages = {386--407},
  numpages = {0},
  year = {1977},
  month = {Jul},
  publisher = {American Physical Society},
  doi = {10.1103/PhysRevB.16.386},
  url = {https://link.aps.org/doi/10.1103/PhysRevB.16.386}
}

@article{Toledano1980_FerroElIso,
  title = {Order parameter symmetries and free-energy expansions for purely ferroelastic transitions},
  author = {Tol\'edano, Jean-Claude and Tol\'edano, Pierre},
  journal = {Phys. Rev. B},
  volume = {21},
  issue = {3},
  pages = {1139--1172},
  numpages = {0},
  year = {1980},
  month = {Feb},
  publisher = {American Physical Society},
  doi = {10.1103/PhysRevB.21.1139},
  url = {https://link.aps.org/doi/10.1103/PhysRevB.21.1139}
}

@article{Toledano1982_NonFerroIso,
  title = {Nonferroic phase transitions},
  author = {Tol\'edano, Pierre and Tol\'edano, Jean-Claude},
  journal = {Phys. Rev. B},
  volume = {25},
  issue = {3},
  pages = {1946--1964},
  numpages = {0},
  year = {1982},
  month = {Feb},
  publisher = {American Physical Society},
  doi = {10.1103/PhysRevB.25.1946},
  url = {https://link.aps.org/doi/10.1103/PhysRevB.25.1946}
}

@article{Jaric1982_GT&PT,
title = {Group theory and phase transitions},
journal = {Physica A: Statistical Mechanics and its Applications},
volume = {114},
number = {1},
pages = {550-556},
year = {1982},
issn = {0378-4371},
doi = {https://doi.org/10.1016/0378-4371(82)90348-X},
url = {https://www.sciencedirect.com/science/article/pii/037843718290348X},
author = {Marko V. Jarić}
}

@book{StokesHatch1989_230SG,
author = {Stokes, H T and Hatch, D M},
title = {Isotropy Subgroups of the 230 Crystallographic Space Groups},
publisher = {WORLD SCIENTIFIC},
year = {1989},
doi = {10.1142/0751},
address = {},
edition   = {},
URL = {https://www.worldscientific.com/doi/abs/10.1142/0751},
eprint = {https://www.worldscientific.com/doi/pdf/10.1142/0751}
}

@article{Petricek2010_SGRepFriends,
author = "Pet{\v{r}}{\'\i}{\v{c}}ek, V{\'{a}}clav and Fuksa, Ji{\v{r}}{\'\i} and Du{\v{s}}ek, Michal",
title = "{Magnetic space and superspace groups, representation~analysis: competing or friendly concepts?}",
journal = "Acta Crystallographica Section A",
year = "2010",
volume = "66",
number = "6",
pages = "649--655",
month = "Nov",
doi = {10.1107/S0108767310030527},
url = {https://doi.org/10.1107/S0108767310030527}
}

@article{StokesHatch2011_SGIso,
  title = {Complete listing of order parameters for a crystalline phase transition: A solution to the generalized inverse Landau problem},
  author = {Hatch, Dorian M. and Stokes, Harold T.},
  journal = {Phys. Rev. B},
  volume = {65},
  issue = {1},
  pages = {014113},
  numpages = {4},
  year = {2001},
  month = {Dec},
  publisher = {American Physical Society},
  doi = {10.1103/PhysRevB.65.014113},
  url = {https://link.aps.org/doi/10.1103/PhysRevB.65.014113}
}

@article{Stokes2017_IsoSuperspace,
author = "Stokes, Harold T. and Campbell, Branton J.",
title = "{A general algorithm for generating isotropy subgroups in superspace}",
journal = "Acta Crystallographica Section A",
year = "2017",
volume = "73",
number = "1",
pages = "4--13",
month = "Jan",
doi = {10.1107/S2053273316017629},
url = {https://doi.org/10.1107/S2053273316017629}
}

@article{Liu2022_SPG,
  title = {Spin-Group Symmetry in Magnetic Materials with Negligible Spin-Orbit Coupling},
  author = {Liu, Pengfei and Li, Jiayu and Han, Jingzhi and Wan, Xiangang and Liu, Qihang},
  journal = {Phys. Rev. X},
  volume = {12},
  issue = {2},
  pages = {021016},
  numpages = {19},
  year = {2022},
  month = {Apr},
  publisher = {American Physical Society},
  doi = {10.1103/PhysRevX.12.021016},
  url = {https://link.aps.org/doi/10.1103/PhysRevX.12.021016}
}

@article{xiao2024ssg,
  title = {{Spin Space Groups: Full Classification and Applications}},
  author = {Xiao, Zhenyu and Zhao, Jianzhou and Li, Yanqi and Shindou, Ryuichi and Song, Zhi-Da},
  journal = {Phys. Rev. X},
  volume = {14},
  issue = {3},
  pages = {031037},
  numpages = {33},
  year = {2024},
  month = {Aug},
  publisher = {American Physical Society},
  doi = {10.1103/PhysRevX.14.031037},
  url = {https://link.aps.org/doi/10.1103/PhysRevX.14.031037}
}

@article{Litvin1977_SpinPoint,
author = "Litvin, D. B.",
title = "{Spin point groups}",
journal = "Acta Crystallographica Section A",
year = "1977",
volume = "33",
number = "2",
pages = "279--287",
month = "Mar",
doi = {10.1107/S0567739477000709},
url = {https://doi.org/10.1107/S0567739477000709},
}

@article{Litvin1974_SpinGroups,
  title={Spin groups},
  author={Litvin, Daniel B and Opechowski, W},
  journal={Physica},
  volume={76},
  number={3},
  pages={538--554},
  year={1974},
  doi={10.1016/0031-8914(74)90157-8},
  publisher={Elsevier}
}

@article{Schiff2025_SPG,
	title={{The crystallographic spin point groups and their representations}},
	author={Hana Schiff and Alberto Corticelli and Afonso Guerreiro and Judit Romhányi and Paul McClarty},
	journal={SciPost Phys.},
	volume={18},
	pages={109},
	year={2025},
	publisher={SciPost},
	doi={10.21468/SciPostPhys.18.3.109},
	url={https://scipost.org/10.21468/SciPostPhys.18.3.109},
}

@article{Chen2024_SSG_Reps,
  title = {{Enumeration and Representation Theory of Spin Space Groups}},
  author = {Chen, Xiaobing and Ren, Jun and Zhu, Yanzhou and Yu, Yutong and Zhang, Ao and Liu, Pengfei and Li, Jiayu and Liu, Yuntian and Li, Caiheng and Liu, Qihang},
  journal = {Phys. Rev. X},
  volume = {14},
  issue = {3},
  pages = {031038},
  numpages = {33},
  year = {2024},
  month = {Aug},
  publisher = {American Physical Society},
  doi = {10.1103/PhysRevX.14.031038},
  url = {https://link.aps.org/doi/10.1103/PhysRevX.14.031038}
}

@article{Lazic2013_spinLine,
  title = {Spin line groups},
  author = {Lazic, Natasa and Milivojevic, Marko and Damnjanovic, Milan},
  journal = "Acta Crystallographica Section A",
year = "2013",
volume = "69",
number = "6",
pages = "611--619",
month = "Nov",
doi = {10.1107/S0108767313022642},
url = {https://doi.org/10.1107/S0108767313022642},
}

@article{jiang2024ssg,
  title = {{Enumeration of Spin-Space Groups: Toward a Complete Description of Symmetries of Magnetic Orders}},
  author = {Jiang, Yi and Song, Ziyin and Zhu, Tiannian and Fang, Zhong and Weng, Hongming and Liu, Zheng-Xin and Yang, Jian and Fang, Chen},
  journal = {Phys. Rev. X},
  volume = {14},
  issue = {3},
  pages = {031039},
  numpages = {25},
  year = {2024},
  month = {Aug},
  publisher = {American Physical Society},
  doi = {10.1103/PhysRevX.14.031039},
  url = {https://link.aps.org/doi/10.1103/PhysRevX.14.031039}
}

@ARTICLE{brinkman1966theory,
       author = {{Brinkman}, W.~F. and {Elliott}, R.~J.},
        title = "{Theory of Spin-Space Groups}",
      journal = {Proceedings of the Royal Society of London Series A},
         year = 1966,
        month = oct,
       volume = {294},
       number = {1438},
        pages = {343-358},
          doi = {10.1098/rspa.1966.0211}
}

@ARTICLE{Brinkman1966b,
       author = {{Brinkman}, W. and {Elliott}, R.~J.},
        title = "{Space Group Theory for Spin Waves}",
      journal = {Journal of Applied Physics},
         year = 1966,
        month = mar,
       volume = {37},
       number = {3},
        pages = {1457-1459},
          doi = {10.1063/1.1708514}
}

@ARTICLE{Kitz1965,
       author = {{Kitz}, A.},
        title = "{{\"U}ber die Symmetriegruppen von Spinverteilungen}",
      journal = {Physica Status Solidi B Basic Research},
         year = 1965,
        month = jan,
       volume = {10},
       number = {2},
        pages = {455-466},
          doi = {10.1002/pssb.2220100206}
}

@article{corticelli2022spin,
  title = {Spin-space groups and magnon band topology},
  author = {Corticelli, A. and Moessner, R. and McClarty, P. A.},
  journal = {Phys. Rev. B},
  volume = {105},
  issue = {6},
  pages = {064430},
  numpages = {28},
  year = {2022},
  month = {Feb},
  publisher = {American Physical Society},
  doi = {10.1103/PhysRevB.105.064430},
  url = {https://link.aps.org/doi/10.1103/PhysRevB.105.064430}
}

@book{coey2021magnetism,
  title={{Handbook of Magnetism and Magnetic Materials}},
  editor={J. M. D. Coey and Stuart S.P. Parkin},
  year={2021},
  publisher={Springer Cham},
  doi={10.1007/978-3-030-63210-6}
}

@article{Messio2011_RMO,
  title = {Lattice symmetries and regular magnetic orders in classical frustrated antiferromagnets},
  author = {Messio, L. and Lhuillier, C. and Misguich, G.},
  journal = {Phys. Rev. B},
  volume = {83},
  issue = {18},
  pages = {184401},
  numpages = {15},
  year = {2011},
  month = {May},
  publisher = {American Physical Society},
  doi = {10.1103/PhysRevB.83.184401},
  url = {https://link.aps.org/doi/10.1103/PhysRevB.83.184401}
}

@article{Etxebarria2025_STENSOR,
author = "Etxebarria, Jesus and Perez-Mato, J. Manuel and Tasci, Emre S. and Elcoro, Luis",
title = "{Crystal tensor properties of magnetic materials with and without spin{--}orbit coupling. Application of spin point groups as approximate symmetries}",
journal = "Acta Crystallographica Section A",
year = "2025",
volume = "81",
number = "4",
pages = "317--338",
month = "Jul",
doi = {10.1107/S2053273325004127},
url = {https://doi.org/10.1107/S2053273325004127}
}

@article{Watanabe2024_Disentangling,
  title = {Symmetry analysis with spin crystallographic groups: Disentangling effects free of spin-orbit coupling in emergent electromagnetism},
  author = {Watanabe, Hikaru and Shinohara, Kohei and Nomoto, Takuya and Togo, Atsushi and Arita, Ryotaro},
  journal = {Phys. Rev. B},
  volume = {109},
  issue = {9},
  pages = {094438},
  numpages = {25},
  year = {2024},
  month = {Mar},
  publisher = {American Physical Society},
  doi = {10.1103/PhysRevB.109.094438},
  url = {https://link.aps.org/doi/10.1103/PhysRevB.109.094438}
}

@article{GonzalezHernandez2021_SpinCond,
  title = {Efficient Electrical Spin Splitter Based on Nonrelativistic Collinear Antiferromagnetism},
  author = {Gonz\'alez-Hern\'andez, Rafael and \ifmmode \check{S}\else \v{S}\fi{}mejkal, Libor and V\'yborn\'y, Karel and Yahagi, Yuta and Sinova, Jairo and Jungwirth, Tom\'a\ifmmode \check{s}\else \v{s}\fi{} and \ifmmode \check{Z}\else \v{Z}\fi{}elezn\'y, Jakub},
  journal = {Phys. Rev. Lett.},
  volume = {126},
  issue = {12},
  pages = {127701},
  numpages = {6},
  year = {2021},
  month = {Mar},
  publisher = {American Physical Society},
  doi = {10.1103/PhysRevLett.126.127701},
  url = {https://link.aps.org/doi/10.1103/PhysRevLett.126.127701}
}

@article{Xiao2026_TensorSymm,
title = {TensorSymmetry: a package to get symmetry-adapted tensors disentangling spin-orbit coupling effect and establishing analytical relationship with magnetic order},
journal = {Computer Physics Communications},
volume = {318},
pages = {109872},
year = {2026},
issn = {0010-4655},
doi = {10.1016/j.cpc.2025.109872},
url = {https://www.sciencedirect.com/science/article/pii/S0010465525003741},
author = {Rui-Chun Xiao and Yuanjun Jin and Zhi-Fan Zhang and Zi-Hao Feng and Ding-Fu Shao and Mingliang Tian}
}

@article{Zelezny2017_spinPolarizedCurrent,
  author  = {{\v{Z}}elezn{\'y}, Jakub and Zhang, Yang and Felser, Claudia and Yan, Binghai},
  title   = {Spin-Polarized Current in Noncollinear Antiferromagnets},
  journal = {Phys. Rev. Lett.},
  volume  = {119},
  issue   = {18},
  pages   = {187204},
  year    = {2017},
  doi     = {10.1103/PhysRevLett.119.187204}
}

@article{Kimata2019_magneticSHE,
  author  = {Kimata, Motoi and Chen, Hua and Kondou, Kouta and Sugimoto, Satoshi and Muduli, Prasanta K. and Ikhlas, Muhammad and Omori, Yasutomo and Tomita, Takahiro and MacDonald, Allan H. and Nakatsuji, Satoru and Otani, Yoshichika},
  title   = {Magnetic and magnetic inverse spin {Hall} effects in a non-collinear antiferromagnet},
  journal = {Nature},
  volume  = {565},
  number  = {7741},
  pages   = {627--630},
  year    = {2019},
  doi     = {10.1038/s41586-018-0853-0}
}

@article{Mook2020_magneticSHE,
  author  = {Mook, Alexander and Neumann, Robin R. and Johansson, Annika and Henk, J{\"u}rgen and Mertig, Ingrid},
  title   = {Origin of the magnetic spin {Hall} effect: Spin current vorticity in the {Fermi} sea},
  journal = {Phys. Rev. Research},
  volume  = {2},
  pages   = {023065},
  year    = {2020},
  doi     = {10.1103/PhysRevResearch.2.023065}
}

@article{Zhang2018_SHEnoSOC,
  author  = {Zhang, Yang and {\v{Z}}elezn{\'y}, Jakub and Sun, Yan and van den Brink, Jeroen and Yan, Binghai},
  title   = {Spin {Hall} effect emerging from a noncollinear magnetic lattice without spin-orbit coupling},
  journal = {New J. Phys.},
  volume  = {20},
  number  = {7},
  pages   = {073028},
  year    = {2018}
}

@article{Bhowal2024_dwave,
  title = {Ferroically Ordered Magnetic Octupoles in $d$-Wave Altermagnets},
  author = {Bhowal, Sayantika and Spaldin, Nicola A.},
  journal = {Phys. Rev. X},
  volume = {14},
  issue = {1},
  pages = {011019},
  numpages = {19},
  year = {2024},
  month = {Feb},
  publisher = {American Physical Society},
  doi = {10.1103/PhysRevX.14.011019},
  url = {https://link.aps.org/doi/10.1103/PhysRevX.14.011019}
}

@article{Bhowal2024_triakontadipolar,
  title = {Nonrelativistic ferromagnetotriakontadipolar order and spin splitting in hematite},
  author = {Verbeek, X. H. and Voderholzer, D. and Sch\"aren, S. and Gachnang, Y. and Spaldin, N. A. and Bhowal, S.},
  journal = {Phys. Rev. Res.},
  volume = {6},
  issue = {4},
  pages = {043157},
  numpages = {10},
  year = {2024},
  month = {Nov},
  publisher = {American Physical Society},
  doi = {10.1103/PhysRevResearch.6.043157},
  url = {https://link.aps.org/doi/10.1103/PhysRevResearch.6.043157}
}

@article{HayamiMultipole1,
author = {Kusunose ,Hiroaki and Oiwa ,Rikuto and Hayami ,Satoru},
title = {Complete Multipole Basis Set for Single-Centered Electron Systems},
journal = {Journal of the Physical Society of Japan},
volume = {89},
number = {10},
pages = {104704},
year = {2020},
doi = {10.7566/JPSJ.89.104704},
url = {https://doi.org/10.7566/JPSJ.89.104704}
}

@article{HayamiMultipole2,
  title = {Bottom-up design of spin-split and reshaped electronic band structures in antiferromagnets without spin-orbit coupling: Procedure on the basis of augmented multipoles},
  author = {Hayami, Satoru and Yanagi, Yuki and Kusunose, Hiroaki},
  journal = {Phys. Rev. B},
  volume = {102},
  issue = {14},
  pages = {144441},
  numpages = {24},
  year = {2020},
  month = {Oct},
  publisher = {American Physical Society},
  doi = {10.1103/PhysRevB.102.144441},
  url = {https://link.aps.org/doi/10.1103/PhysRevB.102.144441}
}

@article{HayamiMultipole3,
author = {Hayami ,Satoru and Kusunose ,Hiroaki},
title = {Unified Description of Electronic Orderings and Cross Correlations by Complete Multipole Representation},
journal = {Journal of the Physical Society of Japan},
volume = {93},
number = {7},
pages = {072001},
year = {2024},
url = {https://doi.org/10.7566/JPSJ.93.072001}
}

@article{Radaelli2024_tensorial_approach,
  title = {Tensorial approach to altermagnetism},
  author = {Radaelli, Paolo G.},
  journal = {Phys. Rev. B},
  volume = {110},
  issue = {21},
  pages = {214428},
  numpages = {19},
  year = {2024},
  month = {Dec},
  publisher = {American Physical Society},
  doi = {10.1103/PhysRevB.110.214428},
  url = {https://link.aps.org/doi/10.1103/PhysRevB.110.214428}
}

@article{clusterMultipoles2017,
  title = {Cluster multipole theory for anomalous Hall effect in antiferromagnets},
  author = {Suzuki, M.-T. and Koretsune, T. and Ochi, M. and Arita, R.},
  journal = {Phys. Rev. B},
  volume = {95},
  issue = {9},
  pages = {094406},
  numpages = {11},
  year = {2017},
  month = {Mar},
  publisher = {American Physical Society},
  doi = {10.1103/PhysRevB.95.094406},
  url = {https://link.aps.org/doi/10.1103/PhysRevB.95.094406}
}

@article{Buiarelli2025_multipolesAM,
  title = {Noncollinear magnetic multipoles in collinear altermagnets},
  author = {Buiarelli, Luca and Fernandes, Rafael M. and Birol, Turan},
  journal = {Phys. Rev. B},
  volume = {112},
  issue = {22},
  pages = {224442},
  numpages = {8},
  year = {2025},
  month = {Dec},
  publisher = {American Physical Society},
  doi = {10.1103/kq6x-7jfc},
  url = {https://link.aps.org/doi/10.1103/kq6x-7jfc}
}

@article{Jahn1949,
author = "Jahn, H. A.",
title = "{Note on the Bhagavantam{--}Suranarayana method of enumerating the physical constants of crystals}",
journal = "Acta Crystallographica",
year = "1949",
volume = "2",
number = "1",
pages = "30--33",
month = "Mar",
doi = {10.1107/S0365110X49000060},
url = {https://doi.org/10.1107/S0365110X49000060},
}

@article{Etxebarria2019_mtensor,
author = "Gallego, Samuel V. and Etxebarria, Jesus and Elcoro, Luis and Tasci, Emre S. and Perez-Mato, J. Manuel",
title = "{Automatic calculation of symmetry-adapted tensors in magnetic and non-magnetic materials: a new tool of the Bilbao Crystallographic Server}",
journal = "Acta Crystallographica Section A",
year = "2019",
volume = "75",
number = "3",
pages = "438--447",
month = "May",
doi = {10.1107/S2053273319001748},
url = {https://doi.org/10.1107/S2053273319001748}
}

@ARTICLE{belashchenko2026,
       author = {{Belashchenko}, K.~D.},
        title = "{Landau theory and exchange instabilities in Mn$_5$Si$_3$: A case against altermagnetism}",
      journal = {arXiv e-prints},
         year = 2026,
        month = aug,
          eid = {arXiv:2608.13483},
        pages = {arXiv:2608.13483},
          doi = {10.48550/arXiv.2608.13483},
archivePrefix = {arXiv},
       eprint = {2608.13483},
 primaryClass = {cond-mat.mtrl-sci}
}

\end{document}